
\magnification = \magstep0

\input eplain.tex

\hsize=18.6truecm   \hoffset=-1.3truecm  \vsize=24.12truecm \voffset=-0.4truecm

 
\def\col#1{\empty} 
\def\bw#1{#1}       

\def\col#1{#1}      
\def\bw#1{\empty}
     

\font\sc=cmr8   \font\sit=cmti8
\font\vsc=cmr7    
\def\mathhexbox#1#2#3{\leavevmode\hbox{$\mathsurround=0pt 
                                       \mathchar"#1#2#3$}}
\def\copyright{{\ooalign{\hfil\raise.07ex\hbox{c}\hfil\crcr\mathhexbox20D}}}
\def\xleft{$\phantom{{\rm Vol.}~0}$} \def\xright{$\phantom{{\rm No.}~0, 199}$}

\def\absbaselines{\baselineskip=11pt \lineskip=0pt \lineskiplimit=0pt}
\def\sglbaselines{\baselineskip=10.4pt \lineskip=0pt \lineskiplimit=0pt}
\def\medbaselines{\baselineskip=10pt \lineskip=0pt \lineskiplimit=0pt}
\def\smlbaselines{\baselineskip=8pt \lineskip=0pt \lineskiplimit=0pt}
\def\vs{\vskip 8pt} \def\vss{\vskip 6pt} \def\vsss{\vskip 2pt}
\parskip = 1pt 
\def\bb{\kern -2pt}  \nopagenumbers
\def\makeheadline{\vbox to 0pt{\vskip-30pt\line{\vbox to8.5pt{}\the
                               \headline}\vss}\nointerlineskip}

\def\footnoterule{\kern-3pt \hrule width \hsize \kern 2.6pt \vskip 3pt}

\def\omit#1{\empty}
\pretolerance=15000  \tolerance=15000
\def\ts{\thinspace}  \def\cl{\centerline}
\def\ni{\noindent}   \def\nhi{\noindent \hangindent=10pt}
                     \def\nnhi{\noindent \hangindent=10pt}
       \def\bk{\kern -0.3em}  \def\b{\kern -0.1em}
\def\r0{$\rho_0$}    
\def\0{\phantom{0}}  \def\1{\phantom{1}}  
\def\etal{{et~al.\ }}  

\def\gapprox{$_>\atop{^\sim}$}  \def\lapprox{$_<\atop{^\sim}$}
\def\ltapprox{\hbox{$<\mkern-19mu\lower4pt\hbox{$\sim$}$}}
\def\gtapprox{\hbox{$>\mkern-19mu\lower4pt\hbox{$\sim$}$}}

\def\mltapprox{\raise2pt\hbox{$<\mkern-19mu\lower5pt\hbox{$\sim$}$}}

\newdimen\sa  \def\sd{\sa=.1em  \ifmmode $\rlap{.}$''$\kern -\sa$
                                \else \rlap{.}$''$\kern -\sa\fi}
              \def\dgd{\sa=.1em \ifmmode $\rlap{.}$^\circ$\kern -\sa$
                                \else \rlap{.}$^\circ$\kern -\sa\fi}
\newdimen\sb  \def\md{\sa=.06em \ifmmode $\rlap{.}$'$\kern -\sa$
                                \else \rlap{.}$'$\kern -\sa\fi}

\def\kms{km~s$^{-1}$}
\def\s{\ifmmode ^{\prime\prime} \else $^{\prime\prime}$ \fi}
\def\min{\ifmmode ^{\prime} \else $^{\prime}$ \fi}
    
\def\m32{M{\ts}32}

\input colordvi
\def\B{\Blue}

\def\R{\Red}

\parindent=0pt

\headline={\leftskip = -0.15in
           \medbaselines\vbox to 0pt{\sc THE ASTROPHYSICAL JOURNAL, 
           000:000 (32pp), 2015 June 31 \hfill\null

           \copyright\/ \vsc 2015. The American Astronomical Society. All Rights
           Reserved. Printed in U.S.A. \hfill\null}}

\cl {\null}  \vs

\sglbaselines

\cl{SCALING LAWS FOR DARK MATTER HALOS IN LATE-TYPE AND DWARF SPHEROIDAL GALAXIES}

\vs

\cl{J{\sc OHN} K{\sc ORMENDY}} \vsss

\cl{\vsc Department of Astronomy, University of Texas at Austin, 
         2515 Speedway, Mail Stop C1400, Austin, TX 78712-1205, USA;
         kormendy@astro.as.utexas.edu} \vss

\cl{\sc AND} \vss

\cl{K.~C.~F{\sc REEMAN}} \vsss

\cl{\vsc Research School of Astronomy and Astrophysics,
         Mount Stromlo Observatory, The Australian National University,} 
         \vskip -2pt
\cl{\vsc Cotter Road, Weston~Creek, Canberra, ACT 2611, Australia;
         Kenneth.Freeman@anu.edu.au} \vsss

\cl{\sit Received 2015, September}
\vs

\parindent = 37pt
\cl {ABSTRACT}
\vss
{\narrower\absbaselines

\ni\quad Published mass models fitted to kinematic data are used to study the
systematic properties of dark matter (DM) halos in Sc -- Im and dwarf spheroidal
(dSph) galaxies.   Halo parameters are derived for \hbox{rotationally supported} 
galaxies by decomposing rotation curves $V(r)$ into visible- 
and dark-matter contributions.  The visible matter potential is calculated
from the surface brightness assuming that the mass-to-light ratio $M/L$ is
constant with radius.  ``Maximum disk'' values of $M/L$ are adjusted to fit 
as much of the inner rotation curve as possible without making the halo have
a hollow core.  Rotation curve decomposition is impossible fainter than
absolute magnitude $M_B \simeq -14$, where $V$ becomes comparable to the
velocity dispersion of the gas.  To increase the luminosity range further,
we include central densities of dSph and dIm galaxies estimated via the Jeans
equation for their stars (dSph) or H{\ts}{\sc I} (dIm).  Combining these data, 
we find that DM halos satisfy well defined scaling laws analogous to the
``fundamental plane'' relations for elliptical galaxies. Halos in less 
luminous galaxies have smaller core radii $r_c$, higher central densities
$\rho_\circ$, and smaller central velocity dispersions $\sigma$.  Scaling laws
provide new constraints on the nature of DM and on galaxy formation and evolution:

\ni\quad 1.~A continuous sequence of decreasing mass extends from the highest-luminosity
Sc{\ts}I galaxies with $M_B \simeq -22.4$ ($H_0 = 70$ km s$^{-1}$ Mpc$^{-1}$) to the 
lowest-luminosity galaxy with sufficient data (an ultrafaint dSph with $M_B \simeq -1$). 
  
\ni\quad 2.~The high DM densities in dSph galaxies are normal for such tiny
galaxies.  Because virialized DM density depends on collapse redshift $z_{\rm coll}$,
$\rho_\circ \propto (1 + z_{\rm coll})^3$, and because the DM densities in the faintest
dwarfs are about $500$ times higher than those in the brightest spirals, therefore the collapse 
redshifts of the faintest dwarfs and the brightest spirals are related by
$(1 + z_{\rm dwarf})/(1 + z_{\rm spiral}) \simeq 8$. 

\ni\quad 3.~The high DM densities in the dSph companions of our Galaxy imply
that they are real galaxies formed from primordial density fluctuations.  They
are not tidal fragments.  Tidal dwarfs cannot retain even the low DM densities
of their giant-galaxy progenitors.  In contrast, dSph galaxies generally have higher DM
densities than those of possible giant-galaxy progenitors.

\ni\quad 4.~We show explicitly that spiral, irregular, and spheroidal galaxies with
$M_V$ \gapprox\ts\ts$-18$ form a sequence of decreasing baryon-to-DM surface density 
with decreasing luminosity.  We suggest that these dS, dIm, and dSph galaxies form a
sequence of decreasing baryon retention (caused by supernova-driven winds) or decreasing
baryon capture (after cosmological reionization) in smaller galaxies.

\ni\quad 5.~In all structural parameter correlations, dS$+$Im and dSph galaxies
behave similarly; any differences between them are small.  We conclude that the difference
between $z \sim 0$ galaxies that still contain gas (and that still can form stars) and
those that do not (and that cannot form stars) is a second-order effect.  The primary
effect appears to be the physics that controls baryon depletion.

\ni\quad 6.~Rotation-curve decompositions reveal a robust, linear correlation between the
maximum rotation velocity $V_{\rm circ,disk}$ of baryonic disks and the outer circular
velocity $V_{\rm circ}$ of test particles in their DM halos.  An important new result is
that $V_{\rm circ,disk} \rightarrow 0$ km s$^{-1}$ at $V_{\rm circ} = 42 \pm 4$ km s$^{-1}$.
Smaller galaxies are dim or dark.  Examples are the dSph companions of our Galaxy and of M{\ts}31.

\ni\quad 7.~ The DM correlations provide a way to estimate the baryon loss and the dynamical
properties of the DM halos of dwarf galaxies.  Our analysis allows us to determine the DM
central density $\rho_\circ$ for the dwarf galaxies but not their DM core radius $r_c$ or DM velocity 
dispersion $\sigma$. If we then assume that dSph$+$dIm galaxies lie on the extrapolation of the 
DM $\rho_\circ$\ts--\ts$M_B$ correlation for brighter galaxies, we find that dSph galaxies were 
originally brighter by $\Delta M_B \simeq 4$ mag and dIm galaxies  by $\Delta M_B \simeq 3.5$ mag.  
To shift them onto the other DM correlations requires that the core radius of the DM is larger than 
the core radius of the visible matter by $\Delta \log{r_c} \simeq 0.70$ for Sph galaxies and by 
$\Delta \log{r_c} \simeq 0.85$ for dIm galaxies.  And the DM particle velocity dispersion is larger 
than the velocity dispersion of the stars in dSph galaxies by $\Delta \sigma \simeq 0.40$, and it 
is larger than the velocity dispersion of the H\ts{\sc I} in dIm galaxies by $\Delta \sigma \simeq 0.50$.   
This means that these almost-dark dwarfs are more massive than we thought.  The typical velocity 
dispersion of their DM halos is $\sigma \sim 30$ km s$^{-1}$, and the typical circular-orbit rotational 
velocity in these halos is $V_{\rm circ} \sim 42$ km s$^{-1}$, in remarkably good agreement with the 
value of $V_{\rm circ}$ where galaxies get dim.
      
\ni\quad 8.~Therefore the range of visible matter content of $\sigma \sim 30$ km s$^{-1}$ DM halos is large -- 
these galaxies range from objects with $M/L_V \sim 10^1$ and rotation curves that are
well enough measured to allow baryon-DM decomposition to the faintest dSphs with $M/L_V \sim 10^{2-3}$. 
This, together with the fact that, as luminosity decreases, dwarf galaxies become much more numerous and 
much more nearly dominated by DM, suggests that there exists a large population of objects that are even darker --
too dark to be discovered by current techniques.  Such objects are a canonical prediction of cold DM theory.  

\ni\quad 9.~The slopes of the DM parameter correlations provide an independent measure on galactic mass scales 
of the slope~$n$ of the power spectrum  $|\delta_k|^2 \propto k^n$ of primordial density fluctuations. 
We derive $n \simeq -2.0 \pm 0.1$.  This is consistent with the theory of cold DM.

\ni\quad 10.~We confirm earlier results that the projected central densities of DM halos,
$\Sigma_{\circ,\rm DM} \propto \rho_\circ r_c \simeq$ constant from $M_B$\ts$\sim$\ts$-5$ to $-22$.
Constant surface density implies a Faber-Jackson law with halo mass $M_{\rm DM}$\ts$\propto$\ts$\sigma^4$.

\vss


\ni {\it Subject{\ts}headings:}  
{\kern -2pt}dark{\ts}matter{\ts}--{\ts}galaxies:{\ts}formation{\ts}--{\ts}galaxies:{\ts}kinematics{\ts}and{\ts}dynamics{\ts}--{\ts}galaxies:{\ts}structure

}

\parindent = 10pt

\doublecolumns\sglbaselines

\cl{\null}
\vskip -10pt
\cl {1.~\sc INTRODUCTION}
\vss

\headline={\vbox to 0pt{\leftskip = -0.15in 
           KORMENDY \& FREEMAN\xleft \hfill DARK MATTER SCALING LAWS \hfill\phantom{0} \hfill\phantom{0} \hfill \xright\folio}}

      The systematic properties of dark matter (DM) can help us to understand
what it is made of and how it affects galaxy formation and evolution.  Best known 
is the imperfect (Casertano \& van Gorkom 1991) conspiracy (van Albada \& Sancisi 1986;
Sancisi \& van Albada~1987; Faber~1987) between visible and dark matter to make almost flat, 
featureless rotation curves. It results from the need, in order to form stars, for baryons to 
dissipate inside their DM halos until they self-gravitate (Blumenthal \etal 1986; Ryden \& Gunn 1987; 
Flores \etal 1993). Rotation curve decompositions have been published for many galaxies.  
We first use these data to investigate the scaling laws for DM halos of late-type disk galaxies.

      We show that DM halos in less luminous galaxies have smaller core radii
$r_c$, higher central densities $\rho_\circ$, and smaller central velocity
dispersions $\sigma$.  These correlations were first found by Athanassoula \etal
(1987, hereafter ABP), by Kormendy (1988, 1990), and by Kormendy \& Freeman (1996). 
Some of the correlations have been confirmed at least qualitatively by, e.{\ts}g.,
Burkert (1995);
Persic \etal (1996, 1997); 
Verheijen (1997);
Salucci \& Burkert (2000); 
Borriello \& Salucci (2001); 
Begum \& Chengalur (2004);
Graham \etal (2006);
Spano \etal (2008);
Kuzio de Naray \etal (2008);
de Blok \etal (2008), and
Plana \etal (2010). 
The most thorough discussion to date is by Kormendy \& Freeman (2004).  This paper
updates that work.

      DM scaling laws are analogous to the fundamental plane relations for
elliptical galaxies (Lauer 1985; Djorgovski \& Davis 1986, 1987; Faber \etal
1987; Dressler \etal 1987; Djorgovski, de Carvalho, \& Han 1988; see Kormendy
\& Djorgovski 1989 for a review), and they are interesting for the same reason:
they provide new constraints on galaxy formation and evolution. Implications 
are discussed in \S\ts9.

      For the giant galaxies, we estimate the halo parameters by decomposing their
H\ts{\sc I} rotation curves $V(r)$ into contributions from their visible matter and their DM halos
(Carignan \& Freeman,~1985; van Albada et al.~1985).  As rotation velocities become smaller
for fainter galaxies (Tully \& Fisher 1977), our faintest galaxies with
absolute magnitudes $M_B \gg -14$  become pressure supported and 
rotation curve decompositions are no longer possible. We can still derive the central
densities of the DM halos from their density and velocity dispersion profiles via
the Jeans equations, but we cannot directly measure the DM core radii $r_c$ and velocity dispersions
$\sigma$ for these faint systems. In this paper, we use both rotation and 
velocity dispersion data to study the properties of DM halos over a large range of galaxy luminosities.

      Only Sc -- Im and dwarf spheroidal (dSph) galaxies~are included, except in Figure 10.
We omit earlier-type (E -- Sbc) galaxies because of the effects of their bulge components.
(1) The rotation curve decompositions for these galaxies involve two visible components with 
different geometries and different unknown mass-to-light ratios.  They are therefore 
less reliable.  (2) Baryonic gravitational compression of the DM can significantly modify 
the halo DM density distribution when the visible mass density is high (e.{\ts}g.,
Sellwood \& McGaugh 2005).  Although many Sa -- Sbc galaxies do satisfy the DM correlations, 
others deviate in the sense of small $r_c$ and large $\rho_\circ$  (ABP; Kormendy 1988, 1990; Jardel \etal 2011).
This is especially true for elliptical galaxies (Thomas \etal 2009; Thomas 2010; Bender \etal 2014). We cannot
be sure that Sc -- Im galaxies are unaffected by baryonic compression, but their halos are 
the most nearly ``pristine'' (that is, uncompressed) that we are able to measure.

\vss\vsss\vsss\vsss
\cl {2.~\sc MEASUREMENT OF DM HALO PARAMETERS}
\cl {\sc BY ROTATION CURVE DECOMPOSITION}
\vss

\vss
\cl {2.1.~\it Technique}
\vss

      The H\ts{\sc I} rotation curves of giant spirals typically extend out to a few tens of kpc
in radius. Over this region, we model the DM halo as a non-singular isothermal sphere
with velocity dispersion $\sigma$, central density $\rho_\circ$, and core radius $r_c$. 
These three parameters are not independent; they are related by
$$ \sigma^2 = (4\pi G \rho_\circ r_c^2)/9                     \eqno(1) $$
(e.{\ts}g.,~King 1966), where $G$ is the gravitational constant. In the core of the isothermal sphere 
($r \ll r_c$) the density is roughly constant and the rotation curve of a massless disk is 
$$V(r) \simeq (4\pi G\rho_\circ/3)^{1/2}~r.      \eqno(2)$$    
For $r \gg r_c$, the density $\rho(r)$ of the isothermal sphere $\propto r^{-2}$ and the rotation curve becomes flat,  
$$V(r) \simeq \sqrt 2 \ts\sigma. $$
If we have rotation data only in the inner 
$V \propto r$ part of the isothermal sphere, then we can measure $\rho_\circ$ but not $r_c$ 
or $\sigma$. Therefore $\rho_\circ$ is often the only halo parameter that we can measure
for low-luminosity galaxies in which the rotation data do not extent to large radii. On
the other hand, if the measurements extend out into the flat part of the rotation curve, 
then all three halo parameters can be estimated. 

     In low-luminosity Sc -- Im galaxies, the visible matter contributes only a small fraction of the total mass. 
In brighter systems, the visible matter appears to dominate the radial potential gradient in the inner parts 
of~the~disk, and a multicomponent mass model (stars{\ts}+{\ts}gas{\ts}+{\ts}DM) is needed.  The shape of the rotation curve 
due to the visible stellar matter is calculated from the observed surface brightness distribution assuming 
that the mass-to-light ratio $M/L$ is constant with radius. Values of $M/L$ are adjusted to fit as much of 
the inner rotation curve as desired.  
The contribution from the H{\ts}{\sc I} gas, including a correction for He, is calculated separately.  
Molecular gas is assumed to follow the light distribution (e.{\ts}g.,~Regan \etal 2001),
so its contribution is included in the adopted $M/L$ ratio.  Given the total rotation curve $V_{\rm vis}$ 
due to the potential of the visible matter, the DM halo rotation curve is $V_{DM}(r)=(V^2 - V_{\rm vis}^2)^{1/2}$. 
A mass model, such as the non-singular isothermal sphere, is then fitted to $V_{DM}$ to derive the halo asymptotic 
velocity $V_\infty= \sqrt2\sigma$, $r_c$,~and~$\rho_\circ$.

      Rotationally supported late-type dwarfs are especially suitable for rotation curve decomposition.  
They are easy to measure:~they are rich in H{\ts}{\sc I}, and the gas extends to large radii.  They are easy to 
interpret: they contain a disk and a halo but not a bulge.  They are interesting: they give leverage to 
the derivation of any correlations between DM parameters and luminosity.  And they are relevant:  the visible 
matter density is low, so it has not greatly modified the halo parameters.

\vss\vskip 1pt
\cl {2.2.~\it Our Assumptions}
\vsss\vskip 1pt

       Lake \& Feinswog (1989) point out that few observations of rotation curves reach large 
enough radii to determine halo parameters if measurement errors are interpreted strictly.  
Further assumptions are required.

      1 -- We assume that rotation curves that become flat stay flat at larger radii.  

      2 --  The mass-to-light ratio of the disk is unknown and cannot be determined
from the rotation curve alone.  As van Albada et al.~(1985) showed, halo parameters and
disk M/L ratios are degenerate.  A wide range of DM and disk parameters can give a 
good fit to the observed rotation curve.  As the amount of visible mass is reduced, the 
required central DM density increases and its core radius decreases.  Their extreme models
(van Albada et al.~1985, Figures 4 and 8) provided almost equally acceptable fits to the rotation curve. 
At one extreme is a maximum disk, with the largest M/L that does not require 
the halo to have a hollow core. This assumption gives the minimum $\rho_\circ$ for the halo and the 
maximum value of $r_c$ consistent with the rotation curve. The other extreme has a disk 
$M/L = 0$.  

      Fortunately, we have additional constraints.~Disk masses cannot be arbitrarily small.  The presence of 
bars and spiral density waves requires that disks be self-gravitating.  ABP used Toomre's (1981) swing 
amplifier theory and the properties of the observed spiral structure (i.{\thinspace}e.,~two arms but not one) 
to constrain the density of the disk. For 18 of 21 Sc\thinspace--{\thinspace}Im galaxies, the outcome 
favored maximum disk decompositions. We adopt maximum disk decompositions, recognizing that this is still
a controversial subject. 
      
      Many authors try to settle the question of maximal~vs. submaximal disks using a range of 
dynamical observations and arguments. Some evidence favors maximum disks (e.{\thinspace}g., Visser 1980; 
Taga \& Iye 1994; Sackett 1997; Bosma 1999; Debattista \& Sellwood 1998, 2000; Sellwood \& Moore 1999; 
Salucci \& Persic 1999; Weiner, Sellwood, \& Williams 2001; Athanassoula 2004;  Weiner 2004). Other evidence 
suggests that some disks are submaximal (e.{\thinspace}g., Bottema 1993, 1997; Courteau \& Rix 1999; 
Herrmann \& Ciardullo 2009; Bershady \etal 2011; van der Kruit \& Freeman\ts2011).  Our choice 
of maximum disks affects only giant galaxies; dwarfs are so dominated by DM that $M/L$ uncertainties 
have little effect.  With submaximal disks, DM parameter correlations with galaxy 
luminosity would be shallower.  See Section 9.11 for further arguments.

      3 -- For the DM halo model, we adopt the non-singular (cored) isothermal sphere (Section\ts2.1). 
Ryder{\ts}et{\ts}al.{\ts}(2004) and de Blok (2010) review the well known collision (Moore 1994) between  the cuspy 
central density profiles $\rho(r)$ seen in CDM simulations and the observational evidence that dwarf galaxies 
(at least) have flat cores. The Navarro \etal (1996, 1997: NFW) profile, with its $\rho \propto r^{-1}$ cusp 
at small radii, has its roots in 
dark-matter-only simulations and is still widely used. See Diemand \& Moore (2011) for a review of cusp 
structure.  CDM simulation papers widely agree that CDM halos are expected to have cuspy centers rather 
than isothermal-like cores (e.{\ts}g., 
Moore \etal 1998, 2001;
Klypin \etal 2001, 2011;
Navarro \etal 2004, 2010;
Col\'\i n \etal 2004;
Hayashi \etal 2004;
Diemand, Moore, \& Stadel 2004;
Diemand \etal 2005, 2008;
Klypin, Trujillo-Gomez, \& Primack 2011).     

      de~Blok (2010) reviews the controversial observational situation. The observed
rotation curves, especially for fainter galaxies, usually indicate that their DM halos have cores.  Some authors 
have tried to ``save cusps'' by emphasizing systematic uncertainties like beam-smearing in the H\ts{\sc I} rotation data 
and non-circular motions in the higher-resolution optical rotation data; these could create the appearance of a core. 
Over the past decade, rotation curve data and mass modeling have improved, and cores in the DM halos of dwarf and
low-surface-brightness galaxies are consistently favored (e.{\ts}g.,
Marchesini \etal 2002;
de Blok \& Bosma 2002;
de Blok \etal 2008;
Weldrake \etal 2003;
Gentile \etal 2004, 2009;
Spano \etal 2008;
Oh \etal 2008, 2011a,{\ts}b;
Donato \etal 2009;
Plana \etal 2010, 
Chen \& McGaugh 2010, and
Walker \etal 2010).
de Blok (2010) concludes that almost all dwarf disk galaxies that have been studied have cored halos.  
The situation is less clear for giant disks like M{\ts}101. A careful review by Sellwood (2009) concludes 
that the core-cusp problem is unsolved for giant galaxies. We note, however, that hydrodynamical 
simulations of galaxy formation with feedback (e.{\ts}g.,~Governato et al.~2010) show that violent feedback from 
rapid star formation in the inner regions of disk galaxies can transform a cusped DM halo into a cored halo.   
We adopt cored isothermal halos in this paper.

      The dSph galaxies are in pressure equilibrium and the dynamical analysis is based on stellar radial 
velocities rather than rotation. Many authors use Jeans equation analyses. Another promising approach is 
Schwarzschild (1979, 1982) orbit superposition modeling of the observed spatial and velocity distributions 
of individual stars in dSph galaxies.  The Nuker team orbit superposition~code (e.{\ts}g.,
Gebhardt \etal 2003;
Richstone \etal 2004;
Thomas \etal 2004, 2005;
Siopis \etal 2009)
has been applied to six dSph companions of our Galaxy by Jardel and Gebhardt.  They conclude that these galaxies 
have a heterogeneous range of DM profiles: dSphs with cores include Fornax (Jardel \& Gebhardt 2012) and probably 
Carina (Jardel \& Gebhardt 2013); dSphs with cuspy, NFW-like profiles include Draco (Jardel \etal 2013) and Sextans 
(Jardel \& Gebhardt 2013),
and Sculptor is uncertain but most consistent with an NFW profile (Jardel \& Gebhardt 2013);
see Jardel (2014) for a summary.  However, (1) the features in the line-of-sight velocity distributions that 
differentiate between best-fit NFW profiles and best-fit cored halos are far from obvious in these papers; (2)~in~the 
formally NFW galaxies, cored halos are excluded by not much more than 1 sigma, and (3) the cored halo machinery 
used in this paper provides an adequate fit to all the data.  Other authors using different analysis techniques 
favor cored halos for dSph galaxies (e.{\ts}g., Amorisco \& Evans 2012).

In Section 6.1, Figure 5 and Appendix A, we show that the baryons in both dSph and dIm galaxies mostly 
have Gaussian density profiles, as expected for an isothermal, isotropic distribution of test particles lying 
within the constant-density core of a cored DM distribution.   The consistency of these assumptions encourages 
us to adopt cored halo models for all galaxies.

      In the present paper, we do not further discuss the issue of halo cores and cusps, other
than to note again that baryonic physics can affect CDM density distributions and produce cores when none were 
present initially.  The most promising mechanism is rapid ejection of most baryons via star-formation feedback 
(Navarro, Eke, \& Frenk 1996;
Gnedin \& Zhao 2002;
Governato \etal 2010; 
Oh \etal 2011a;
de Souza \etal 2011;
Madau, Shen, \& Governato 2104;
De Cintio \etal 2014).  
The idea is especially compelling for tiny dwarfs.  
Surface brightnesses decrease rapidly with 
decreasing galaxy luminosity.  Galaxies like Draco and UMi are so DM-dominated that if the present stars 
were spread out into primordial gas, that gas would not self-gravitate enough to make those stars.  The 
suggestion has long been that these galaxies contained more baryons when the stars formed and then blew 
most of the baryons away via the first supernovae (see Kormendy 1985, 1987b; Kormendy \etal 2009; Kormendy \& Bender 2012
for the observations and Dekel \& Silk 1986 for the theoretical concepts). Observational evidence that is 
consistent with this picture is presented in Section\ts8.

     For the purposes of this paper, the difference between cored and cusped DM profiles is 
relatively benign.  The history of work on the luminous density profiles of elliptical galaxies is 
relevant here.  Ground-based observations of the central parts of ellipticals were made during the 1980s 
and revealed results that include the fundamental plane correlations (Kormendy 1984; Lauer 1985; Kormendy 1987b,{\thinspace}c). 
Then the {\it Hubble Space Telescope\/} (HST) showed that the brightest ellipticals have cores with surface 
brightnesses $\Sigma(r) \propto r^{-m}$ at small radii, with
$m \simeq 0$ to 0.25 (e.{\thinspace}g., Lauer et al.~1995).  Earlier ground-based observations of core radii 
and central densities proved to probe the same physics as HST observations of break radii and densities 
(Kormendy et al.~1994).  Thus, most of the ground-based results remained valid.  In the future, when the details 
of halo density profiles are better known, we expect that our DM parameters will continue to measure the 
relevant physics.  Our core radius may turn out to be a measure of a break radius in the halo density profile, 
and the central density will probably be a measure of some average density inside the break radius, independent 
of whether the halos have flat cores or not.

\vss\vsss
\cl {2.3.~\it Galaxy Selection Criteria}    
\vss

      For rotation-dominated galaxies, DM halo parameters are taken from rotation curve decompositions 
in the literature. While some authors use the non-singular isothermal
sphere model for their dark halos, most others use a simple analytic (pseudo-isothermal) 
sphere. These models are significantly different, and we need to derive the transformations between DM halo 
parameters from the two models (Section\ts3), using the non-singular isothermal 
sphere as our reference model.  Other than this transformation, we have made as few changes as possible, 
adjusting the data to a uniform distance scale and adopting the following selection criteria:

   1 -- As mentioned above, we include only late-type galaxies (Sc -- Im) and dSph systems.  It now seems 
likely that dSph galaxies are related to late-type galaxies. The structural
parameters of their visible matter are similar to those of the lowest-luminosity Im galaxies 
(Kormendy 1985, 1987c; 
Binggeli \& Cameron 1991; 
Ferguson \& Binggeli 1994;
Kormendy \& Bender 2012;
see also Figure 9 of this paper).  
Most of the classical nearby dSph galaxies show evidence of star formation in the past 1 -- 8 Gyr (e.{\ts}g., 
Aaronson \& Mould 1985; 
Mighell 1990; 
Mighell \& Butcher 1992; 
Lee \etal 1993; 
Beauchamp \etal 1995; 
Weisz 2011; see 
Da Costa 1994;
Mateo 1998, and
Tolstoy, Hill, \& Tosi 2009
for reviews). At the time of their most recent star formation, they would have been classified
as irregular galaxies (Kormendy \& Bender 1994, 2012).  See also conclusion 5 in the Abstract
and Section\ts9.6 in this paper.  We therefore include \hbox{Sc -- Im} and dSph galaxies in the same
parameter correlation diagrams.

      2 --  Most galaxies with inclinations $i$\ts$<$\ts$40^\circ$ are excluded. As 
Broeils (1992) remarks,  H{\thinspace}I rotation curves are less accurate for galaxies
that are nearly face-on.  Similar criteria were used by, e.{\ts}g., Begeman (1987) and 
de Blok \etal (2008). Oval distortions 
(Bosma 1978; 
Kormendy 1982;
Kormendy \& Kennicutt 2004)
can lead to incorrect estimates of inclinations for the more face-on galaxies. We retained
four $i < 40^\circ$ galaxies from ABP (M{\thinspace}101, NGC 5236, NGC 6946, and IC 342) because they 
provide leverage at high luminosities. Although we did apply this inclination cut to our sample, it turns 
out that most nearly face-on galaxies satisfy the DM parameter correlations.

      3 -- The rotation curves must extend to large enough radii to constrain the DM parameters.  
The adopted criterion is that the observed rotation curve extends to at least 4.5 exponential scale lengths
of the disk. For a self-gravitating exponential disk, the rotation curve peaks at 2.2 scale lengths (Freeman 1970), 
so the rotation curve contribution from the disk drops significantly over 4.5 scale lengths.   A flat 
observed rotation curve then gives useful constraints on the DM parameters.  The particular choice of 4.5 scale
lengths allows us again to keep a few galaxies that provide leverage at high luminosities.  This radius criterion 
was not rigidly applied: we retained a few galaxies whose rotation data extend out to 3.5 -- 4.4 scale lengths,
because the DM halo was dominant enough so that halo parameters were well determined.  These galaxies are
DDO 127, DDO~154, DDO 168, NGC 247, NGC 925, and IC 2574.  We found that the radius cut is important.  
Without it, the sample is larger and mostly consistent with the correlations, but with larger scatter than 
seen in Figures 2 -- 4 and 6 -- 7. This selection criterion is an important difference between the present work 
and previous investigations of DM scaling laws.

      4 -- In the late stages of the selection process, various ``sanity checks''
on physically plausible decompositions resulted in the omission of additional
objects:

      (i) Some recent papers do not use maximum disk $V(r)$ decompositions but rather 
use disk mass-to-light ratios based on optical or infrared colors (e.{\ts}g., using Bell \& de 
Jong 2001).  These decompositions are usually plausible, but we need to be consistent in using 
maximum disk decompositions if we hope to measure the scatter in the DM parameter correlations.  

      (ii) A few galaxies (e.{\ts}g., UGC 5716: van Zee \etal 1997) were omitted because the adopted  
mass-to-light ratio was much too large for a blue, star-forming disk.  

      (iii) Several authors model the DM halo with the pseudo-isothermal-sphere (PITS: see
Section 3). The PITS has a well defined central density but differs significantly in its structure 
and rotation curve from the nonsingular isothermal sphere. Galaxies were omitted when the published PITS 
$\rho_\circ$ and $r_c$ values implied asymptotic outer rotation velocities that are much larger than the 
outermost velocities that are actually observed.  

      (iv) NGC 1705 (Meurer \etal 1998) and NGC 2915 (Meurer \etal 1996; Elson \etal 2010) 
were omitted because (1) intense starbursts drive strong winds that may affect the inner 
rotation curves; (2) H{\ts}{\sc I} beam smearing in NGC 1705 affects the central $V(r)$ point that, 
in essence, determines $M/L_B$, and (3) both galaxies are sometimes classified as amorphous (I0) 
rather than as Magellanic irregulars (Im). See however Elson, de Blok \& Kraan-Korteweg (2010, 2013).

      (v) The edge-on galaxy NGC 801 is classified Sc in de Vaucouleurs \etal (1991, hereafter RC3), 
but photometry in Kormendy \& Bender (2011, see Figures S6 and S7) shows that it has a pseudobulge  
that makes up 23\ts\% of the total light.  This is clearly evident in the photometry 
of Kent (1986), Andredakis \& Sanders (1994), and Courteau (1996).  Including the pseudobulge 
in any decomposition makes the results very uncertain.  NGC 801 was the lowest-$\rho_\circ$, 
largest-$r_c$ point in Figure 4 of Kormendy \& Freeman (2004).  Here, we conclude that NGC 801
is a misclassified Sbc, and (except in Figure 8) we discard it.  

      (vi) Finally, the ultrathin, edge-on Sd galaxy UGC 7321 (O'Brien \etal 2010 and references
therein) was omitted because its rotation curve is better fitted with a NFW DM profile than 
by a cored halo. 

In addition to the selection cuts, we adopt the following procedures
to make parameters from different sources be as consistent with each other
as possible.

      More accurate distances are available now than the Kraan-Korteweg (1986)
Virgocentric flow model distances used in Kormendy \& Freeman (2004).  
We adopt distance measurements and sources as listed in Table 1 (Section~4).
Accurate distances based on primary standard candles (e.{\ts}g., Cepheid variables)
are available for 62\ts\% of our sample.  Another 12\ts\% of our sample galaxies are 
far enough away ($D = 37$ Mpc to 84 Mpc) that distances based on recession velocities 
and a Hubble constant of $H_0 = 70$ km s$^{-1}$ Mpc$^{-1}$ (Komatsu \etal 2011)
are accurate.  For the remainder of our sample, a more complicated strategy
was required (see Notes to Table\ts1).

      Dark matter parameters were corrected to our adopted distances by assuming
that $r_c \propto D$ and that $\rho_\circ \propto D^{-2}$.  This is not strictly correct, 
because gas and dynamical masses scale differently with distance.  But the errors 
are small on the scales of Figures 2 -- 4 and 6 -- 7.  Estimates of $\sigma$ are independent
of galaxy distance.

      Galactic absorption corrections are from Schlegel \etal (1998) as
listed in NED.  Total magnitudes $B_T$ are generally means of RC3 values 
listed in NED, values listed in the main catalog of Hyperleda, and values
given in the ``Integrated photometry'' link at Hyperleda.  For galaxies with
rotation curve decompositions from van Zee \etal (1997), that paper is also
the source of $B_T$.    Absolute magnitudes listed in Table 1
are corrected for internal absorption as in Tully \& Fouqu\'e (1985). 

\vfill\eject

\vss\vsss\vsss\vsss
\cl {3.~\sc MATCHING PSEUDO-ISOTHERMAL MODELS}
\cl {\sc TO THE NON-SINGULAR ISOTHERMAL SPHERE}
\vss

      For rotation curve decompositions, a DM halo model is needed. The discussion
in Section 2.1 was in terms of a non-singular isothermal sphere (ITS) model. Some 
authors use this model; then we use their derived parameters $r_c$, $\rho_\circ$, and 
$\sigma$ without modification. The density distribution and rotation curve of the 
isothermal sphere need to be calculated numerically, so some authors use the ``pseudo-isothermal sphere'' 
(the PITS) model for which the density has the simple analytic form, 
$$\rho\thinspace(r) = \rho_\circ/(1 + r^2/a^2)~. \eqno{(3)} $$
No strong argument favors either the PITS or the ITS, other than that the ITS comes
from a simple isotropic distribution function (e.{{\ts}g.,~King 1966).  However, if we wish to 
use DM halo parameters from different sources, we need to work out how to scale the 
parameters for PITS decompositions to the system of parameters from ITS decompositions.  
While the PITS and ITS have a similar $r^{-2}$ density distribution at large $r$, the PITS 
is otherwise not a good approximation to the ITS, and there is no exact scaling. For example,
Figure\ts1 ({\it left lower panel\/}) shows how the rotation curve of the PITS tends to its value at large $r$ 
from below, while the rotation curve of the ITS approaches its asymptotic value from above.  

Figure 1 shows three possible ways to scale the rotation and velocity dispersion curves 
of the PITS to the ITS. In all three scalings, the central densities of the two models are the
same.  The panels differ in how radii and velocities are scaled.  We will argue that the best 
scaling is intermediate between the middle and right panels of Figure 1. 

In the left panels, the scaling in radius and velocity forces the rotational
velocity and velocity dispersion to be the same at large radii. With this scaling,  the core radius 
$r_c$ of the isothermal is $r_c = 3a/\sqrt{2}$. The velocity dispersions match only at large $r$,
because an isotropic PITS is not isothermal. Its velocity dispersion is
$$
  \sigma^2(r) = V^2_\infty{\thinspace}(1 + x^2) \biggl({{\pi^2}\over{8}} - 
                {{\tan^{-1} x}\over{x}} -
                {{(\tan^{-1} x)^2}\over{2}}\biggr){\thinspace},  \eqno{(4)}
$$
where $x = r/a$.  For most galaxies, the H{\thinspace}{\sc I} distribution does not extend beyond 
about $2.5 r_c$, and in this region the velocity dispersion of the PITS is less than that of 
the ITS. Similarly, the rotation curves match poorly in this region.  We discard this scaling. 

It seems more realistic to adopt a scaling of the PITS such that its rotation curve matches the 
rotation curve of the ITS most closely in the region where we have rotation data ($r\ts\ltapprox\ts 2.5 r_c$). 
In the right panels of Figure 1, we make the velocity dispersion of the ITS equal to the central dispersion 
$\sigma(0)$ of the PITS, where $\sigma(0) = 0.4834 V_\infty$. Then
 $r_c = (0.6837)3a/\sqrt{2} = 1.4503{\thinspace}a$ and $\sigma = 0.4834{\thinspace}V_\infty$.  
This scaling preserves Equation (1) for the isothermal sphere, 
$\sigma^2 = 4 \pi G \rho_\circ r_c^2 / 9$.

\singlecolumn

\cl{\null}\vfill 

\vfill

\includegraphics{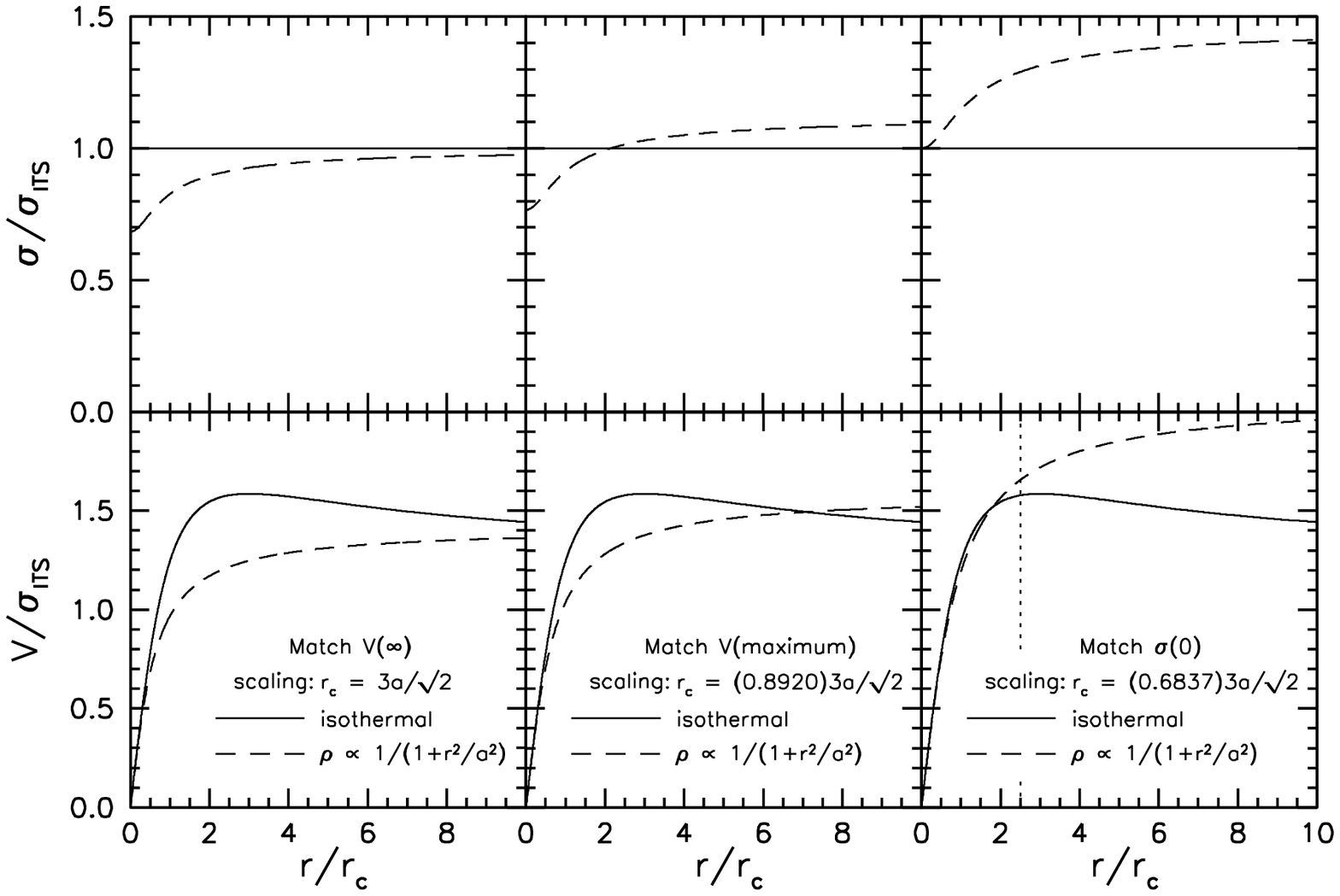} 

      Fig.~1~-- Rotation curves ({\it bottom\/}) and velocity dispersion
profiles ({\it top\/}) for the isothermal and analytic (PITS) halos normalized
to the velocity dispersion $\sigma_{\rm ITS}$ of the isothermal
sphere. The analytic halo is scaled in $r$ and $V$ so that both
models have the same central density and ({\it left\/}) asymptotic outer
rotation velocity, ({\it middle\/}) maximum rotation velocity, and
({\it right\/}) central velocity dispersion.  In the bottom-right
panel, the analytic and isothermal halos have similar rotation curves
out to $r/r_c \simeq 2.5$ ({\it vertical dotted line\/}), i.{\thinspace}e.,
over the radius range of typical HI rotation curves.  However, the two models
extrapolate very differently as $r \rightarrow \infty$.

\cl{\null}

\eject

\doublecolumns

      Although the PITS and ITS rotation curves are similar in the region where
they are fitted to rotation data (Figure\ts1, {\it bottom right\/}), this scaling gives
a value of $V_\infty$ for the PITS that is much larger than the observed velocity
of the flat part of the rotation curve ({\it upper right panel\/}).  We see from the literature 
that authors rarely extrapolate to such large (unobserved) maximum rotational velocities.
This suggests that we should use a scaling as in the middle panel, in which
$V(\infty)$ for the PITS is scaled to the maximum, not the asymptotic, rotation
velocity of the ITS.  Then, when rotation curve decomposition is carried out using Equation (3)
to derive the PITS parameters $\rho_\circ$, $a$, and $V(\infty)$, we would adopt, for the
corresponding ITS decomposition, $r_c = 1.8918{\thinspace}a$, the PITS value of 
$\rho_\circ$, and a halo velocity dispersion $\sigma = 0.6306{\thinspace}V(\infty)$.  
Again, these parameters are related through Equation (1) as they should be for the ITS.

      In the end, after some experimentation, we adopted none of the above scalings:

      Instead, we found enough rotation curve decompositions using the PITS and enough other
decompositions using the ITS so that we could construct the DM parameter correlations 
separately for each model.  By comparing these correlations, we empirically derive the best 
scaling between the two models.  Using this method, Kormendy \& Freeman (2004 derived the scalings 
$\rho_\circ = 0.9255{\thinspace}\rho_{0,\rm PITS}$; 
$r_c = 1.6154{\thinspace}a$,
and 
$\sigma = 0.7334{\thinspace}\sigma_{\rm PITS} 
        = 0.5186{\thinspace}V_{\infty,\rm PITS}$.
In Section 5, we begin with this scaling to construct Figure 2.  We then derive
corrections to the 2004 scaling based on the present galaxy sample 
(equations 14 -- 16).  Combining the 2004 and 2011 results provides the final
scaling between PITS and ITS adopted in this paper: 
$\rho_\circ = 0.8554{\thinspace}\rho_{0,\rm PITS}$; 
$r_c = 1.6591{\thinspace}a$,
and 
$\sigma = 0.7234{\thinspace}\sigma_{\rm PITS} 
        = 0.5115{\thinspace}V_{\infty,\rm PITS}$
(equations 17 -- 19).

\vfill\eject

\singlecolumn

\vskip -10pt

\cl {4.~\sc PARAMETERS OF DM HALO CORES}
\vss

      DM parameters are listed in Table 1.  Columns 2 and 4 are the galaxy distance 
and its absolute magnitude corrected for internal extinction.  The next three columns
give the DM parameters; those derived using the PITS are scaled to the ITS via 
Equations 17 -- 19.  The last two columns give the DM model used and the source of
the rotation curve decomposition.

\vskip -12pt \cl{\null}

%
%
\def\endtable{\endgroup}
\def\tableheight{\vrule width 0pt height 8.5pt depth 3.5pt}
{\catcode`|=\active \catcode`&=\active 
    \gdef\tabledelim{\catcode`|=\active \let|=\vbar
                     \catcode`&=\active \let&=\nobar} }
\def\table{\begingroup
    \def\twidth{\hsize}
    \def\tablewidth##1{\def\twidth{##1}}
    \def\defaultheight{\vrule width 0pt height 8.5pt depth 3.5pt}
    \def\heightdepth##1{\dimen0=##1
        \ifdim\dimen0>5pt 
            \divide\dimen0 by 2 \advance\dimen0 by 2.5pt
            \dimen1=\dimen0 \advance\dimen1 by -5pt
            \vrule width 0pt height \the\dimen0  depth \the\dimen1
        \else  \divide\dimen0 by 2
            \vrule width 0pt height \the\dimen0  depth \the\dimen0 \fi}
    \def\spacing##1{\def\defaultheight{\heightdepth{##1}}}
    \def\nextheight##1{\noalign{\gdef\tableheight{\heightdepth{##1}}}}
    \def\end{\cr\noalign{\gdef\tableheight{\defaultheight}}}
    \def\zerowidth##1{\omit\hidewidth ##1 \hidewidth}    
    \def\hline{\noalign{\hrule}}
    \def\skip##1{\noalign{\vskip##1}}
    \def\bskip##1{\noalign{\hbox to \twidth{\vrule height##1 depth 0pt \hfil
        \vrule height##1 depth 0pt}}}
    \def\header##1{\noalign{\hbox to \twidth{\hfil ##1 \unskip\hfil}}}
    \def\bheader##1{\noalign{\hbox to \twidth{\vrule\hfil ##1 
        \unskip\hfil\vrule}}}
    \def\spanloop{\span\omit \advance\mscount by -1}
    \def\extend##1##2{\omit
        \mscount=##1 \multiply\mscount by 2 \advance\mscount by -1
        \loop\ifnum\mscount>1 \spanloop\repeat \ \hfil ##2 \unskip\hfil}
    \def\vbar{&\vrule&}
    \def\nobar{&&}
    \def\hdash##1{ \noalign{ \relax \gdef\tableheight{\heightdepth{0pt}}
        \toks0={} \count0=1 \count1=0 \putout##1\end 
        \toks0=\expandafter{\the\toks0 &\end} \xdef\piggy{\the\toks0} }
        \piggy}
    \let\e=\expandafter
    \def\putspace{\ifnum\count0>1 \advance\count0 by -1
        \toks0=\e\e\e{\the\e\toks0\e&\e\multispan\e{\the\count0}\hfill} 
        \fi \count0=0 }
    \def\putrule{\ifnum\count1>0 \advance\count1 by 1
        \toks0=\e\e\e{\the\e\toks0\e&\e\multispan\e{\the\count1}\leaders\hrule\hfill}
        \fi \count1=0 }
    \def\putout##1{\ifx##1\end \putspace \putrule \let\next=\relax 
        \else \let\next=\putout
            \ifx##1- \advance\count1 by 2 \putspace
            \else    \advance\count0 by 2 \putrule \fi \fi \next}   }
\def\tablespec#1{
    \def\vdimens{\noexpand\tableheight}
    \def\tabby{\tabskip=0pt plus100pt minus100pt}
    \def\r{&################\tabby&\hfil################\unskip}
    \def\c{&################\tabby&\hfil################\unskip\hfil}
    \def\l{&################\tabby&################\unskip\hfil}
    \edef\templ{\noexpand\vdimens ########\unskip  #1 
         \unskip&########\tabskip=0pt&########\cr}
    \tabledelim
    \edef\body##1{ \vbox{
        \tabskip=0pt \offinterlineskip
        \halign to \twidth {\templ ##1}}} }

\input colordvi
\def\B{\Blue}

\def\R{\Red}

\def\0{$\phantom{0}$}
\def\dot{$\phantom{.}$}

$$
\table
\tablewidth{18.5truecm}
\tablespec{\l\c\l\c\c\c\c\c\l}
\body{
\header{TABLE 1}
\skip{5pt}
\header{DARK MATTER HALO PARAMETERS FOR ROTATIONALLY SUPPORTED GALAXIES}
\skip{10pt}
\hline \skip{0.001truein} \hline 
\skip{.2truecm}\hline \skip{0.001truein} \hline 
\skip{5pt}
& Galaxy     &  \0$D$&Source&\0\dot$M_{B}$& $\log{\sigma}$ &\0$\log{\rho_0}$& $\log{r_c}$       & DM    & Source      & \end
&            &  \0[Mpc] &     &          &  [km s$^{-1}$] &\dot[$M_\odot~{\rm pc}^{-3}$]& [kpc] & Model &             & \end
& (1)        & (2)      & (3) & (4)      & (5)            & (6)                         & (7)   & (8)   & (9)         & \end
\skip{5pt}
\hline \skip{0.001truein} \hline
\skip{5pt}
\B{& NGC   24   &  \08.1\0 & 1   & $-18.10$ &  1.8129   & $-1.8020$ &  0.8242  &  \0ITS  &  Chemin et al.~2006        &} \end \skip{7pt}
\B{& NGC   45   &  \07.1\0 & 1   & $-18.54$ &  1.7404   & $-2.0444$ &  0.8728  &  \0ITS  &  Chemin et al.~2006        &} \end \skip{7pt}
\B{& NGC   55   &  \02.17  & 1   & $-19.25$ &  1.7559   & $-2.4073$ &  1.0719  &  \0ITS  &  Puche et al.~1991         &} \end
\R{& NGC   55   &  \02.17  & 1   & $-19.25$ &  1.8732   & $-2.5263$ &  1.2498  &   PITS  &  Broeils 1992              &} \end \skip{7pt}
\B{& NGC  247   &  \03.65  & 1   & $-19.17$ &  1.8785   & $-2.5158$ &  1.2471  &  \0ITS  &  ABP                       &} \end
\R{& NGC  247   &  \03.65  & 1   & $-19.08$ &  1.8424   & $-2.5836$ &  1.2476  &   PITS  &  Broeils 1992              &} \end \skip{7pt}
\B{& NGC  253   &  \03.94  & 1,2 & $-21.03$ &  2.0734   & $-2.3110$ &  1.3396  &  \0ITS  &  ABP                       &} \end \skip{7pt}
\R{& NGC  925   &  \09.2\0 & 1,2 & $-20.17$ &  1.8813   & $-1.9606$ &  0.9719  &   PITS  &  de Blok et al.~2008       &} \end \skip{7pt}
\B{& NGC  300   &  \02.00  & 1   & $-18.43$ &  1.7348   & $-2.1715$ &  0.9348  &  \0ITS  &  ABP                       &} \end
\R{& NGC  300   &  \02.00  & 1   & $-18.43$ &  1.8294   & $-2.2403$ &  1.0650  &   PITS  &  Broeils 1992              &} \end \skip{7pt}
\B{& NGC  598   &  \00.85  & 1,2 & $-19.00$ &  1.8075   & $-2.1518$ &  0.9937  &  \0ITS  &  ABP                       &} \end
\R{& NGC  598   &  \00.85  & 1,2 & $-19.00$ &  1.9417   & $-2.0961$ &  1.0971  &   PITS  &  Corbelli 2003             &} \end \skip{7pt}
\R{& NGC 1003   &  \09.2\0 & 1   & $-19.15$ &  1.8327   & $-2.3112$ &  1.0986  &   PITS  &  Broeils 1992              &} \end \skip{7pt}
\R{& NGC 1560   &  \03.45  & 1,2 & $-17.36$ &  1.8327   & $-2.2712$ &  1.1131  &   PITS  &  Broeils 1992              &} \end \skip{7pt}
\R{& NGC 2366   &  \03.19  & 1,2 & $-16.53$ &  1.4521   & $-1.6732$ &  0.3990  &   PITS  &  Oh et al.~2011b           &} \end \skip{7pt}
\B{& NGC 2403   &  \03.22  & 1   & $-19.39$ &  1.9025   & $-1.8747$ &  0.9502  &  \0ITS  &  ABP                       &} \end
\R{& NGC 2403   &  \03.22  & 1   & $-19.39$ &  1.8964   & $-2.0379$ &  1.0288  &   PITS  &  Broeils 1992              &} \end \skip{7pt}
\R{& NGC 2998   &   71.1\0 & 3,4 & $-21.76$ &  2.0926   & $-2.8692$ &  1.6375  &   PITS  &  Broeils 1992              &} \end \skip{7pt}
\B{& NGC 3109   &  \01.34  & 1,2 & $-16.44$ &  1.6884   & $-1.9226$ &  0.7600  &  \0ITS  &  Jobin et al.~1990         &} \end
\R{& NGC 3109   &  \01.34  & 1,2 & $-16.44$ &  1.8581   & $-2.1709$ &  1.0561  &   PITS  &  Broeils 1992              &} \end \skip{7pt}
\B{& NGC 3198   &   13.6\0 & 1,5 & $-20.53$ &  1.8976   & $-2.4601$ &  1.2379  &  \0ITS  &Blais-Ouellette et al.~1999 &} \end
\R{& NGC 3198   &   13.6\0 & 1,5 & $-20.53$ &  1.9020   & $-2.4966$ &  1.2611  &   PITS  &  Broeils 1992              &} \end \skip{7pt}
\R{& NGC 3274   &  \06.5\0 & 1,2 & $-16.46$ &  1.6360   & $-0.9209$ &  0.2067  &   PITS  &  de Blok et al.~2002       &} \end \skip{7pt}
\B{& NGC 3359   &   19.5\0 & 3,4 & $-20.90$ &  1.9128   & $-2.6141$ &  1.3300  &  \0ITS  &  ABP                       &} \end \skip{7pt}
\R{& NGC 3621   &   \06.64 & 1   & $-19.99$ &  1.9020   & $-1.9591$ &  0.9919  &   PITS  &  de Blok et al.~2008       &} \end \skip{7pt}
\R{& NGC 3726   &   18.6\0 & 1,6 & $-21.19$ &  1.9367   & $-2.2067$ &  1.1503  &   PITS  &  Verheijen 1997            &} \end \skip{7pt}
\R{& NGC 3741   &  \03.19  & 1   & $-13.86$ &  1.4928   & $-1.7555$ &  0.3958  & Burkert &  Gentile et al.~2007       &} \end \skip{7pt}
\R{& NGC 3917   &   18.6\0 & 1,6 & $-19.82$ &  1.8023   & $-1.8857$ &  0.8554  &   PITS  &  Verheijen 1997            &} \end \skip{7pt}
\R{& NGC 4183   &   18.6\0 & 1,6  & $-19.56$ &  1.7300   & $-2.3661$ &  1.0233  &   PITS  &  Verheijen 1997            &} \end \skip{7pt}
\skip{5pt}
\hline \skip{0.001truein} \hline
\skip{5pt}
}
\endtable
$$

\vfill\eject

$$
\table
\tablewidth{18.5truecm}
\tablespec{\l\c\l\c\c\c\c\c\l}
\body{
\header{TABLE 1 --- \it Continued}
\skip {5pt}
\header{DARK MATTER HALO PARAMETERS FOR ROTATIONALLY SUPPORTED GALAXIES}
\skip{10pt}
\hline \skip{0.001truein} \hline 
\skip{.2truecm}\hline \skip{0.001truein} \hline 
\skip{5pt}
& Galaxy     &  \0$D$&Source&\0\dot$M_{B}$& $\log{\sigma}$&\0$\log{\rho_0}$              & $\log{r_c}$&  DM   & Source &  \end
&            &  \0[Mpc] &     &          &  [km s$^{-1}$] &\dot[$M_\odot~{\rm pc}^{-3}$] & [kpc]      & Model &        &  \end
& (1)        & (2)      & (3) & (4)      & (5)            & (6)                          & (7)        & (8)   & (9)    &  \end
\skip{5pt}
\hline \skip{0.001truein} \hline
\skip{5pt}
\B{& NGC 4244   &  \04.49  & 1,2  & $-18.77$ &  1.7528   & $-2.4294$ &  1.0771  &  \0ITS  &  ABP                       &} \end \skip{7pt}
\B{& NGC 4395   &  \04.61  & 1,2  & $-18.19$ &  1.7889   & $-2.4741$ &  1.1362  &  \0ITS  &  ABP                       &} \end \skip{7pt}
\R{& NGC 4455   &  \07.8\0 & 1    & $-17.49$ &  1.7245   & $-2.0936$ &  0.8815  &   PITS  &  de Blok et al.~2002       &} \end \skip{7pt}
\B{& NGC 5033   &   17.2\0 &3,4,7 & $-21.06$ &  2.0885   & $-2.5221$ &  1.4598  &  \0ITS  &  ABP                       &} \end
\R{& NGC 5033   &   17.2\0 &3,4,7 & $-21.06$ &  1.9393   & $-2.2059$ &  1.1507  &   PITS  &  Broeils 1992              &} \end \skip{7pt}
\B{& NGC 5204   &  \04.65  & 1,2  & $-16.90$ &  1.6435   & $-1.3596$ &  0.4176  &  \0ITS  &  Sicotte et al.~1997       &} \end
\R{& NGC 5204   &  \04.65  & 1,2  & $-16.90$ &  1.6190   & $-1.0419$ &  0.2502  &   PITS  &  Swaters et al.~2003       &} \end \skip{7pt}
\B{& NGC 5236   &  \04.47  & 1,2  & $-20.63$ &  1.9890   & $-2.0209$ &  1.1125  &  \0ITS  &  ABP                       &} \end \skip{7pt}
\B{& NGC 5457   &  \07.2\0 & 2,9  & $-21.41$ &  1.9450   & $-2.4479$ &  1.2775  &  \0ITS  &  ABP                       &} \end \skip{7pt}
\B{& NGC 5585   &  \07.2\0 &1,2,10& $-18.47$ &  1.7202   & $-1.4656$ &  0.5633  &  \0ITS  & Blais-Ouellette et al.~1999&} \end
\R{& NGC 5585   &  \07.2\0 &1,2,10& $-18.47$ &  1.7045   & $-1.4426$ &  0.5360  &   PITS  &  Broeils 1992              &} \end \skip{7pt}
\B{& NGC 5907   &   15.15  & 11   & $-20.82$ &  2.0693   & $-2.4750$ &  1.4170  &  \0ITS  &  ABP                       &} \end
\R{& NGC 5907   &   15.15  & 11   & $-20.82$ &  2.0473   & $-2.1380$ &  1.2246  &   PITS  &  Miller et al.~1995        &} \end \skip{7pt}
\B{& NGC 6015   &   17.7\0 & 3,4  & $-20.12$ &  1.9619   & $-2.3587$ &  1.2511  &  \0ITS  &Verdes-Montenegro et al.~1997&}\end \skip{7pt}
\B{& NGC 6503   &  \05.27  & 1,2  & $-18.57$ &  1.8407   & $-1.7056$ &  0.8057  &  \0ITS  &  ABP                       &} \end
\R{& NGC 6503   &  \05.27  & 1,2  & $-18.57$ &  1.7695   & $-1.3832$ &  0.5714  &   PITS  &  Broeils 1992              &} \end \skip{7pt}
\R{& NGC 6822   &  \00.50  & 1,2  & $-15.67$ &  1.5596   & $-1.5577$ &  0.4488  &   PITS  &  Weldrake et al.~2003      &} \end \skip{7pt}
\B{& NGC 6946   &  \05.9\0 & 1,2  & $-21.06$ &  2.0885   & $-2.0322$ &  1.2166  &  \0ITS  &  ABP                       &} \end \skip{7pt}
\B{& NGC 7793   &  \03.91  & 1,2  & $-18.86$ &  1.6107   & $-1.5467$ &  0.4946  &  \0ITS  &  Carignan et al.~1990      &} \end \skip{7pt}
\B{& IC   342   &  \03.28  & 2,9  & $-20.90$ &  2.0488   & $-2.2782$ &  1.2988  &  \0ITS  &  ABP                       &} \end \skip{7pt}
\B{& IC  2574   &  \04.02  & 1,2  & $-18.24$ &  1.6990   & $-2.4480$ &  1.0302  &  \0ITS  &  Martimbeau et al.~1994    &} \end \skip{7pt}
\R{& UGC  191   &   17.8\0 & 3,4  & $-18.23$ &  1.6773   & $-2.4867$ &  1.0310  &   PITS  &  van Zee et al.~1997       &} \end \skip{7pt}
\B{& UGC 2259   &    \09.2\0 & 1    & $-16.63$ &  1.7404   & $-1.9768$ &  0.8390  &  \0ITS  &  Carignan et al.~1988      &} \end
\R{& UGC 2259   &    \09.2\0 & 1    & $-16.71$ &  1.8456   & $-2.0352$ &  0.9778  &   PITS  &  Broeils 1992              &} \end \skip{7pt}
\R{& UGC 2885   &     84.0\0 & 3,4  & $-22.50$ &  2.2909   & $-3.0104$ &  1.9004  &   PITS  &  Broeils 1992              &} \end \skip{7pt}
\R{& UGC 3174   &    \08.1\0 & 1    & $-15.29$ &  1.5539   & $-1.1367$ &  0.2326  &   PITS  &  van Zee et al.~1997       &} \end \skip{7pt}
\R{& UGC 4499   &     17.2\0 &13,12 & $-18.08$ &  1.6365   & $-2.0805$ &  0.7871  &   PITS  &  Swaters et al.~2003       &} \end \skip{7pt}
\R{& UGC 5005   &     56.0\0 & 3,4  & $-18.00$ &  1.6717   & $-1.9561$ &  0.7572  &   PITS  &  de Blok et al.~1997       &} \end \skip{7pt}
\R{& UGC 5764   &    \07.1\0 & 1    & $-14.47$ &  1.4797   & $-1.3038$ &  0.2418  &   PITS  &  van Zee et al.~1997       &} \end \skip{7pt}
\R{& UGC 6446   &     18.6\0 & 1,6  & $-18.49$ &  1.5781   & $-1.4200$ &  0.4130  &   PITS  &  Verheijen 1997            &} \end \skip{7pt}
\R{& UGC 6983   &     18.6\0 & 1,6  & $-18.76$ &  1.7844   & $-2.6574$ &  1.2233  &   PITS  &  Verheijen 1997            &} \end \skip{7pt}
\R{& UGC 11820  &     18.8\0 & 3,4  & $-17.44$ &  1.6383   & $-2.3430$ &  0.9200  &   PITS  &  van Zee et al.~1997       &} \end \skip{7pt}
\B{& UGC A442   &    \04.27  & 1,2  & $-15.16$ &  1.5441   & $-1.7091$ &  0.5112  &  \0ITS  &  C\^ot\'e et al.~2000      &} \end \skip{7pt}
\skip{5pt}
\hline \skip{0.001truein} \hline
\skip{5pt}
}
\endtable
$$

\vfill\eject

\cl{\null} \vskip -30pt

$$
\table
\tablewidth{18.5truecm}
\tablespec{\l\c\l\c\c\c\c\c\l}
\body{
\header{TABLE 1 --- \it Continued}
\skip{5pt}
\header{DARK MATTER HALO PARAMETERS}
\skip{10pt}
\hline \skip{0.001truein} \hline 
\skip{.2truecm}\hline \skip{0.001truein} \hline 
\skip{5pt}
& Galaxy     &  \0$D$&Source&\0\dot$M_{B}$& $\log{\sigma}$&\0$\log{\rho_0}$              & $\log{r_c}$   &  DM   & Source & \end
&            &  \0[Mpc] &     &          &  [km s$^{-1}$] &\dot[$M_\odot~{\rm pc}^{-3}$] & [kpc]         & Model &        & \end
& (1)        & (2)      & (3) & (4)      & (5)            & (6)                          & (7)           & (8)   & (9)    & \end
\skip{5pt}
\hline \skip{0.001truein} \hline
\skip{5pt}
\R{& ESO 287-G13&     37.5\0 & 3,4  & $-20.90$ &  2.0962   & $-2.4740$ &  1.4429  &   PITS  &  Gentile et al.~2004        &} \end \skip{6pt}
\B{& ESO 381-G20&    \05.44  & 1    & $-15.11$ &  1.5315   & $-2.0820$ &  0.6829  &  \0ITS  &  C\^ot\'e et al.~2000       &} \end \skip{6pt}
\B{& ESO 444-G84&    \04.61  & 1,2  & $-13.82$ &  1.6021   & $-1.2755$ &  0.3501  &  \0ITS  &  C\^ot\'e et al.~2000       &} \end \skip{6pt}
\R{& F568-V1    &     84.1\0 & 3,4  & $-18.40$ &  1.7844   & $-2.3460$ &  1.0677  &   PITS  &  Swaters et al.~2000        &} \end \skip{6pt}
\R{& F583-1,D584-4&   36.9\0 & 3,4  & $-17.35$ &  1.6861   & $-2.0362$ &  0.8145  &   PITS  &  de Blok et al.~2001        &} \end
\R{& F583-1,D584-4&   36.9\0 & 3,4  & $-17.35$ &  1.7049   & $-2.0454$ &  0.8117  &   PITS  &  Kuzio de Naray et al.~2008 &} \end
   & F583-1,D584-4&   36.9\0 & 3,4  & $-17.35$ &  1.6955   & $-2.0408$ &  0.8131  &   PITS  &  \dots                      &  \end \skip{6pt}
\B{& DDO  127   &    \06.9\0 & 1,2  & $-14.98$ &  1.4230   & $-1.9212$ &  0.4941  &  \0ITS  &  Bosma: Kormendy et al.~2004&} \end \skip{6pt}
\B{& DDO  154   &    \04.3\0 & 1,2  & $-14.75$ &  1.4624   & $-1.9142$ &  0.5263  &  \0ITS  &  Carignan et al.~1998       &} \end 
\R{& DDO  154   &    \04.3\0 & 1,2  & $-14.75$ &  1.4797   & $-1.9460$ &  0.5523  &   PITS  &  Broeils 1992               &} \end \skip{6pt}
\B{& DDO  161   &    \07.2\0 & 3,4  & $-16.84$ &  1.6128   & $-2.2648$ &  0.8573  &  \0ITS  &  C\^ot\'e et al.~2000       &} \end \skip{6pt}
\R{& DDO  168   &    \04.33  & 1,2  & $-16.27$ &  1.7001   & $-1.8688$ &  0.7437  &   PITS  &  Broeils 1992               &} \end \skip{6pt}
\R{& DDO  170   &     16.1\0 & 3,4  & $-16.15$ &  1.5839   & $-2.0398$ &  0.7093  &   PITS  &  Broeils 1992               &} \end
\skip{5pt}
\hline \skip{0.001truein} \hline
\skip{5pt}
}
\endtable
$$

\pretolerance=15000  \tolerance=15000
\lineskip=0pt \lineskiplimit=0pt

\vskip -10pt

      NOTES ON DISTANCES -- Adopted distances (Column 2) are from sources listed in Column (3).  To save space, most of our 
sources are compilations for many galaxies rather than the papers in which distance measurements are reported.  Our primary 
source is the Kennicutt \etal (2008) list of galaxies in the local 11-Mpc volume.  For $D \leq 11$ Mpc, we use his distances
whenever possible.  Our second main source is the Karachentsev \etal (2004) Catalog of Neighboring Galaxies.
These two sources almost always agree exactly; for our sample, they never disagree significantly.
For 37 of our 60 galaxies (i.{\ts}e., 62\ts\%), 
we use direct distance measurements in order of preference, as derived from Cepheid variables, from the tip of the red giant 
branch in the color-magnitude diagram, from brightest stars, or from membership in groups whose distances were measured using 
either the above direct methods or (in the case of UGC 4499) using surface brightness fluctuations in NGC 5866.  Beyond this
volume, five galaxies are in the Ursa Major cluster; we adopt $D = 18.6$ Mpc from Kennicutt \etal (2008) and Tully \&
Pierce (2000).  When no direct measurement is available, we base $D$ on a derivation of the large-scale flow field 
of galaxies as given in the listed sources.  For four nearby galaxies, we adopt Kennicutt's flow-field distances based on a Hubble 
constant of $H_0 = 75$ km s$^{-1}$ Mpc$^{-1}$.  We then have a significant gap in distance between these objects at $D = 7$ to 8 
Mpc and galaxies at $D \geq 15$ Mpc for which we use flow-field distances from References (3) and (4).  For the latter galaxies, we
adopt $H_0 \simeq 70$ km s$^{-1}$ Mpc$^{-1}$ consistent with WMAP determinations (Komatsu \etal 2009, 2011). 
It may seem unrealistic to use different Hubble constants for nearby and distant galaxies, but Karachentsev \& Makarov (1996) show
that the apparent value of $H_0$ reaches a maximum at $D \sim 2$ Mpc and then decreases out to larger distances.  Kennicutt 
\etal (2008) adopt the Karachentsev \& Makarov flow-field solution, having engineered the currently best possible consistency between 
direct distance measurements and ones determined from flow fields.  On the other hand, at large distances (seven galaxies have 
$D = 37$ Mpc to 84 Mpc), $H_0$ is robustly smaller; we adopt the WMAP value.  In between 
are 7 galaxies at $D = 15$ Mpc to 19 Mpc
 for which flow-field results are particularly uncertain.  
\vs

      DISTANCE REFERENCES -- (1) Kennicutt \etal (2008); \par
\hskip 135.8pt               (2) Karachentsev \etal (2004); \par
\hskip 135.8pt               (3) Mould \etal (2000) Virgocentric flow field solution as implemented in Reference (4); \par
\hskip 135.8pt               (4) NED ``D (Virgo Infall only)'' with $H_0 = 70$ km s$^{-1}$ Mpc$^{-1}$;\par
\hskip 135.8pt               (5) Freedman \etal (2001); \par
\hskip 135.8pt               (6) Tully \& Pierce (2000); \par
\hskip 135.8pt               (7) Flow field solution (3,4) applied to the Can Ven Spur (group 43 $-1$ in Reference 8); \par
\hskip 135.8pt               (8) Tully (1988); \par
\hskip 135.8pt               (9) Kormendy \etal (2010); \par
\hskip 131.6pt               (10) Drozdovsky \& Karachentsev (2000); \par
\hskip 131.6pt               (11) Flow-field $D = 14.9$ Mpc for group 44 $-1$ (Reference 8) averaged with SBF $D = 15.4$\par
\hskip 131.6pt      \phantom{(12)} Mpc for group member NGC 5866 (Reference~12); \par
\hskip 131.6pt               (12) Tonry \etal (2001); \par
\hskip 131.6pt               (13) In group 15 $-10$ (Ref.~8) with NGC 2681 which has SBF $D = 17.2$ Mpc (Ref.~12). \par
\vss

      NOTES ON DM MODELS -- The dark matter model used in rotation curve decomposition is given in Column (8).  ITS~means the 
nonsingular isothermal sphere; these lines are encoded in blue to match the point colors in Figures 2\ts--\ts4 and 6\ts--\ts8.  PITS means the 
pseudo-isothermal sphere (Equation 3); these lines are in red to match the point colors in \hbox{Figures 2\ts--\ts4 and 6\ts--\ts8.}   NGC 3741 was
analyzed (Gentile \etal 2007) using a Burkert (1995) analytic dark matter profile.  The Burkert profile has a constant-density core 
and a volume density $\rho(r)$ which closely resembles that of the PITS over a factor of $> 20$ in density.  We scaled
its parameters to PITS parameters and then used our standard scaling to convert PITS parameters to ITS equivalents.  The resulting
parameters for NGC 3741 are listed and plotted in red.  All decompositions used in this paper are listed in the table, one per line.  
Their sources are given in Column (9).  When there are multiple sources, the last line for that galaxy (in black) gives the mean 
$\log {r_c}$, $\log {\rho_0}$, and $\log {\sigma}$ for the decompositions listed in the previous lines.  These are illustrated in Figure 3,
but the ITS decomposition results are adopted (see the discussion of Figure 3).
For the 15 galaxies with two independent decompositions -- almost always ones with different DM models -- comparison of the results
gives estimates of the 1 s.~d.~measurement errors in the parameters; their averages are
$\epsilon(\log {\sigma}) = 0.033$, $\epsilon(\log {\rho_0}) = 0.070$, and $\epsilon(\log {r_c}) = 0.063$.  These include errors 
in calibrating PITS to ITS results, but they do not include systematic errors that result if some disks are sub-maximal.  These 
mean parameter errors for the 1/4 of our galaxies that have two decompositions are representative of the errors for galaxies
that have only one decomposition.



\doublecolumns

\vfill\eject

\cl {5.~\sc DM HALO SCALING LAWS}
\cl {\sc FROM ROTATION-CURVE DECOMPOSITION}
\vss

      The correlations between halo $r_c$, $\rho_\circ$, and $\sigma$ and~galaxy
absolute magnitude $M_B$ are shown in Figures 2 -- 4 for the rotationally supported
Sc -- Im galaxies.   Figure~2 updates the Kormendy \& Freeman (2004) calibration that 
we use to convert the results of PITS rotation curve decompositions to our best estimates 
of equivalent parameters derived using isothermal halos.  Figure 3 shows why we adopt 
ITS results when decompositions using ITS and PITS halos are both available.  Figure 4 combines 
all of the data as discussed below.  Figure 6 adds measurements of the faint, pressure-supported
Sph and Im galaxies as derived in Section 6.

      Isothermals and pseudo-isothermals fitted to DM halos measure astrophysically 
equivalent parameters.  However, they differ in detail.  As discussed in Section 3,
we need to calibrate parameters derived using PITS-based rotation curve decompositions 
to those given by decompositions that were made using isothermals.  This calibration
was carried out in Kormendy \& Freeman (2004) by exactly the same method that we 
use below to update the calibration.  We begin with the 2004 calibration,   
\vskip -19pt
$$\eqalignno{
  \rho_{0,\rm ITS,2004} &= 0.9255\;\rho_{0,\rm PITS}\;;                                &(5)\cr
  r_{c,\rm ITS,2004}    &= 1.6154\;a_{\rm PITS}\;;                                     &(6)\cr
  \sigma_{\rm ITS,2004} &= 0.7334\;\sigma_{\rm PITS} = 0.5186\;V_{\infty,\rm PITS}\;.  &(7)\cr
}$$
Here we explicitly identify the parameters derived with the different DM models.
This scaling is intermediate between the middle and right panels of Figure 1.

      Applied to the PITS-based DM parameters derived by the references listed
in Table 1, the 2004 calibration yields the correlations shown in red in the right panels 
of Figure 2.  The ITS-based decompositions yield the correlations shown in blue in the 
left panels of Figure 2.  The ITS-based parameters are as listed in Table 1.

      The scatter about the correlations is slightly smaller in the left panels 
than in the right panels.  This is an early sign (confirmed in Figure\ts3) that decompositions 
based on the isothermal sphere are better behaved.  Such a conclusion is not surprising, given 
the slow rise of the PITS rotation curve to its asymptotic value (Figure\ts1).  However, 
rotation curve decompositions based on isothermal and PITS models separately give essentially 
the same correlations.  We have 32 galaxies with rotation curve decompositions based on 
isothermal DM halos and 41 with decompositions based on the PITS.  Only 14 galaxies are common
to both samples.  So the correlations shown in Figure 2 ({\it left}) and ({\it right}) are 
largely independent.  We want to combine the two samples.  We do so by updating the scaling 
in the correlations in Figure 2 ({\it right}), as follows:

      Least-squares fits to the correlations in Fig.~2 ({\it left}) give:
$$\eqalignno{
  \log{\rho_\circ} &= ~\;\,0.1219\,(M_B + 18) - 2.0320~({\rm \hbox{rms:}}\ts0.30~{\rm dex});~ \rlap{(8)}\cr
\,\log{r_c}\,      &=     -0.1576\,(M_B + 18) + 0.8702~({\rm \hbox{rms:}}\ts0.18~{\rm dex});~ \rlap{(9)}\cr
  \log{\sigma}~    &=     -0.0927\,(M_B + 18) + 1.7462~({\rm \hbox{rms:}}\ts0.08~{\rm dex}).~ \rlap{(10)}\cr
}$$

      Least-squares fits to the correlations in Fig.\ts2 ({\it right})~are:
$$\eqalignno{
  \log{\rho_{0,\rm ITS,2004}}   &= ~\;\,0.1535\,(M_B + 18) - 2.0024~({\rm \hbox{rms:}}\ts0.34~{\rm dex}); \rlap{(11)}\cr
\,\log{r_{c,\rm ITS,2004}}\,    &=     -0.1775\,(M_B + 18) + 0.8587~({\rm \hbox{rms:}}\ts0.23~{\rm dex}); \rlap{(12)}\cr
  \log{\sigma_{\rm ITS,2004}}~  &=     -0.0933\,(M_B + 18) + 1.7522~({\rm \hbox{rms:}}\ts0.10~{\rm dex}). \rlap{(13)}\cr
}$$
The two samples have essentially the same average absolute magnitude: 
${<}M_B{>} = -18.55$ for objects analyzed with isothermals and 
${<}M_B{>} = -18.30$ for those analyzed with the PITS.
In the above, we therefore symmetrized the least-squares fits around $M_B = -18$.  As in Tremaine \etal (2002),
we also symmetrized the other variables, i.{\ts}e., $\log {\rho_\circ}$~around $-2$, $\log {r_c}$ around 0.9,
and $\log {\sigma}$ around 1.8.  Then, requiring that the correlations agree at $M_B = -18$ provides 
a correction to the 2004 scaling of the PITS results to those measured with the isothermal sphere,
$$\eqalignno{
  \rho_{0,\rm ITS} &= 0.9340\;\rho_{0,\rm ITS,2004}\;;                                         &\rlap{(14)}\cr
  r_{c,\rm ITS}    &= 1.0270\;r_{c,\rm ITS,2004}\;;                                            &\rlap{(15)}\cr
  \sigma_{\rm ITS} &= 0.9863\;\sigma_{\rm ITS,2004} = (0.9863) (0.5186)\;V_{\infty,\rm PITS}\;.&\rlap{(16)}\cr
}$$
The change from the 2004 calibration is small.  Combining Equations 5\ts--\ts7 with Equations 14\ts--\ts16 
gives our final calibration,
$$\eqalignno{
  \rho_{0,\rm ITS} &= 0.8554\;\rho_{0,\rm PITS}\;;                                &(17)\cr
  r_{c,\rm ITS}    &= 1.6591\;a_{\rm PITS}\;;                                     &(18)\cr
  \sigma_{\rm ITS} &= 0.7234\;\sigma_{\rm PITS} = 0.5115\;V_{\infty,\rm PITS}\;.  &(19)\cr
}$$
This scaling is also intermediate between the middle and right panels of Figure 1.  Equation 17 has
been tweaked by about 1\ts\% to make the final ITS parameters satisfy exactly the usual relation 
for an isothermal (dropping the ``ITS'' subscripts),
$$
\sigma^2 = {{4 \pi G \rho_\circ r_c^2} \over 9 }~. \eqno{(20)}
$$

\vfill\eject

\singlecolumn

\cl{\null}\vfill

\includegraphics{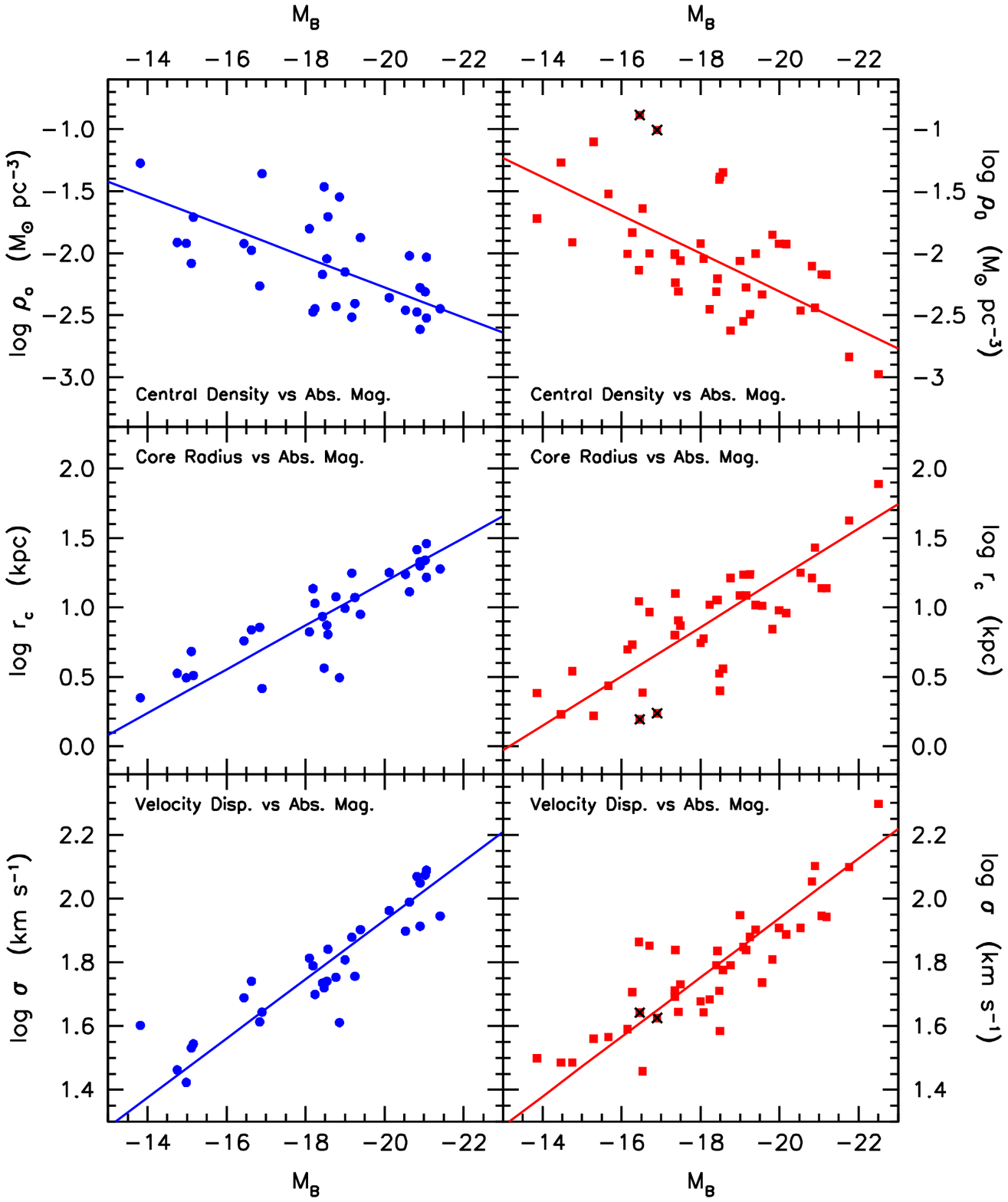}

      Fig.~2~-- Correlations of DM central density $\rho_\circ$ ({\it top panels}),
core radius $r_c$ ({\it middle}), and particle velocity dispersion~$\sigma$
({\it bottom}) with galaxy absolute magnitude $M_B$.  The left panels 
show results derived using isothermal halos (``ITS'') in rotation curve decompositions
of Sc -- Im galaxies ({\it blue points, from blue lines in Table 1\/}).  The right 
panels show results ({\it red points}) derived using pseudo-isothermal 
halo models (``PITS'': Equation 3).  The PITS points are calibrated to the ITS points as in
Kormendy \& Freeman (2004; Equations 5 -- 7 here).  The purpose of this figure is
to correct the 2004 calibration to provide the final calibration that we apply to PITS 
decomposition results to get the parameters that are listed in the red lines in 
Table 1 and that are plotted in Figures 3 -- 8.  To derive the corrections, we calculate 
symmetric, least-squares fits (Tremaine \etal 2002) in each panel ({\it blue and red lines}) 
and then shift the PITS points vertically until the fits agree at $M_B = -18$.  NGC 3274 and
NGC 5204 ({\it overplotted with black X}) are omitted from the PITS fits here to keep the slopes 
of the red and blue lines similar.  They are retained in the final correlation plots and  
fits.  The recalibrations resulting from this Figure are Equations (14\ts--\ts16).
The final, combined 2004 and 2011 calibrations are Equations (17\ts--\ts19).

\eject

\cl{\null}\vfill

\doublecolumns

\cl{\null} \vskip 8.1truein

\includegraphics{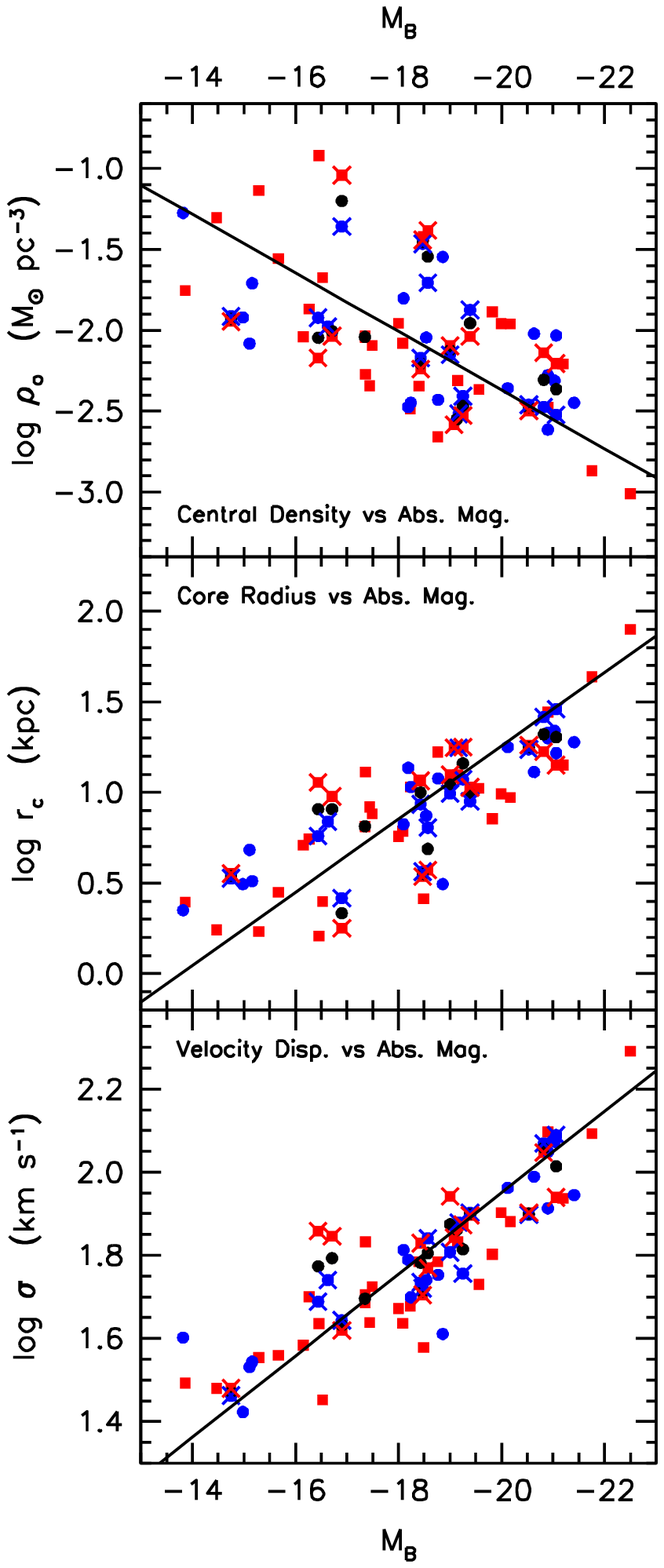}

      Fig.~3~-- Superposition of Figure 2 correlations derived from ITS 
decompositions ({\it blue points}) and from~PITS decompositions ({\it red points}) 
after recalibration of the latter using Equations (14\ts--\ts16).  For 14 galaxies, 
we have both ITS and PITS decompositions.  They are identified with 
blue and red crosses, and their averages are shown in black. 

      Figure 3 superposes the ITS and PITS correlations from Figure 2 after recalibration
of the latter.   For 14 galaxies, we have rotation curve
decompositions available using both halo models.  Their parameters are identified with 
red and blue crosses.  For each of these galaxies, the average of the ITS and rescaled PITS
parameters are shown in black.  It is therefore easy to identify galaxies with two decompositions
by finding triples of red, black, and blue points with the same $M_B$ and with the red and blue
points overplotted with crosses.

      One galaxy, F583-1, has two PITS decompositions.  They agree well.
Their means are also plotted in black.

      Our initial intention was to use the above averages in our further analysis.  But 
Figure 3 shows that, {\it in almost every case when corresponding red and blue points disagree, 
the PITS points deviate more from the correlations.}  (The lines shown are symmetric least-squares 
fits to all the points using the mean parameters when there are two measurements for one galaxy. 
We do not adopt these fits, but we note here that they are almost identical to the fits adopted
in Figure 4.)

      In agreement with our initial impression on the previous page, we therefore conclude that isothermal 
halos provide more reliable decompositions.  So we adopt the ITS results~--~not the mean parameters --
for the above 14 galaxies.  For F583-1, we adopt the means of the two PITS decompositions.

      We do, however, compare the ITS and scaled PITS parameters for the above 14 galaxies plus the two results on F583-1 to
estimate the relative internal measurement errors in the parameters.  Then, the 15 sets of mean
parameters that are plotted in black in Figure 3 have mean standard deviations of 
$\epsilon(\log {\rho_\circ}) = 0.073$,
$\epsilon(\log {r_c})    = 0.064$, and
$\epsilon(\log {\sigma}) = 0.033$.  
These values are used only as measurement errors that are input into the least-squares fits in
Figure 4.  Only their relative values are important.

      Figure 4 then shows the final DM correlation results, combining the ITS and scaled PITS
parameters.  When both are available, we use results based on isothermal halos.  
Based on 59 galaxies spanning 9 magnitudes~in~$M_B$, the correlations are quite robust.  
Uncertainties in rotation curve decomposition are unlikely to threaten~them, although their
slopes are still uncertain.  The Figure~4 correlations are essentially identical to those found
in Kormendy \& Freeman (2004).  Present results are more reliable for two reasons.  New rotation
curve decompositions published since 2004 have been added to our sample.  More importantly,
accurate distances based on primary standard candles (Cepheid variables, the tip of the red giant 
branch in the color-magnitude diagram, brightest stars, and the zero point of surface brightness 
fluctuations) are now available for 2/3 of our galaxies.

      As in Figures 2 and 3, we fitted straight lines to the correlations in Figure 4 following
the precepts of Tremaine \etal (2002).  The two variables in each fit are treated symmetrically.
Each variable is normalized approximately around its mean; the actual values used are given above
Equations 14 -- 16.  Internal measurement errors as derived above are input, and the external
scatter is estimated and iterated until the reduced $\chi^2 = 1$.  This scatter (Figure 4 caption)
is more likely to reflect heterogeneous data and decomposition procedures than it is likely to be 
a measure of any intrinsic, astrophysical scatter.

\eject

\cl{\null}\vfill

\headline={\vbox to 0pt{\leftskip = -0.15in 
           KORMENDY \& FREEMAN\xleft \hfill DARK MATTER SCALING LAWS \hfill\phantom{0} \hfill\phantom{0} \hfill \xright\folio}}

\singlecolumn

\includegraphics{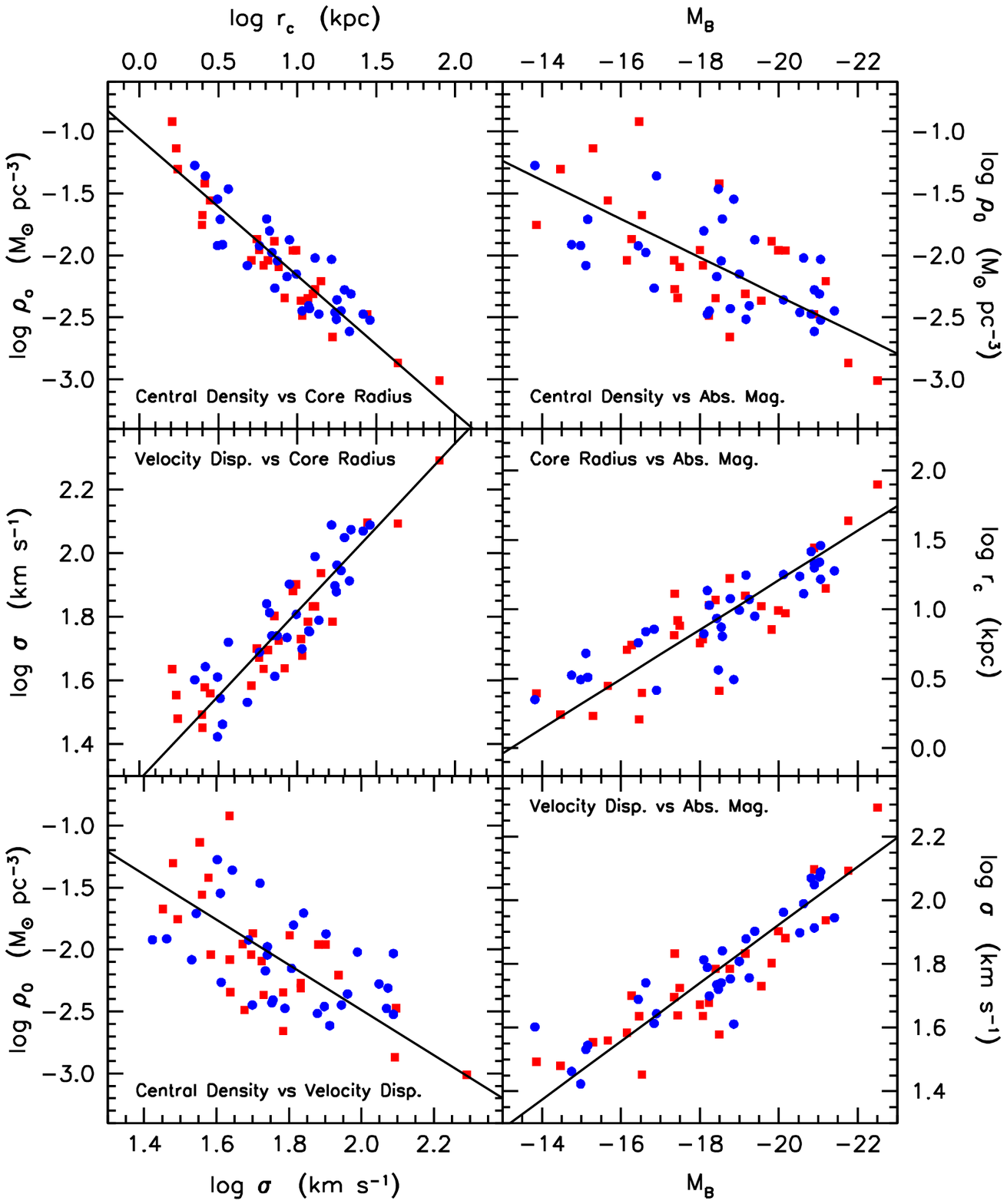} 

      Fig.~4~-- Final dark matter parameter correlations for the rotationally supported
Sc{\thinspace}--{\thinspace}Im
galaxies after recalibration of PITS results ({\it red points}) to those derived using 
isothermal halos ({\it blue points}).  The lines are symmetric, least-squares fits (Tremaine 
\etal 2002) to all of the points.  We estimate measurement errors by comparing the DM parameters 
for 15 galaxies with two independent decompositions (see text).  The mean standard deviations are
$\epsilon(\log {r_c})    = 0.064$, 
$\epsilon(\log {\rho_\circ}) = 0.073$, and 
$\epsilon(\log {\sigma}) = 0.033$.  
The intrinsic scatter in each parameter is then estimated and iterated until we get the most 
consistent possible values that imply a reduced $\chi^2$ = 1 in all six fits.  These intrinsic scatter values are
$\sigma(M_B)           = 0.77 \pm 0.03$,
$\sigma(\log {r_c})    = 0.09 \pm 0.02$, 
$\sigma(\log {\rho_\circ}) = 0.24 \pm 0.05$, and 
$\sigma(\log {\sigma}) = 0.06 \pm 0.01$. 
These values are probably dominated by the different assumptions (e.{\ts}g., on disk
mass-to-light ratios) made in the heterogeneous collection of rotation curve decomposition papers. 
Nevertheless, they are upper limits on the true intrinsic scatter in the DM correlations.
The final fits are Equations (21\ts--\ts26).

\eject

 As derived, the least-squares fits to the points in Figure 4 are:
$$\eqalignno{
  \log{\rho_\circ}   + 2.0  &=     -(1.1089 \pm 0.0659)\,(\log{r_c} - 0.9)     \;- (0.0535 \pm 0.0235)~~~~({\rm rms}  = 0.18~{\rm dex})\;; &(21)\cr
  \log{\sigma}\; - 1.8  &= ~\;\,(0.5280 \pm 0.0347)\,(\log{r_c} - 0.9)     \;- (0.0369 \pm 0.0119)~~~~({\rm rms}  = 0.09~{\rm dex})\;; &(22)\cr
  \log{\rho_\circ}   + 2.0  &=     -(1.8212 \pm 0.2738)\,(\log{\sigma}~ - 1.8)\ts- (0.1225 \pm 0.0461)~~~\;({\rm rms} = 0.35~{\rm dex})\;; &(23)\cr
  \log{\rho_\circ}   + 2.0  &= ~\;\,(0.1554 \pm 0.0230)\,(M_B + 18)~~\ts\;       - (0.0163 \pm 0.0447)~~~~({\rm rms}  = 0.34~{\rm dex})\;; &(24)\cr
\,\log{r_c}\,    - 0.9  &=     -(0.1782 \pm 0.0142)\,(M_B + 18)~~\ts\;       - (0.0461 \pm 0.0277)~~~~({\rm rms}  = 0.21~{\rm dex})\;; &(25)\cr
  \log{\sigma}~  - 1.8  &=     -(0.0915 \pm 0.0058)\,(M_B + 18)~~\ts\;       - (0.0604 \pm 0.0118)~~~~({\rm rms}  = 0.09~{\rm dex})\;, &(26)\cr
}$$
where $r_c$ is measured in kpc, $\rho_\circ$ is measured in $M_\odot$ pc$^{-3}$, and $\sigma$ is measured in km s$^{-1}$. \vs

      In physically more transparent terms,
$$\eqalignno{
  \rho_\circ &= \biggl(0.0136^{+0.0017}_{-0.0015}~M_{\odot}~{\rm pc}^{-3}\biggr)\biggl({L_B \over 10^9~L_{B\odot}}\biggr)^{-0.388 \pm 0.057};      &(27)\cr
  \noalign{\vskip 5pt}
   r_c   &= \biggl(4.80^{+0.35}_{-0.33}~{\rm kpc}\biggr)\biggl({L_B \over 10^9~L_{B\odot}}\biggr)^{0.446 \pm 0.035};                           &(28)\cr
  \noalign{\vskip 5pt}
  \sigma &= \biggl(44.8^{+1.4}_{-1.3}~{\rm km~s}^{-1}\biggr)\biggl({L_B \over 10^9~L_{B\odot}}\biggr)^{0.229 \pm0.014};~{\rm i.{\thinspace}e.,}&(29)\cr
  \noalign{\vskip 10pt}
   L_B   &\propto \sigma^{4.37 \pm 0.28};                                                                           &(30)\cr
  \noalign{\vskip 6pt}
%
%
  \rho_\circ &= \biggl(0.00685^{+0.00040}_{-0.00037}~M_{\odot}~{\rm pc}^{-3}\biggr)\biggl({r_c \over 10~{\rm kpc}}\biggr)^{-1.109 \pm 0.066};           &(31)\cr
  \noalign{\vskip 5pt}
  \rho_\circ &= \biggl(0.00326^{+0.00058}_{-0.00050}~M_{\odot}~{\rm pc}^{-3}\biggr)\biggl({\sigma \over 100~{\rm km~s}^{-1}}\biggr)^{-1.821 \pm 0.274}; &(32)\cr
  \noalign{\vskip 5pt}
  \sigma &= \biggl(65.5^{+1.9}_{-1.8}~{\rm km~s}^{-1}\biggr)\biggl({r_c \over 10~{\rm kpc}}\biggr)^{0.528 \pm 0.035}\;.                             &(33)\cr
}$$

Here, we have normalized parameters in terms of convenient round numbers that are, however, slightly different 
from the symmetrization values used in Equations (21) -- (26).  For example (Cox 2000), 
$10^9\;L_{B\odot}$ is $M_B = -17.03$, not $M_B = -18$.  The result is that we give up a little of the advantage 
(to the zeropoint errors) in symmetrization; that is, the zero point errors in Equations (27) -- (33) are slightly 
larger than those in Equations (21) -- (26) (see Tremaine \etal 2002 for a discussion).  On the other hand, the equations 
are much easier to use, and the formal fitting errors are still small and in any case smaller than the systematic 
effects of different possible choices (e.{\ts}g., non-maximal disks) in the machinery of rotation curve decomposition.     

\doublecolumns

\vss\vsss\vsss\vsss\vskip 1.3pt
\cl {6.~\sc CENTRAL DM DENSITIES OF DWARF GALAXIES}
\vss

       The faint dwarf spheroidal (dSph) and dwarf Magellanic irregular (dIm) galaxies 
provide an opportunity to increase the range of galaxy luminosities over which we can examine
at least the central densities of DM halos. If we can include the ultrafaint dSph 
galaxies with absolute magnitudes as faint as $M_B = -1$, then the luminosity baseline for our DM 
scaling laws increases to about 22 magnitudes. These tiny dwarfs do not rotate significantly.
Instead, their stars are supported by their random velocities.  So our tracers of the dynamics are the
distribution and velocity dispersion of the stars.  Similarly, in the smallest dIm galaxies, the HI is 
supported by its velocity dispersion and provides the tracer. 
We will see that the method for estimating the central densities of the DM distribution in these two 
kinds of dwarfs is similar, although the tracers are quite different.

      Early work by Aaronson (1983) and by Aaronson \& Olszewski (1987) on stars in the Draco and UMi 
dSph galaxies showed that their velocity dispersions are much larger than expected if they are 
equilibrium systems that consist only of old, metal-poor stars.  Their high velocity dispersions imply 
mass-to-light ratios of $\sim$\ts100, not the globular-cluster-like $M/L$ ratios of $\sim$\ts2 that are
expected for an old, metal-poor stellar population.  It was already apparent at the time that their 
central DM densities $\rho_\circ$\ts$\sim$\ts0.6{\ts}to\ts1\ts$M_\odot${\ts}pc$^{-3}$ were ``shockingly high.  
Indeed, these are the highest central DM densities seen in any galaxy so far" (Kormendy 1987a).  This was an 
early indication of the DM correlations shown in this paper. 

      Much effort went into trying to find an escape from the conclusion that dSph galaxies are dark matter dominated. 
Early observational concerns about measurement errors, atmospheric effects in the most luminous stars, and unrecognized 
binary stars were gradually laid to rest.  Now we have large samples of accurate radial velocities in many dSph 
galaxies: see Walker et al (2009) for a compilation of dispersion data for classical and ultrafaint dSph galaxies.  

      Concerns were also raised about possible dynamical effects that might mimic the effects of dark matter:  

      Some authors argue that dwarf spheroidal galaxies may be descendants of dark-matter-free tidal 
dwarfs that were made in interactions of larger galaxies (Yang{\ts}et{\ts}al.\ts2014).  This discussion continues.~We 
do not review it here.  We adopt the conventional interpretation that dSph galaxies are dominated by DM when
their velocity dispersions are too high to be consistent with old stellar populations.

As the number of dSph galaxies with dispersion measurements has increased, escape routes that depend on 
rare events have become implausible. These include the suggestion (Kuhn \& Miller 1989; Kuhn 1993) that 
the stars formerly in dSph galaxies are unbound because of Galactic tides, so we overestimate the masses 
of systems that are far from equilibrium.  The required orbital resonance works best if the dispersion is
 only marginally larger than the escape velocity and if not too many systems need special engineering. 
 But $M/L_V$ ratios of 10 -- 100 (not~2 !) imply velocities that are inflated by factors of $\sim 2 - 6$.  
Piatek \& Pryor (1995),  Oh, Lin, \& Aarseth (1995), Sellwood \& Pryor (1998), Klessen, Grebel, \& Harbeck (2003),
and Wilkinson \etal (2004) argue convincingly that tides do not inflate velocity dispersions this much,
especially not without producing detectable velocity gradients across the galaxies. This remains true even 
though apparently extratidal stars have been seen in some dSphs (e.{\thinspace}g., Irwin \& Hatzidimitriou 1995; 
Piatek et al. 2001, 2002; Palma et al.~2003).  And some dSphs are too far from our Galaxy to be affected 
by tides (Mateo \etal 1998; Mateo 1998; Koch \etal 2007).  All dSph galaxies with dynamical analyses now 
appear to be DM dominated.  It has become difficult to argue that we get fooled by special circumstances.

      One DM alternative is not addressed by the above arguments -- Modified Newtonian Dynamics (MOND: 
Milgrom 1983a, b, c; 
Milgrom \& Bekenstein 1987). 
While MOND continues to be debated (see
Sanders \& McGaugh 2002, 
Sellwood 2004 
for reviews), we adopt conventional Newtonian gravity and treat measurements of high velocity dispersions 
in dSph galaxies as detections of DM.

      Large samples of accurate stellar velocities are now available for many dSph galaxies. We use them to 
estimate central DM densities via the Jeans equation.  For the smallest dIm galaxies, rotation is negligible 
and the HI gas is supported mainly by its turbulent velocity dispersion.  An HI study of the faint system GR8 
with the VLA by Carignan, Beaulieu, \& Freeman (1990) used the radial distribution of HI velocity dispersion 
and surface density to estimate the central density of its DM halo.  The HI in many dIm galaxies has now been 
observed with adequate spatial resolution.  We follow the methods of Carignan and collaborators to derive the 
central density of their halos.

\vss\vsss
\cl {6.1.~\it Dwarf Spheroidal Galaxies}
\vss

      For dSph galaxies, the data that we can use to estimate the central dark matter densities are the 
projected stellar velocity dispersion profiles $\sigma(R)$ and the surface density distributions $I(R)$ 
of the stars. Even if we assume that the galaxies are spherically symmetric, these observables alone do not 
provide strong constraints on the central DM density. For example, we do not know the form 
of the underlying DM distribution or how the anisotropies of the stellar velocity distributions change with 
radius (see Walker et al. 2009 for more details). Some assumptions are needed.  These should be consistent 
with the assumptions that we used in Section\ts2 for the spiral galaxies.  

\cl{\null}

       We use the spherical Jeans equation to estimate central DM densities from 
$\sigma(R)$ and $I(R)$. The measurable $I(R)$ distributions for these low-surface-brightness dSphs 
do not usually extend over a large range of surface density, and the star counts are often represented by 
simple approximations such as Plummer and King models.  Because the Jeans equation involves 
derivatives of the density distribution, the outcome can be sensitive to details of how $I(R)$ is modelled 
(e.{\ts}g., Evans et al.~2009).  We estimate the form of the baryon distribution $I(R)$ expected 
in faint dwarf galaxies under a set of simple assumptions:

      We assume that DM halos have constant-density cores.  The issue is not settled (Section\ts2.2),
though some authors support this assumption with their dynamical analyses (e.{\ts}g., Amorisco \& Evans 2012).
We adopt this assumption to be consistent with our analysis of the Sc\ts--{\ts}Im galaxies. 
      
      How do the radial extents of the stars in dSph galaxies compare with the likely sizes of their DM cores? 
Typical half-light radii $r_h$ for dSph galaxies are a few hundred~pc. From Table 1, the core radii 
for the DM halos in the faintest of our Sc\ts--{\ts}Im galaxies are about $1.6 - 2$ kpc. Although the 
DM core radii of dSph galaxies may be smaller than those of faint, rotation-dominated Sc\ts--{\ts}Im systems, 
it seems reasonable to consider an approximation in which the stars are immersed entirely within the uniform-density 
core of the DM halo.  Section 7 shows that this produces results that are internally consistent.

      The projected velocity dispersion profile $\sigma(R)$ also appears in the Jeans equation.  
The observed $\sigma(R)$ profiles for the classical dSphs are almost isothermal well beyond their 
half-light radii $r_h$, out to $R \sim 1000$ pc. Even for the ultrafaint system Seg 1, with $r_h \simeq 30$ pc, 
the $\sigma(R)$ profile is approximately isothermal out to about 60 pc.  We assume 
(i) that each system is spherically symmetric, 
(ii) that the stars in a dSph galaxy lie entirely in the constant-density DM core, 
(iii) that the $\sigma(r)$ distribution is isothermal, 
(iv) that the velocity distribution is isotropic, and 
(v) that the contribution of the stars to the total density is negligible.
Then the spherical Jeans equation takes the simple form 
$$ 
{\sigma_*^2 \over \rho_*} {d\rho_* \over dr} = -{GM(r) \over r^2} = -{4 \pi G \rho_\circ \over 3}\, r~, \eqno{(34)}
$$ 
\noindent where $\rho_*$ and $\sigma_*$ are the volume density and velocity dispersion of the stars, $M(r)$ 
is the total mass enclosed within radius $r$, and $\rho_\circ$ is the central density of the DM.  Then the
stellar volume density is Gaussian, 
\vskip -3pt
$$\rho_*(r) \propto \exp(-r^2/a_*^2)~, \eqno{(35)} $$
\vskip -3pt
\noindent and the volume density of the DM core is 
\vskip -3pt
$$\rho_\circ = 3\sigma_*^2/2\pi G a_*^2~. \eqno{(36)} $$ 
The projected surface density $\Sigma_*(R)$ of a Gaussian volume density distribution $\rho_*(r)$ is also  
Gaussian with the same scale length $a_*$.  In fact, the star-count distributions of the classical dSph galaxies and 
some of the ultrafaint systems are well represented by Gaussians over most of their radial extent.  
As an example, Figure 5 shows the Gaussian radial profiles of the projected star counts for the Carina dSph system.
Similar plots for all dSph galaxies used in this paper are shown in Figure\ts11 in Appendix A.

\phantom{0000000000000000000000000000000000000}

\eject

\cl{\null}

\vskip 2.75truein  



 \includegraphics{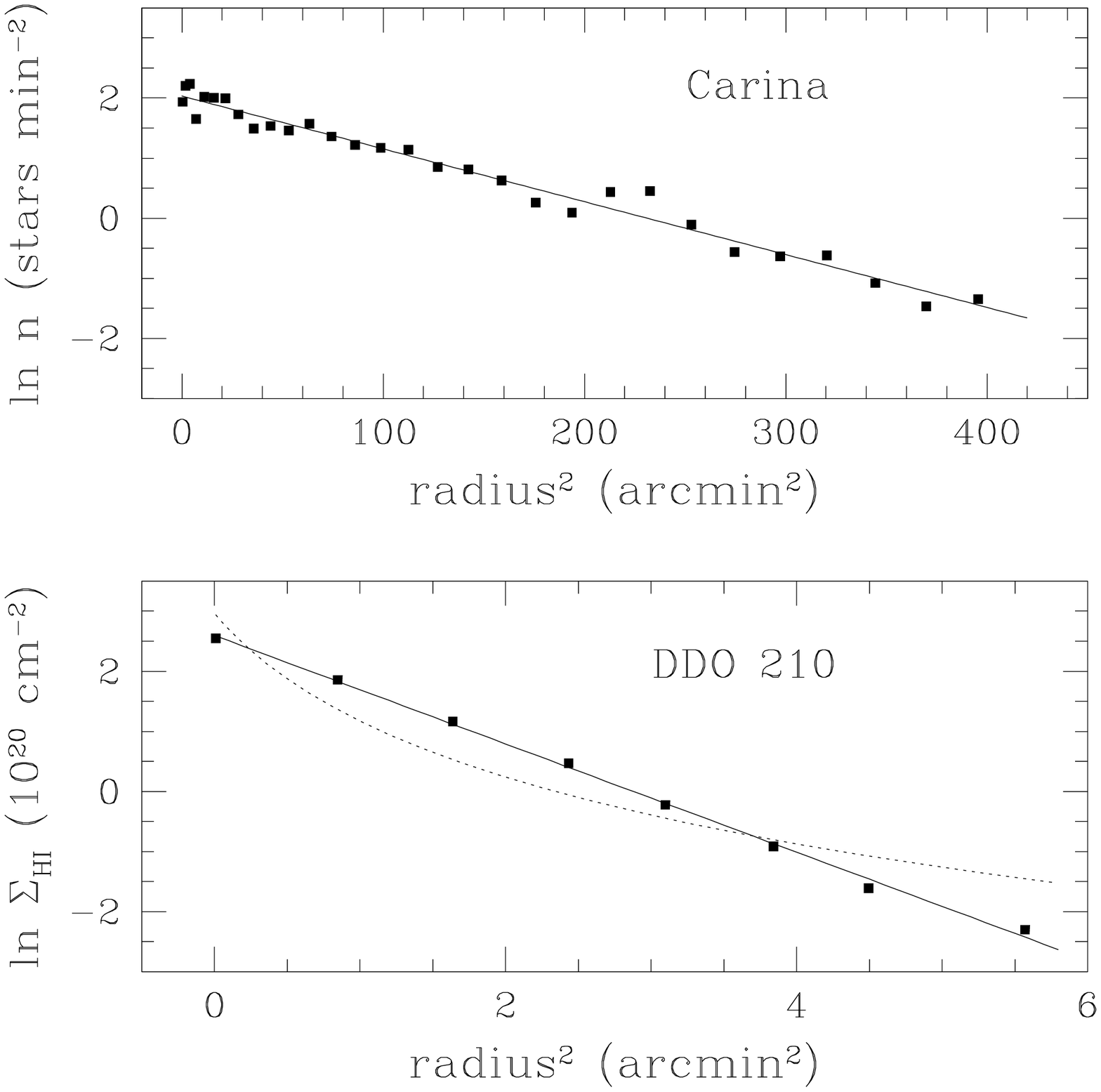} 

      Fig.~5~-- Gaussian distributions of surface density in two dwarf galaxies. The upper panel
shows the radial distribution of star count density $n$  against major-axis radius for the Carina dSph 
galaxy (data from Irwin \& Hatzidimitriou 1995).   The lower panel shows the radial distribution of 
HI surface density $\Sigma_{HI}$ against major-axis radius for the DDO 210 dIm galaxy (derived from 
Young \etal 2003). In this (ln $\Sigma_{HI}$\ts--\ts$r^2)$ plane, a Gaussian distribution is a straight line.  
The solid straight lines show Gaussian fits to these two galaxies. Because 
Plummer models are often used to represent the surface densities of dwarf galaxies, the dotted 
curve in the lower panel shows the best-fit Plummer model for comparison.

\vskip 10pt

      To compare the DM parameters of dSph galaxies with those for the rotationally supported disk galaxies,~we~use 
the halo density~$\rho_\circ$ derived from the Gaussian \hbox{star-count} distribution and $\sigma_*$ via Equation (36) 
uncorrected for any contribution from the stars.  While the near-isothermality of the velocity dispersion makes the 
analysis simpler, it does mean that there is little structure in the radial distribution of the kinematics 
from which one could estimate the core radius of the DM.  At the present time, the central density 
is the only halo parameter that we can directly measure for dSph galaxies.  
 
      For the halo size parameter in rotationally supported disk galaxies, we used the 
core radius $r_c$ of the DM distribution derived from the shape of the rotation curve, following the King model 
scaling $r_c^2 = 9 \sigma^2 /4 \pi G \rho_\circ$.  Because we cannot derive a size parameter for the DM in dSph 
galaxies from the stellar kinematics, we begin by tabulating the corresponding $r_c$ value for the 
stellar distribution.  To do this, we need to derive empirically the relation between the King $r_c$ and the Gaussian 
$a_*$ using eight dSph galaxies for which both lengths have been measured.  The two lengths are tightly correlated: 
\vskip-20pt
\cl{\null}
$$\log r_c = 0.752 \log a_* - 0.273~. \eqno{(37)}$$ 
\cl{\null}
\vskip -15pt

     With this calibration,  Table\ts2 lists the distance, absolute magnitude, stellar velocity dispersion, 
DM core density, and {\it stellar\/} $r_c$ for eleven dSph companions of the Milky Way for which the star-count
profiles are close to Gaussian.  For CVnI and Coma\ts1, only a mean value of $\sigma$ is available, and we have
to assume that these systems are isothermal.  We exclude two faint systems (Her and Wil 1) for which
the star-count profiles are not Gaussian.

\vss
\cl {6.2.~\it Dwarf Irregular Galaxies:}
\cl{\it H{\ts}{\sit I} Equilibrium in a Uniform-Density DM Core}
\vss

In the faintest dIm galaxies, the H{\ts}{\sc I} is supported by its velocity dispersion.
We use the H{\ts}{\sc I} as our mass tracer.  The H{\ts}{\sc I} velocity dispersion is 
usually observed to be close to isothermal and is likely to be isotropic. Assuming that the 
H{\ts}{\sc I} lies within the uniform-density core of the DM, arguments like the ones that we used for 
dSph galaxies lead us to expect that the distribution of projected H{\ts}{\sc I} surface density
is again Gaussian.  This is observed.  The observed Gaussian length scale $a_*$ for the H{\ts}{\sc I} 
distribution and the observed H{\ts}{\sc I} velocity dispersion $\sigma$ then give the central volume 
density of the DM halo core, $\rho_\circ = 3\sigma^2/2\pi G a_*^2$, as above.  As an example, Figure 5 
shows the Gaussian profile of H{\ts}{\sc I} surface density for the dIm galaxy DDO\ts210.   
Because Plummer models are a convenient and often-used represention of the surface densities 
in dwarf systems, the figure also shows the best Plummer model fitted to the H{\ts}{\sc I} surface density 
profile.  The Gaussian provides a better fit when the surface density data extend over more 
than a few e-foldings.  The near-Gaussian surface density distributions for the other dIm galaxies used 
in this paper are shown in Figure 12 (Appendix A).

      As we did for dSph galaxies, we consider the equilibrium of a spherical, isothermal, isotropic 
system of test particles (H{\ts}{\sc I}) in a uniform-density DM core.  For the test particles, 
$\rho_*(r)$ is the volume density, $\Sigma_*(R)$ is the surface density, and
$c_*$ is the isothermal sound speed or velocity dispersion.  Some dIm galaxies
show slow, solid-body H{\ts}{\sc I} rotation that we can readily include in our derivation. 
We take $\Omega_*$ to be the angular velocity, assumed constant. For the DM, $M(r)$ is the 
enclosed mass and $\rho_\circ$ is the core density. 

      The equation of hydrostatic equilibrium is then
$$ {{c_*^2} \over {\rho_*}} {{d\rho_*} \over {dr}} - \Omega_*^2 r = -{{GM(r)} \over {r^2}}~, \eqno{(38)} $$
and the H{\ts}{\sc I} volume density again is Gaussian, 
$$\rho_*(r) \propto \exp(-r^2/a_*^2)~. \eqno{(39)} $$
\noindent The central volume density of the halo is
$$\rho_\circ = {{3 c_*^2} \over {2\pi G a_*^2}} + {{3\Omega_*^2} \over {4 \pi G}}  \eqno{(40)} $$
\noindent The correction to the dispersion $c_*$ from the channel spacing of the telescope ($\sim 1.5$ km s$^{-1}$) 
is negligible.  We estimate the error in the DM density $\rho_\circ$ derived from Equation~(40) as follows. 
The contribution to $\rho_\circ$ from the $\Omega_*^2$ term is small. Then the error in $\rho_\circ$ is 
$$\biggl({{\sigma_{\rho_\circ}} \over {\rho_\circ}}\biggr)^2 = 4\biggl({{\sigma_{c_*}} \over {c_*}}\biggr)^2 + 4\biggl({{\sigma_{a_*}} \over {a_*}}\biggr)^2~, \eqno{(41)}
$$
and the error in $\log \rho_\circ$ is
$$ \sigma_{\log \rho_\circ} = {{1} \over {2.303}} {{\sigma_{\rho_\circ}} \over {\rho_\circ}}~. \eqno{(42)}$$

      The projected H{\ts}{\sc I} surface density is Gaussian with the same scale length $a_*$ as that of $\rho_*$ 
(Figures 5 and 12),
$$\Sigma(R) = \Sigma(0) \exp(-R^2/a_*^2)~. \eqno{(43)}$$
We need to correct $a_*$ for the telescope beam width.
For a Gaussian telescope beam of width $W$\ts={\ts}FWHM}/2.355, the corrected value
of $a_*$ for the galaxy is
$$a_{*_c}^2 = a_*^2 - 2W^2~. \eqno{(44)}$$ 
If the error in $a_*$ from the least-squares fit is $\sigma_{a_*}$, then the error in $a_{*_c}$ is
given by
$${{\sigma_{a_{*_c}}} \over {a_{*_c}}} = {{a_*^2} \over {a_*^2 - 2 W^2}} \biggl({{\sigma_{a_*}} \over {a_*}}\biggr)~. \eqno{(45)}$$

\cl{\null}

      As we did for dSph galaxies, we first use the core radius of the visible matter.  We will see that 
this is much smaller than the core radius of the DM halo.  We transform the beam-corrected H{\ts}{\sc I} Gaussian
scale length $a_*$ (Equation 44, dropping the subscript $c$ from here on) using the relation (Section 6.1), 
$\log r_c = 0.752 \log a_* - 0.273$. 
The error in $\log r_c$ is $0.327 (\sigma_{a_*}/a_*)$.

      Table 2 lists the distance, absolute magnitude, stellar velocity dispersion, DM core density, and $r_c$ of the 
H\ts{\sc I} distribution for twelve nearby dIm galaxies for which the dynamical contribution of the H\ts{\sc I} rotation is 
negligible and the H\ts{\sc I} surface density distribution is close to Gaussian. 

\singlecolumn

\vskip -30pt
\cl{\null}

\input colordvi
\def\B{\Blue}

\def\R{\Red}

\def\0{$\phantom{0}$}
\def\dot{$\phantom{.}$}

$$
\table
\tablewidth{18.5truecm}
\tablespec{\l\c\l\r\c\c\c\c\l}
\body{
\header{TABLE 2}
\skip{5pt}
\header{DARK MATTER CENTRAL DENSITIES AND PARAMETERS FOR SPHEROIDAL AND IRREGULAR GALAXIES}
\skip{10pt}
\hline \skip{0.001truein} \hline
\skip{.2truecm}\hline \skip{0.001truein} \hline
\skip{5pt}
& Galaxy      &   \0$D$     & Source & $M$\rlap{$_{B}$}~ & Source & $\log{\sigma}$&      \0$\log{\rho_0}$       & $\log{r_c}$   & Source                  &  \end
&             &   \0[Mpc]   &      &     &        & [km s$^{-1}$]&\dot[$M_\odot~{\rm pc}^{-3}$]&    [kpc]      &                         &  \end
& (1)         &  \0(2)      &\0(3) &    (4)      &    (5)       &            (6)              &     (7)       &  (8)   & \0(9)  \end
\skip{5pt}
\hline \skip{0.001truein} \hline
\skip{10pt}
&    Carina & 0.094 & NED &$ -8.0 $&  W09   & $0.820 \pm 0.079$ & $-1.245 \pm 0.158$ & $-0.675 \pm 0.005$ & IH95   &  \end \skip{7pt}
&     Draco & 0.076 & NED &$ -8.1 $&  W09   & $0.959 \pm 0.057$ & $-0.688 \pm 0.292$ & $-0.780 \pm 0.101$ & IH95   &  \end \skip{7pt}
&    Fornax & 0.138 & NED &$-12.4 $&  W09   & $1.068 \pm 0.033$ & $-1.557 \pm 0.069$ & $-0.371 \pm 0.006$ & C05    &  \end \skip{7pt}
&     Leo I & 0.270 & NED &$-10.9 $&  W09   & $0.964 \pm 0.066$ & $-1.142 \pm 0.133$ & $-0.606 \pm 0.007$ & IH95   &  \end \skip{7pt}
&    Leo II & 0.205 & NED &$ -9.0 $&  W09   & $0.820 \pm 0.046$ & $-0.952 \pm 0.094$ & $-0.785 \pm 0.007$ & C07    &  \end \skip{7pt}
&       Scl & 0.088 & NED &$ -9.9 $&  W09   & $0.964 \pm 0.052$ & $-1.098 \pm 0.104$ & $-0.622 \pm 0.004$ & IH95   &  \end \skip{7pt}
&   Sextans & 0.086 & NED &$ -8.6 $&  W09   & $0.898 \pm 0.071$ & $-1.647 \pm 0.145$ & $-0.466 \pm 0.010$ & IH95   &  \end \skip{7pt}
&       UMi & 0.069 & NED &$ -7.8 $&  W09   & $0.978 \pm 0.055$ & $-1.101 \pm 0.111$ & $-0.611 \pm 0.007$ & IH95   &  \end \skip{7pt}
&     CVn I & 0.218 & M08 &$ -8.0 $&  W09   & $0.881 \pm 0.023$ & $-1.728 \pm 0.061$ & $-0.448 \pm 0.015$ & M08    &  \end \skip{7pt}
&     Seg I & 0.023 & M08 &$ -0.9 $&  W09   & $0.633 \pm 0.121$ & $ 0.332 \pm 0.247$ & $-1.408 \pm 0.018$ & M08    &  \end \skip{7pt}
&      Coma & 0.044 & NED &$ -3.5 $&  W09   & $0.663 \pm 0.076$ & $-0.784 \pm 0.153$ & $-0.967 \pm 0.009$ & M08    &  \end \skip{7pt}
\skip{5pt}
\hline \skip{0.001truein} \hline                                   
\skip{10pt}
&       GR8 &  2.13 & TRG &$ -12.0 $&  B06  &  $0.954 \pm 0.039$ & $-1.270 \pm 0.080$ & $-0.565 \pm 0.007$ & B03A  &  \end \skip{7pt}
&      SDIG &  3.21 & TRG &$ -11.2 $&  B06  &  $0.954 \pm 0.082$ & $-1.576 \pm 0.165$ & $-0.450 \pm 0.006$ & C00   &  \end \skip{7pt}
&     Cam B &  3.34 & TRG &$ -11.8 $&  B03B &  $0.845 \pm 0.062$ & $-2.103 \pm 0.090$ & $-0.281 \pm 0.002$ & B03B  &  \end \skip{7pt}
&     Leo T &  0.42 &  HB &$  -6.5 $&  I07  &  $0.839 \pm 0.063$ & $-0.900 \pm 0.132$ & $-0.791 \pm 0.015$ & R08   &  \end \skip{7pt}
&      LGS3 &  0.76 & TRG &$  -9.4 $&  M98  &  $0.929 \pm 0.051$ & $-1.114 \pm 0.114$ & $-0.642 \pm 0.018$ & Y97   &  \end \skip{7pt}
&     Leo A &  0.75 & TRG &$ -11.4 $&  B06  &  $0.978 \pm 0.059$ & $-1.853 \pm 0.125$ & $-0.328 \pm 0.014$ & Y96   &  \end \skip{7pt}
&   DDO 210 &  0.99 & TRG &$ -11.0 $&  B06  &  $0.813 \pm 0.067$ & $-1.163 \pm 0.107$ & $-0.675 \pm 0.003$ & Y03   &  \end \skip{7pt}
&   DDO 216 &  1.10 & TRG &$ -13.0 $&  Z00  &  $1.021 \pm 0.041$ & $-1.544 \pm 0.076$ & $-0.396 \pm 0.004$ & Y03   &  \end \skip{7pt}
&  UGCA 292 &  3.60 & TRG &$ -11.7 $&  Z00  &  $0.978 \pm 0.046$ & $-1.742 \pm 0.079$ & $-0.342 \pm 0.008$ & Y03   &  \end \skip{7pt}
&    DDO 53 &  3.60 & TRG &$ -13.4 $&  B06  &  $0.978 \pm 0.073$ & $-1.723 \pm 0.133$ & $-0.358 \pm 0.010$ & B06   &  \end \skip{7pt}
&  UGC 7298 &  4.20 & TRG &$ -12.3 $&  B06  &  $0.929 \pm 0.066$ & $-1.670 \pm 0.135$ & $-0.431 \pm 0.009$ & B06   &  \end \skip{7pt}
&     KK230 &  1.90 & TRG &$  -9.5 $&  B06  &  $0.875 \pm 0.029$ & $-1.426 \pm 0.064$ & $-0.566 \pm 0.010$ & B06   &  \end \skip{7pt}
\skip{5pt}
\hline \skip{0.001truein} \hline
}
\endtable
$$

\noindent NOTES FOR dSph GALAXIES:~Column (2) lists distances D based on stellar indicators from NED or M08 (Column~3).
Column (4) absolute magnitudes $M_B$ are from W09 (Column 5).  Column (6) velocity dispersions are from W09.  Column (7) 
$\log \rho_\circ$ is from Equation (36).  Errors do not include distance errors.  Column (8) lists $\log r_c$ {\it for the
stars, not the DM\/} derived from Gaussian lengths $a_*$ via Equation (37).  Errors are from fits to Gaussian surface density 
distribution and do not include distance errors. Column (9) gives the sources of the star-count data.  References:

\vss

\noindent 
C05 = Coleman \etal (2005); 
C07 = Coleman \etal (2007); 
IH95 = Irwin \& Hatzidimitriou (1995); 
M08 = Martin \etal (2008); 
W09 = Walker et al (2009).

\vss

\noindent NOTES FOR dIm GALAXIES: Column (2) lists distances D based on stellar indicators from NED (mostly TRG stars)
given in Column (3).  Column (4) absolute magnitudes $M_B$ are adopted from sources in Column (5) with some adjustment 
for distance. Column (6) H{\ts}{\sc I} velocity dispersions are adopted from sources in Column (9) with some adjustment for rotation.
Column (7) $\log \rho_\circ$ is from Equation (40) including the small contribution from rotation (not separately tabulated). 
Errors do not include distance errors.  Column (8) lists $\log r_c$ {\it for the H{\ts}{\sit I} gas, not the DM\/} derived 
from Gaussian scale lengths~$a_*$ via Equation (37).  Errors are from fits to Gaussian surface density distributions and 
do not include distance errors.  Column (9) gives the sources of the H{\ts}{\sc I} data.  References:

\vss

\noindent
B03A = Begum \etal (2003); 
B03B = Begum \& Chengalur (2003);
B06 = Begum \etal (2006); 
C00 = Cote \etal (2000); 
I07 = Irwin \etal (2007); 
M98 = Mateo (1998); 
R08 = Ryan-Weber \etal (2008); 
Y96 = Young \& Lo (1996); 
Y97 = Young \& Lo (1997);
Y03 = Young \etal (2003); 
Z00 = van Zee (2000).

\vskip 0.5truein

\doublecolumns

\vss
\cl {7.~\sc DM HALO SCALING LAWS INCLUDING DWARF GALAXIES:}
\cl {\sc ESTIMATING THE BARYON LOSS IN DWARF GALAXIES}
\vss

      Figure 6 shows the combined scaling laws for the DM halos of late-type giant and dwarf galaxies.  
For the dwarfs, we here use the $r_c$ and $\sigma$ values for the baryons, i.{\ts}e., stars or H{\ts}{\sc I} 
({\it light and dark green symbols\/}).  With these values, the dwarfs lie below several of the scaling laws defined 
by the bright Sc -- Im galaxies.  This is not surprising, because there is no reason why the baryonic $r_c$ 
and $\sigma$ values should be appropriate for the DM halos of the dwarfs.  It is also likely that the dwarfs 
have lost a larger fraction of their baryons during their evolution than did the bright galaxies, because 
their potential wells are much shallower (e.{\ts}g., Dekel \& Silk 1986).  Baryon loss is probably reflected in their 
faint $M_B$ values. This provides us with a way to estimate the baryon loss in the dwarfs and the relative 
values of the core radii and velocity dispersions of their baryons and DM, as follows:

     We assume that dSph and dIm galaxies lie on the extrapolation of the DM scaling laws for bright galaxies.
We further assume that the central densities $\rho_\circ$ of the DM halos of all of our galaxies have been 
correctly measured. We then shift the $M_B$, $r_c$ and $\sigma$ values for the dwarfs to move them onto the 
scaling laws for bright galaxies. The key to this procedure is the $M_B$\ts--\ts$\log \rho_\circ$ 
relation in the \hbox{top-right} panel of Figure\ts6.  
The $\log \rho_\circ$ values for the dwarfs are direct measurements of the DM, as explained above.  We assume 
that they are correct, so they do~not~require~a~shift.  The shift in $M_B$ required to move the dwarfs on to 
the straight-line fit to the scaling law for bright galaxies is therefore determined.  Similarly, keeping the 
values of $\log \rho_\circ$ fixed, we determine the shifts between the visible and DM values of $\log r_c$ and 
$\log \sigma$ from the two panels that show the $\log r_c$ vs.~$\log \rho_\circ$ and the $\log \sigma$ vs.~$\log \rho_\circ$ 
correlations.  These shifts in $M_B$, $r_c$ and $\sigma$ are then applied to the three remaining panels.
If the procedure is valid, then the dwarfs should have shifted consistently to the brighter galaxy correlations 
in all six panels. 

\cl{\null}
\cl{\null}
\cl{\null}
\cl{\null}
\cl{\null}
\cl{\null} 
\vskip -3pt

      The shifted version of the scaling laws is shown in Figure 7. The derived shift in $M_B$ is $-4.0$ mag for 
the dSph galaxies and $-3.5$ mag for the dIm galaxies.    These shifts imply that the dwarfs have lost $25$ to $40$ 
times more of their baryons relative to the brighter galaxies, consistent with the much larger $M/L$ ratios observed 
in the dwarfs. Although we describe the shift in $M_B$ as a relative loss of baryons by the dwarfs, this may not 
be correct.  Some or all of the scarcity of baryons in extreme dwarfs could be a result of the reionization of the
Universe and the subsequent difficulty that tiny halos have in capturing gas (Klypin \etal 1999;
Bullock \etal 2000; Cattaneo \etal 2011).

     For the core radius, the adopted shift in $\log r_c$ is $0.70$ for the dSph galaxies and $0.85$ for the dIm galaxies.  
These shifts give the ratio of the core radius of the DM halo to the core radius of the baryonic distribution 
(stars or H{\ts}{\sc I}). If this is correct, then the baryon distributions are much less extended than the DM 
distributions in the dwarfs, by factors of about $5$ to $7$.

     The adopted shift in $\log \sigma$ is $0.40$ for the dSph galaxies and $0.50$ for the dIm galaxies.  That is,
the velocity dispersions of the baryons are about $1/3$ of the velocity dispersions of the DM halo particles for these 
dwarfs.  If this is correct, then the potential wells of their DM halos are much deeper than one would infer from their 
baryonic velocity dispersions. From Table 2, the typical baryonic velocity dispersion of the dwarfs is about 
$10$ \kms.  Therefore the velocity dispersions of their DM halos are about $30$ \kms.  This is similar to the velocity dispersions 
for the halos of the fainter Sc -- Im galaxies in Table 1.  We explore this similarity further in Section 8.2.

\vfill\eject

\singlecolumn

\cl{\null}\vfill

\singlecolumn


  \includegraphics{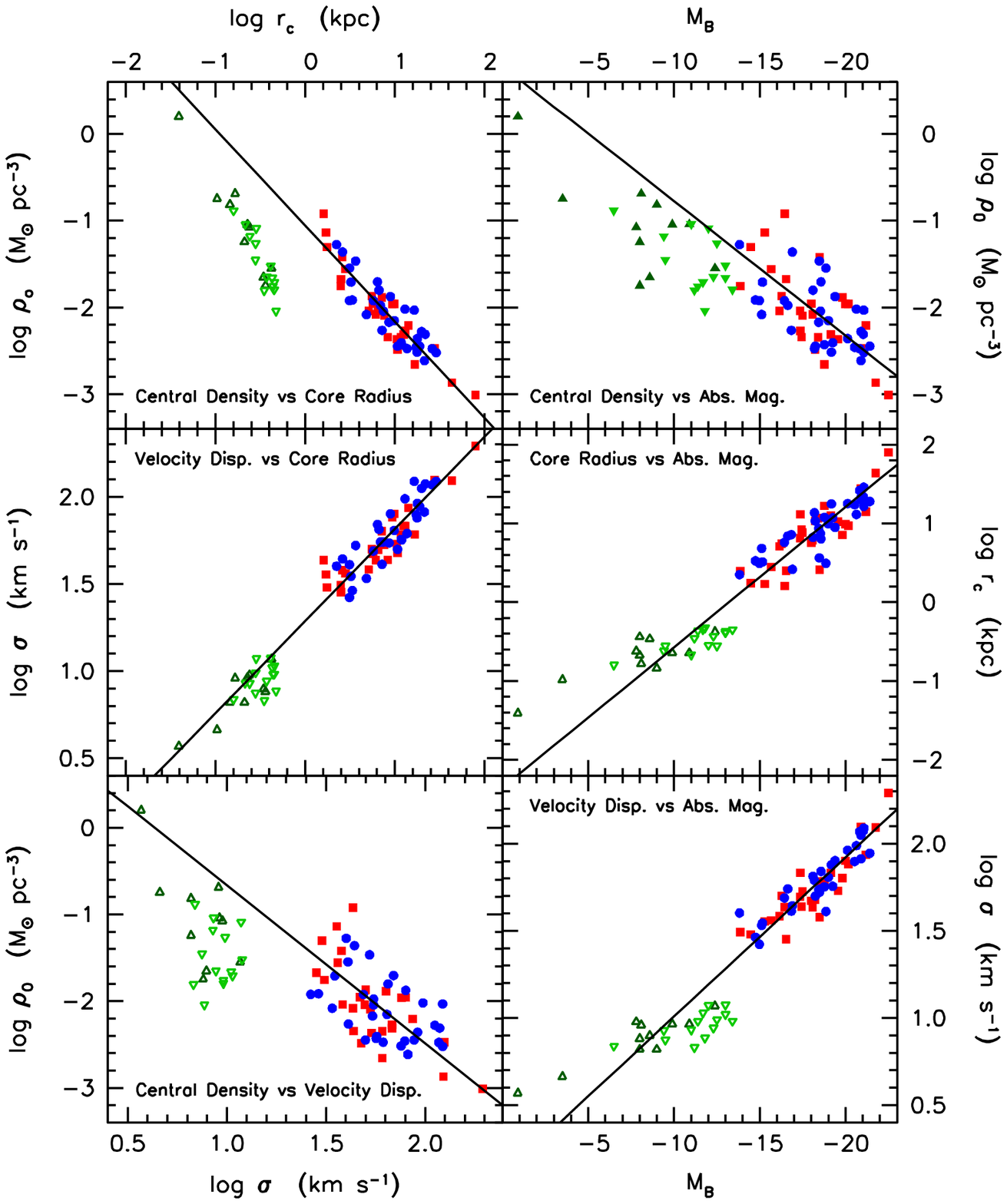}

      Fig.~6~-- Dark matter parameter correlations for 
Sc{\thinspace}--{\thinspace}Im galaxies ({\it red and blue points} from Figure 4 and Table 1), 
dwarf spheroidal (dSph) galaxies ({\it green triangles} from Table 2), and 
dIm galaxies ({\it upside-down green triangles} from Table 2).  
For dSph and dIm galaxies, $\rho_\circ$ is a meaure of the DM and hence is plotted with filled
symbols in the top-right panel.  However, $r_c$ and $\sigma$ are visible-matter parameters
and so are plotted with open symbols in the other panels.  The straight lines (from Figure 4)
are least-squares fits to the Sc -- Im galaxies only.  Note: Gilmore \etal (2007) also find
that dSph $\rho_\circ$ values fall below the extrapolation of the correlation for brighter 
galaxies (see their p.~949, left column).

\eject

\cl{\null}\vfill

\singlecolumn


  \includegraphics{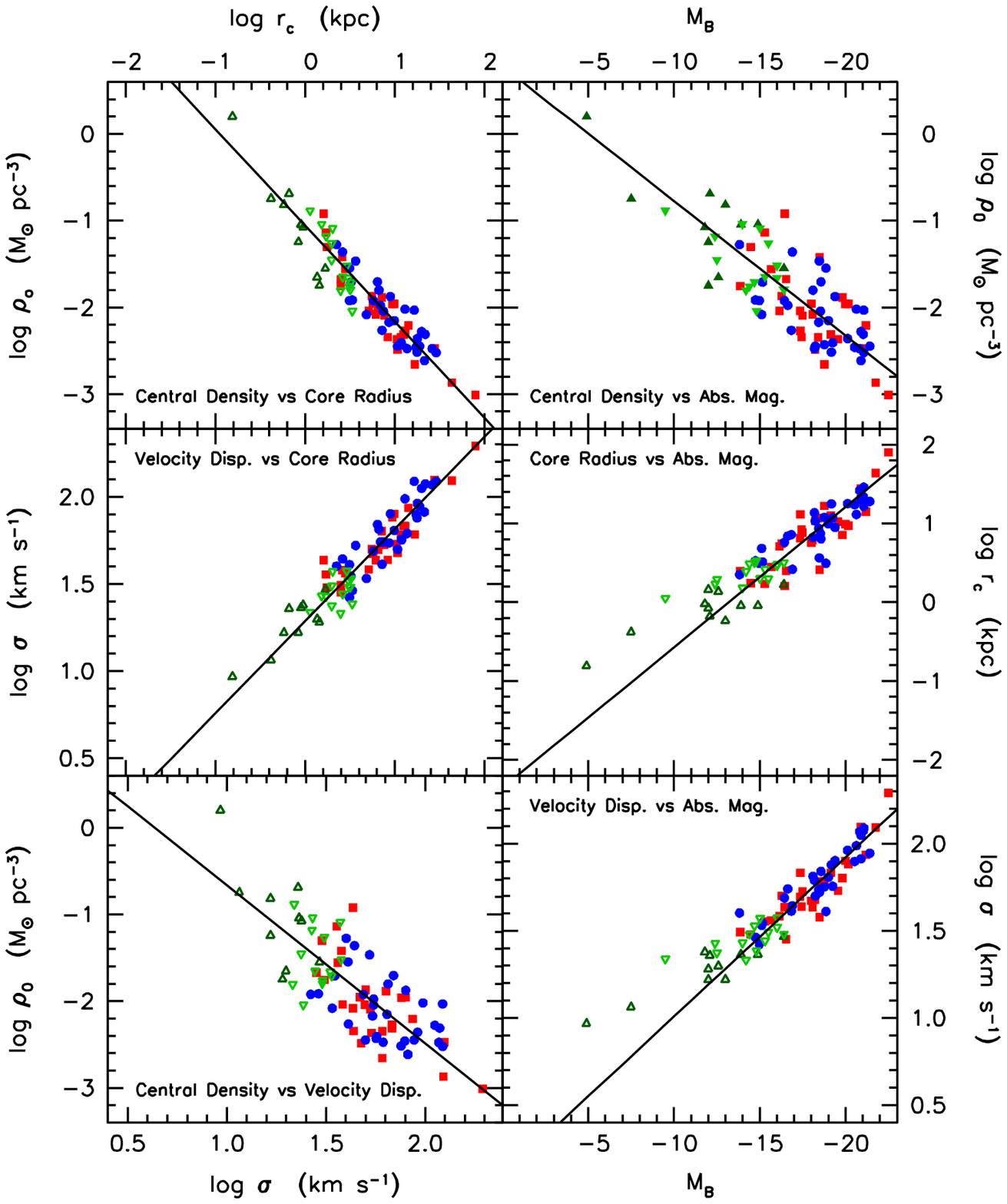}

      Fig.~7~-- Dark matter parameter correlations for  Sc{\thinspace}--{\thinspace}Im 
galaxies, dSph galaxies ({\it green triangles\/}), and dIm galaxies ({\it upside-down green triangles\/}) 
as in Figure 6 but with the dSph and dIm galaxies shifted in $M_B$,
$\log r_c$ and $\log \sigma$ to bring them on to the scaling laws for the  
Sc{\thinspace}--{\thinspace}Im galaxies.  The goal is to estimate (1) the likely effect 
of baryon loss from these faint galaxies and (2) the typical ratios of the baryonic to 
DM values of $\log r_c$ and $\log \sigma$, assuming that the DM halos of the dwarfs follow the scaling
relations for the  Sc{\thinspace}--{\thinspace}Im galaxies. The shifts for the dSph and dIm
galaxes are, respectively, 
$M_B \rightarrow M_B - 4.0$ and $M_B \rightarrow M_B - 3.5$;
$\log r_c \rightarrow \log r_c + 0.70$ and $\log r_c \rightarrow \log r_c + 0.85$; and 
$\log \sigma \rightarrow \log \sigma + 0.40$ and $\log \sigma \rightarrow \log \sigma + 0.50$.
These shifts lead to DM central surface densities that are independent of luminosity (Figure~8).

\eject

\doublecolumns

\cl{\null}

\cl{\null}

\vskip 5.0truecm

\includegraphics{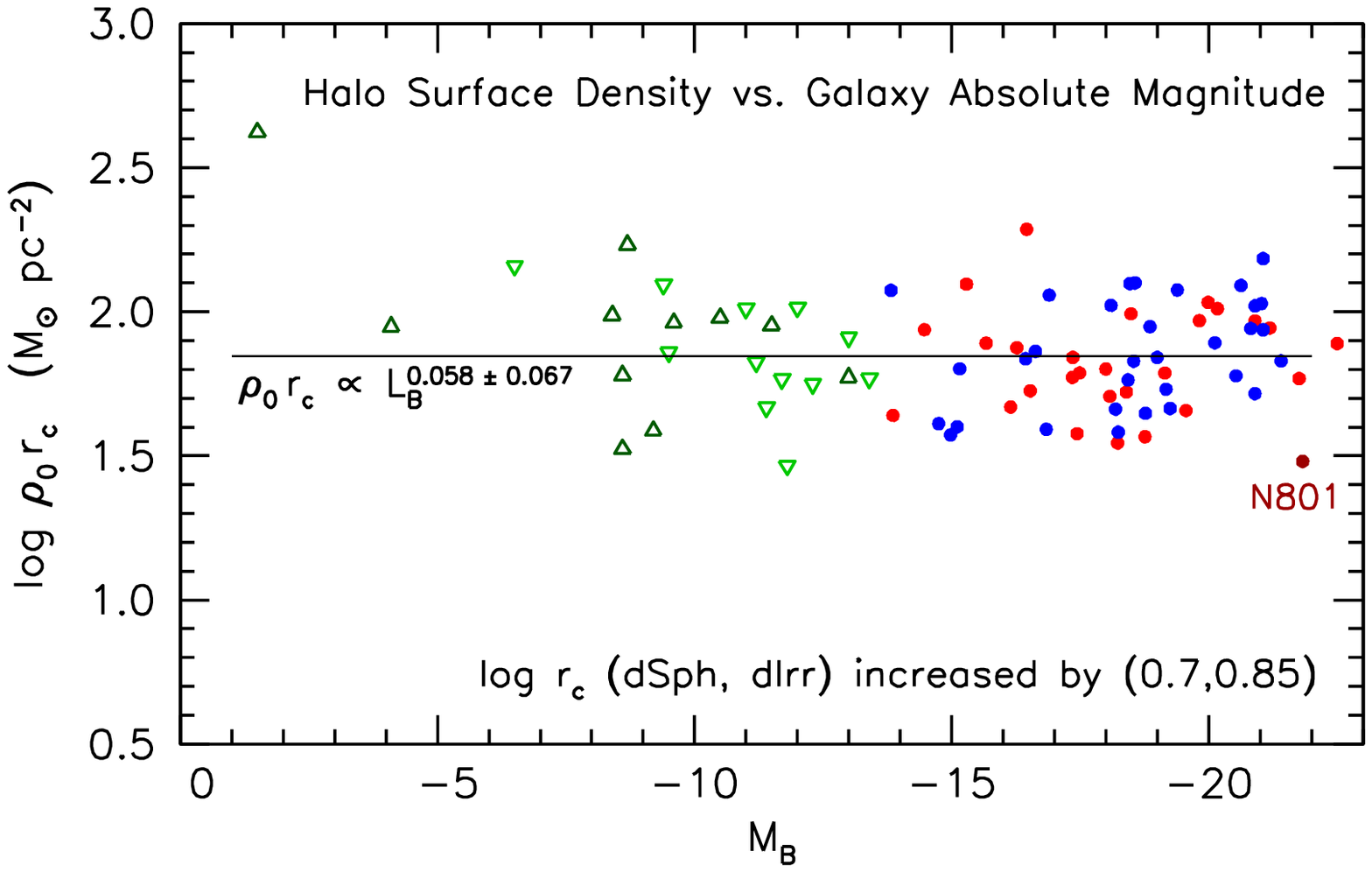}

      Fig.~8 -- The product $\rho_\circ r_c \propto$ surface density 
of DM halos as a function of observed (not shifted) absolute magnitude.  Symbols
are as in Fig.~2\ts--\ts7.~The~straight~line~shows~the almost-constant  
surface density for the  \hbox{Sc\ts--{\ts}Im} galaxies.  The dark and light green points are 
for dSph and dIm galaxies.~The $\log r_c$ values for their DM halos are derived 
from the $\log r_c$ values for their stellar or H\ts{\sc I} distributions by applying shifts 
as given in the key and explained above.


\vskip 10pt

      Equations 27 and 28 imply that $\rho_\circ\,r_c \propto L_B^{0.058 \pm 0.067}$, i.{\ts}e., that the 
central surface densities of the DM halos of \hbox{Sc\ts--{\ts}Im} galaxies are almost independent of galaxy  
luminosity.  This near-independence is illustrated explicitly in Figure 8.  Constant surface density directly 
implies a Faber-Jackson law of the form $M_{DM} \propto \sigma^4$ for the dark halo, where $M_{DM}$ is the 
DM halo mass and $\sigma$ is its velocity dispersion (Kormendy \& Freeman 2004).~The~dwarf~galaxies~are~also 
shown in Figure 8, after applying the increases $\Delta\log r_c = 0.70$ for dSph galaxies and $\Delta\log r_c = 0.85$
for dIm galaxies that we use to estimate the true core radii of their DM halos.  We see that the central DM halo surface 
densities for the dwarfs are then very similar to those of the brighter galaxies.  This consistency is of 
interest but provides little independent information, because the core radii of the dwarfs were adjusted to 
fit on the $\log r_c - \log \rho_\circ$ scaling law for the brighter galaxies in Figures 6 and 7.
      
      The product G $\times$ (surface density) is an acceleration: the near-constant surface density of the 
DM halos over a very large range in $M_B$ means that the characteristic acceleration of DM halos is almost 
the same for halos over this entire range in luminosity. 

      The conclusion that the surface densities of DM halos are essentially independent of luminosity was first noted
by Kormendy \& Freeman (2004).  It is also seen by 
Spano et al.~(2008); 
Gentile et al.~(2009); 
Donato et al. (2009); 
Plana et al.~(2010) and 
Salucci et al (2012) and, slightly more indirectly, by 
Walker \etal (2010).  
Within $r_e$ of the visible matter rather than within $r_c$ of the DM,  Napolitano \etal (2010) derive a related 
result, finding that the mean DM density inside $r_e$ decreases from giant to dwarf galaxies but is constant 
for dwarfs over a large range in halo mass. 

\vs\vskip 4pt
\cl {8.~\sc COMPARISON OF THE SCALING RELATIONS}
\cl {\sc FOR VISIBLE GALAXIES AND DM HALOS}

\vss\vsss
\cl {8.1.~\it Smaller Dwarf Galaxies}
\cl {\it Form a Sequence of Decreasing Baryon Retention}
\vss

      Figure 9 compares the baryonic and dark matter surface densities of galaxies as functions of
luminosity.  It shows quantitatively a conclusion that is well known in the literature: the ratio of baryonic mass 
to DM mass within the radial extent of the baryons decreases strongly with decreasing galaxy luminosity.  Like 
Kormendy et al.~(2009, hereafter KFCB) and Kormendy \& Bender (2012), we conclude that at $M_V$\ts\gapprox\ts$-18$, 
the dwarf spiral, Im, and Sph galaxies in Figures 2{\ts}--{\ts}9 form a sequence of decreasing baryon retention 
in smaller galaxies.  

      In contrast, the lower-luminosity bulges and elliptical galaxies have progressively higher surface densities 
at or inside the effective radius.  If we ignore baryonic pulling corrections, the ratio of visible to DM surface
density increases toward lower bulge luminosity.  Like KFCB, Kormendy (1989), and many other authors, we conclude 
that bulges and ellipticals form a sequence of increasing dissipation during the formation of smaller galaxies.   

\cl{\null}

\vskip 3.52truein

\includegraphics{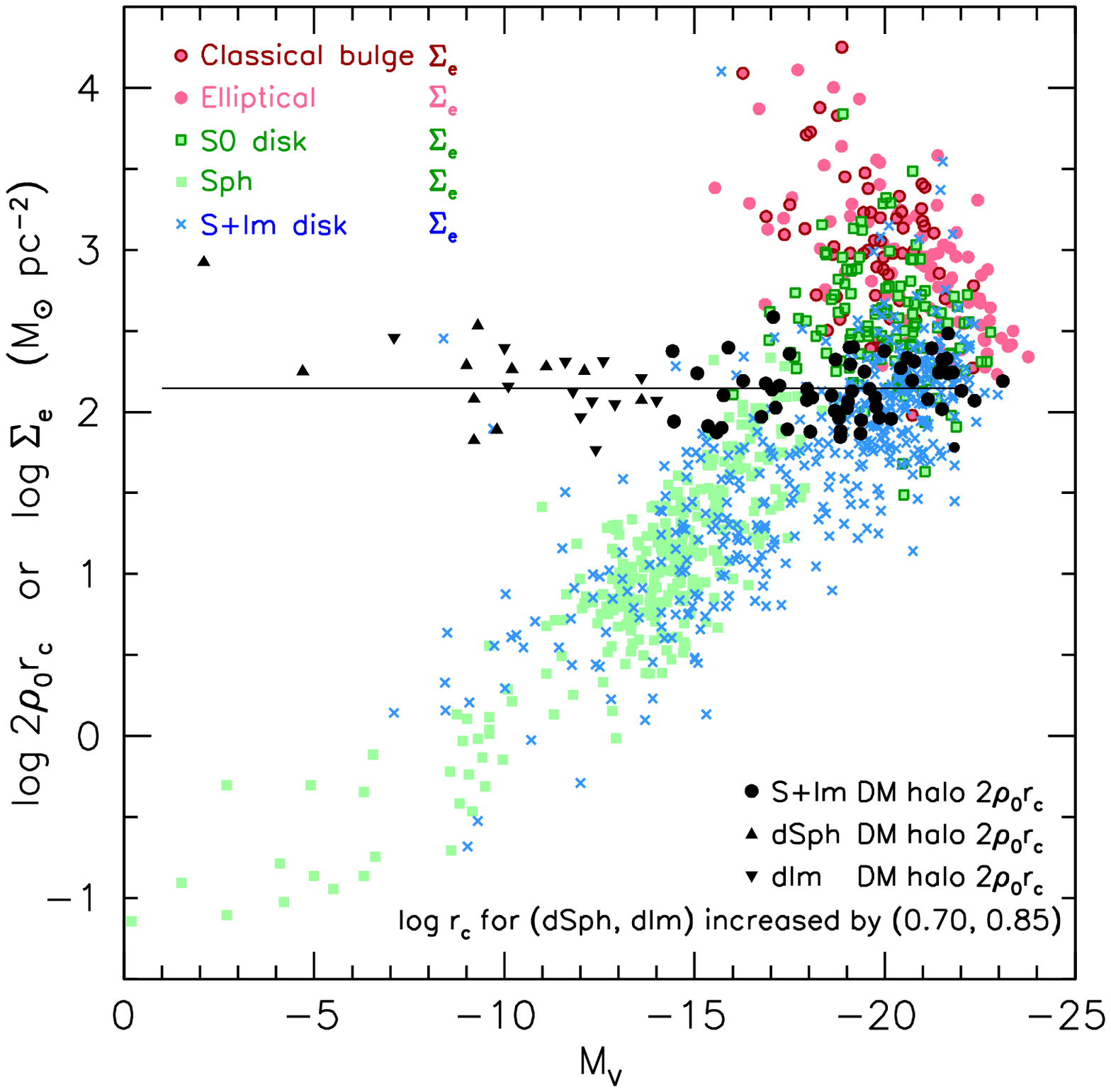}

      Fig.~9 -- Comparison of DM halo parameters from Figure 7 with visible matter structural parameters 
from Kormendy \& Bender (2012).  Central projected densities are plotted for DM halos; effective surface 
densities $\Sigma_e = \Sigma(r_e)$ are shown for visible components.  Here $r_e$ is the radius that encloses 
half of the light of the component. Surface brightnesses are converted to stellar surface densities using 
mass-to-light ratios $M/L_V = 8$ for ellipticals, 5 for classical bulges and S0 disks, and 2  or spiral galaxy 
disks, Im galaxies, and Sph galaxies.  These are typical values that approximate the results of dynamical 
or stellar population studies; their uncertainties are not large enough to affect our conclusions.  
For example, mass-to-light ratios for the stellar parts of spiral and irregular galaxies can be smaller than 2, 
but we neglect the contribution of H{\ts}{\sc I} gas, and this partly compensates for the presence of 
young stars.

\vskip 10pt

      For the brighter $M_V$\ts$<$\ts$-18$ galaxies of all kinds, the effective density in stars is similar to 
the DM density at or inside the same radius.  For Sc\ts--{\ts}Im systems, this is at least partly a consequence 
of using maximum-disk decompositions and of the rotation curve conspiracy (Bahcall\ts\&{\ts}Casertano\ts1985; 
van Albada \& Sancisi 1986; Sancisi \& van Albada 1987), i.{\ts}e., the observation that rotation curves of 
giant galaxies are roughly flat and featureless, so the parts of galaxies that are controlled by DM are not 
easily distinguished from the parts that are controlled by visible matter.
  
      There is a caveat.  For bulges and ellipticals, the high baryon densities at $r \ll r_e$ may pull gravitationally
on DM halos by enough to increase their central densities over the values for Sc--Im galaxies that are shown in 
Figure 9.  But bulges and ellipticals have {\it central\/} projected densities that are more than 3 dex higher 
than the effective densities shown in Figure 9.  So the central parts of early-type galaxies are {\it very\/} 
baryon-dominated.  The central densities of disks are 0.7 dex (for an exponential) higher than the effective 
densities shown in Figure 9.  So even pure disks are moderately dominated by visible matter near their centers.  
Both results are qualitatively as expected: Visible matter needs to dissipate, sink inside the DM, and become 
self-gravitating enough to form stars and visible galaxies (e.{\ts}g., Gunn 1987; Ryden \& Gunn 1987).  Disks
also need to be self-gravitating in order to be unstable to the formation of bars and spiral density waves (ABP). 
And a great deal of dissipation happens in the wet mergers that make coreless-disky-rotating ellipticals 
(KFCB):~their densities rise above the DM densities by larger amounts at fainter $M_V$.

      At $M_V$\ts$>$\ts$-18$, tinier~dwarfs are more DM dominated.  By $M_V$\ts\gapprox\ts$-10$, they are 
almost-dark galaxies with just enough of a frosting of stars~so~that~they can be detected.  We conclude: 
(1) {\it The differences between dIm and dSph galaxies in all parameter correlations shown in 
this paper are small.  Whether or not a galaxy retains cold gas and can still form stars in today's
Universe is a second-order effect.}  This \hbox{argues{\ts}--{\ts}as Dekel\ts\&{\ts}Silk\ts(1986)} 
emphasized{\ts}--{\ts}that the baryon retention sequence seen for the fainter galaxies in Figure 9 is 
generated primarily by supernova-driven baryon loss or another process (such as a failure to capture 
baryons after reionization) that has the same effect.  (2)~{\it We~suggest that there exists a large 
population of tiny halos that~are essentially dark{\ts}--{\ts}that the discoverable galaxies at 
$M_V$\ts\gapprox\ts$-13$ represent a smaller and smaller fraction of tinier and tinier halos.}  
This has been suggested as the solution to the problem that the fluctuation spectrum of cold dark 
matter predicts more dwarfs than are observed in environments like the Local Group (e.{\ts}g., 
Moore et al.~1999; Klypin et al.~1999).

\vss\vsss
\cl {8.2.~\it Galaxies Are Dim When Halo $V_{\rm circ} < 42 \pm 4$ km s$^{-1}$}
\vss

      Figure 10 shows one of the most important new results in this paper:~The baryon content of 
galaxies goes to nearly zero robustly at halo $V_{\rm circ} = 42 \pm 4$ km s$^{-1}$.  This is another 
representation of the galaxy baryon retention sequence.  The inference is that still-lower-mass 
halos are mostly dark; they contain a discoverable frosting of baryons only if they manage at 
least slightly to avoid the physical processes that created the baryon retention sequence.

      Figure 10 shows that rotation curve decompositions reveal a linear correlation 
between the maximum rotation velocities $V_{\rm circ,disk}$ of galaxy disks and 
the outer, circular-orbit rotation velocities  $V_{\rm circ}$ of test particles in their halos.  
Baryons become unimportant where this extrapolates to  $V_{\rm circ,disk} = 0$, i.{\ts}e., at 
$V_{\rm circ} \simeq 42 \pm 4$ km s$^{-1}$.

     The black and red points in Figure 10 show the {\it maximum} rotation velocities given
by the rotation curve decompositions for the disk and bulge, respectively.  The decompositions 
results are from van Albada \etal (1985);
\phantom{0000000}

\vskip 7.7truecm


\includegraphics{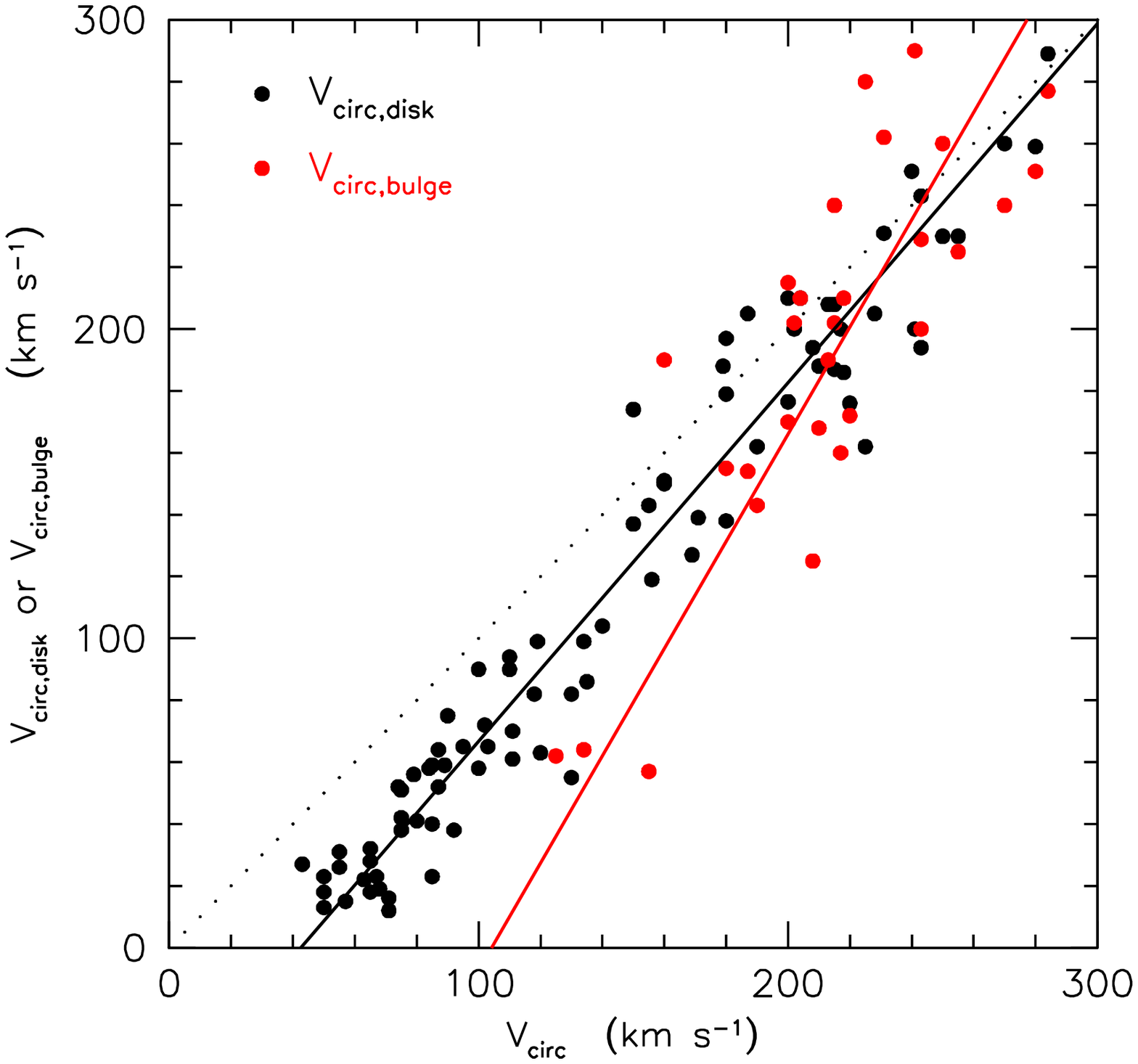}

      Fig.~10 -- Maximum rotation velocity of the bulge $V_{\rm circ,bulge}$ ({\it red points})
and disk $V_{\rm circ,disk}$ ({\it black points}) given in bulge-disk-halo decompositions of 
observed rotation curves whose outer, DM rotation velocities are $V_{\rm circ}$.  The dotted line
indicates equality of the maximum rotation velocities of the dark and visible matter.  Every 
red point has a corresponding black point, but many galaxies are bulgeless, and then
only a disk was included in the decomposition; for these, the plot shows only a black point.  
This figure illustrates that the ``rotation~curve~conspiracy'', essentially:
$V_{\rm circ,bulge} \simeq V_{\rm circ,disk} \simeq V_{\rm circ}$ for the halo
(Bahcall \& Casertano 1985; van Albada \& Sancisi 1986; Sancisi \& van Albada~1987), happens 
mostly for galaxies with $V_{\rm circ} \sim 200$~km~s$^{-1}$.    The lines are least-squares 
fits with each variable symmetrized around 200 km s$^{-1}$ (Equations 46 and 47). 
The correlation for bulges is steeper than the one for disks; thus bulges disappear at
$V_{\rm circ} \sim 104 \pm 16$ km s$^{-1}$.  This figure is updapted from Figure S2 in Kormendy \& Bender (2011).  
\lineskip=-20pt \lineskiplimit=-20pt

\vskip 10pt

\noindent  
Athanassoula \etal (1987);
Kent (1987, 1989);
Carignan \& Freeman (1988);
Jobin \& Carignan (1990);
Begeman \etal (1991);
Puche \& Carignan (1991);
Broeils (1992);
Martimbeau \etal (1994);
Miller \& Rubin (1995);
Rhee (1996);
Sofue (1996);
de Blok \& McGaugh (1997);
Sicotte \& Carignan (1997);
van Zee \etal (1997);
Verdes-Montenegro \etal (1997);
Verheijen (1997);
Carignan \& Purton (1998);
Meurer \etal (1998);
Blais-Ouellette \etal (1999);
C\^ot\'e \etal (2000);
Swaters \etal (2000);
de Blok \& Bosma (2002);
Corbelli (2003);
Weldrake \etal (2003);
Chemin \etal (2006);
Gentile \etal (2007);
Noordermeer \etal (2007);
de Blok \etal (2008);
Dicaire \etal (2008);
Noordermeer (2008);
Yoshino \& Ichikawa (2008);
Sofue \etal (2009);
Elson \etal (2010);
Puglielli \etal (2010); and
Oh \etal (2011a,{\ts}b).

      The disk and bulge fits in Figure 10 are, respectively
\vskip -25pt
\cl{\null}
$$\eqalignno{
\noalign{\vskip -20pt}
V_{\rm circ,disk}  &= (1.16 \pm 0.03)(V_{\rm circ} - 200) + (183 \pm 3);\quad      &(46) \cr
\noalign{\vskip 3pt}
\R{V_{\rm circ,bulge} &= (1.73 \pm 0.29)(V_{\rm circ} - 200) + (166 \pm 9),\quad   &(47)}\cr
\noalign{\vskip -0pt}
}$$
\cl{\null}
\vskip -15pt
\noindent
where all quantities are in km s$^{-1}$.
The important result is that both fits robustly have intercepts at \hbox{$V_{\rm circ} > 0$}.  Little extrapolation is needed: 
even baryonic disks disappear at $V_{\rm circ} \simeq 42 \pm 4$  km s$^{-1}$.  It is well known that smaller galaxies 
are more completely dominated by dark matter (Figure\ts9), but Figure 10 provides a measure of the halo mass at which 
baryons become dynamically unimportant.  Dwarf spheroidal and irregular galaxies with mass-to-light ratios of 
$\sim 10$ to $10^2$ (i.{\ts}e., the objects in Table\ts2) are, of course, missing from Figure 10; they do not rotate 
rapidly enough for rotation curve decomposition.  But they are examples of objects with trace amounts of baryons
that~lie~below the black points in Figure 10.  The message is that dark galaxies are likely to have
$V_{\rm circ}$\lapprox\ts42\ts$\pm$\ts4{\ts}km{\ts}s$^{-1}$ and \hbox{$\sigma$ \lapprox \ts$30 \pm 3$ km s$^{-1}$}. 
{\it These values are remarkably similar to the values that we derive for the halos of dSph and dIm galaxies 
from their baryonic velocity dispersions} (Section~7).

      In fact, Figures 9 and 10 together make a strong case that the sequence of almost-empty halos may extend to smaller 
$V_{\rm circ}$ and~$\sigma$ than the above values.  So we again suggest that, at lower luminosities, the galaxies 
that are plotted are increasingly the ``tip of an iceberg'' of undiscovered, still-darker dwarfs, as predicted 
by the CDM density fluctuation spectrum.

      Important: These extreme dwarfs apparently remain luminous in the Virgo cluster.  There is
no ``missing dwarfs problem'' in the Virgo cluster (Moore \etal 1999).

\headline={\vbox to 0pt{\leftskip = -0.15in \textBlack
           KORMENDY \& FREEMAN\xleft \hfill DARK MATTER SCALING LAWS \hfill\phantom{0} \hfill\phantom{0} \hfill \xright\folio}}

\vss\vsss\vsss\vsss
\cl {9.~\sc CONCLUSIONS}
\vss 

      This section lists our observational results in black.  \textBlue Inferences and
theoretical conclusions are listed in blue. \textBlack

\cl {9.1.~\it Dark Matter Halos Satisfy Well Defined Scaling Laws}
\vss

      The main observational result of this paper is that dark matter halos of Sc\ts--{\ts}Im galaxies satisfy 
well defined scaling laws.  Halos in less luminous galaxies have smaller core radii, higher central densities, 
and smaller velocity dispersions.  This confirms the results of previous analyses of smaller samples 
(Kormendy 1988, 1990; Kormendy \& Freeman 1996, 2004).  Scaling laws provide new constraints on the nature 
of DM and on galaxy formation and evolution.  Most of these remain to be explored.  
\textBlue Simple implications \textBlack and other correlation results include:

\textBlue
\vss\vsss
\cl {9.2.~\it Dwarf Spheroidals Are Real Galaxies, Not Tidal Fragments}
\vss

\textBlack 

The surprisingly high DM densities in dwarf spheroidals are normal for galaxies of such 
low luminosity (Kormendy 1987a). 
\textBlue 
This implies that dSphs are real galaxies that collapsed out of the DM fluctuation spectrum.  
They are not tidal fragments. Tides plausibly can pull bound fragments out of 
more luminous galaxies, but they cannot retain even the relatively low DM
densities in those progenitors (Barnes \& Hernquist 1992), much less increase $\rho_\circ$
to the high values characteristic of dwarf spheroidals. \textBlack

Figure 6 shows that the DM halo density of the faintest dwarfs can exceed $1 M_\odot$ pc$^{-3}$. This halo density is
much higher than the density of the Galactic disk near the Sun ($\sim 0.1 M_\odot$ pc$^{-3}$).\textBlue\ Although the low surface
brightnesses of the faintest dwarfs may make them appear fragile, their high DM densities make them in fact among the most 
robust to tidal disruption by their parent galaxy.  We could expect that accreted faint dwarfs would survive orbit decay 
and tidal disruption and would end up, possibly intact, in the inner regions of the parent system. \textBlack

In this context, we can compare the location of the DM-dominated dSph galaxies in the plane of half-light radius 
$r_h$ vs.~luminosity $L$ with the location of the baryon-dominated globular clusters, non-dwarf spheroidals (sometimes
called dwarf ellipticals) and normal, small (i.{\ts}e., compact) ellipticals (e.~g., 
Brodie \& Romanowsky 2014;
Kormendy \& Bender 2012). The dSph galaxies have $r_h$ values that are one to two orders 
of magnitude larger than the baryon-dominated objects at similar $L$.\textBlue\ Again, this is an indication that the 
baryons in the dSph systems are a minor tracer component, almost irrelevant to the equilibrium of these galaxies.

\vss\vsss
\cl {9.3.~\it Dwarf Spheroidal, Spiral, and Irregular Galaxies}
\cl {\it Form a Single Physical Sequence in their DM Parameters}
\vss

      \textBlack  Dwarf spheroidal galaxies are not included in the least-squares fits in 
Figure 7 because only $\rho_\circ$ can be derived for their halos.  However, we show that, 
with a simple shift in baryonic $M_B$, $\log r_c$, and $\log \sigma$, the dwarfs lie
on the extrapolation of the DM scaling laws for brighter galaxies in all panels of Figure 7. \textBlue 
Section 9.4 interprets the shifts.  If it is correct, then DM halos of dSph and Sc -- Im galaxies 
form a single physical sequence as a function of DM core density, velocity dispersion, or -- by  inference -- DM mass.

\vss\vsss
\cl {9.4.~\it The DM Correlations Provide a Way}
\cl {\it To Estimate the Baryon Loss and the}
\cl {\it Dynamical Properties of the DM Halos of Dwarf Galaxies}
\vss

      \textBlack The central DM densities $\rho_\circ$ of dwarf galaxies are, within the approximations of our 
analysis that $\rho(r) \simeq$ constant, derived correctly, independent of the unknown DM core radius $r_c$ and 
velocity dispersion $\sigma$.
\textBlue This provides us with a way to estimate both the baryon loss for these galaxies and the true values of 
$r_c$ and $\sigma$ for their DM halos:

      To shift the dwarf galaxies onto the extrapolation of the $\rho_\circ$\ts--\ts$M_B$ correlation for 
brighter galaxies, we need to assume that dSph galaxies were originally brighter by $\Delta M_B \simeq 4$ mag 
and that dIm galaxies were brighter by $\Delta M_B \simeq 3.5$ mag, relative to brighter galaxies which 
themselves have probably lost a significant fraction of their baryons.  To shift the dwarfs onto the other 
DM correlations then implies that the core radius of the DM is larger than the core radius of the visible matter 
by $\Delta \log{r_c} \simeq 0.70$ for Sph galaxies and by $\Delta \log{r_c} \simeq 0.85$ for dIm galaxies.  
And the DM particle velocity dispersion is larger than the velocity dispersions of the stars in dSph galaxies
by $\Delta \log \sigma \simeq 0.40$ and larger than the velocity dispersion in the gas in dIm galaxies by 
$\Delta \log \sigma \simeq 0.50$.  With these shifts, dwarfs of both kinds lie on the extrapolation of the 
DM correlations that we derived for galaxies with rotation curve decomposition in all panels of Figure\ts7.
      
      The above shifts are consistent with the hypothesis~that fainter dwarfs have lower visible matter densities 
(Fig.\ts9) and higher mass-to-light ratios $M/L_V$\ts$\sim$\ts$10^2\ts-\ts10^3$ because they lost more of their 
baryons early.  One possible \hbox{reason} is galactic winds (e.{\thinspace}g.,{\ts}Dekel\ts\&{\ts}Silk\ts1986). 
In the absence of DM, the loss of most baryons would unbind the few stars that had already formed.  But since 
faint \hbox{galaxies} have dense DM halos, the systems remain bound despite any baryon loss. An alternative to
baryon blowout is the difficulty of capturing or holding on to baryons in low-mass DM potential wells when the 
Universe was reionized (Klypin{\ts}et{\ts}al.\ts1999;{\ts}Bullock{\ts}et{\ts}al.\ts2000;{\ts}Cattaneo{\ts}et{\ts}al.\ts2011).  
\hfuzz=20pt 

      The inferred shifts of $\Delta \sigma \simeq 0.40$ in dSphs and 0.50 in dIms imply
that these almost-dark dwarfs have higher \textBlue DM masses than we thought.  They imply that the typical
velocity dispersion of their DM halos are $\sigma$ $\sim$ 30 km s$^{-1}$.  Thus these galaxies lie 
near the $V_{\rm disk}$\ts=\ts0{\ts}km{\ts}s$^{-1}$ intercept at $V_{\rm circ} \simeq 42 \pm 4$ km s$^{-1}$ 
of the black line in Figure\ts10.  It~is~not clear that we have discovered any galaxies with DM 
$V_{\rm circ} \ll 40$ km s$^{-1}$.

\textBlack

\vss\vss
\cl {9.5.~\it Spiral, Irregular, and Spheroidal Galaxies}
\cl {\it With $M_V$\ts\gapprox\ts$-18$ Form a Sequence of}
\cl {\it Decreasing Baryon-to-DM Ratio at Decreasing Luminosity}
\vss

      Figure\ts9 robustly illustrates the well known result that the ratio of baryon surface
density to DM surface density decreases dramatically at $M_V$ \gapprox \ts$-18$.  By $M_V \sim -10$, 
galaxies typically have mass-to-light ratios $M/L_V \sim 10^2$.  The most extreme dwarfs barely contain 
enough baryons to be discoverable.  Specifically, Figure 10 shows that the baryon content of 
spiral and irregular galaxies goes to zero at DM halo $V_{\rm circ} > 0$.  Instead, the 
$V_{\rm circ,disk} = 0$ km s$^{-1}$ intercept of the remarkably linear and very well determined 
correlation of the black points in Figure 10 happens at halo $V_{\rm circ} = 42 \pm 4$ km s$^{-1}$.  
Smaller galaxies are robustly faint. \textBlue In Section 9.4, we suggest that fainter dSph and dIm 
galaxies form a sequence of decreasing baryon retention (or capture) in tinier galaxies. 

\textBlack

      As DM $V_{\rm circ}$ decreases in Figure 10, bulges get less massive more rapidly than disks.  
Bulges disappear entirely at $V_{\rm circ} \sim 104 \pm 16$ km s$^{-1}$.  This is consistent with the
 well known observation that galaxies such as M{\ts}33 usually do not have bulges.  Much fainter galaxies 
never have bulges.  However, we note that there exist bulgeless galaxies at all $V_{\rm circ}$ up to 
at least 300 km s$^{-1}$ (see Kormendy \etal 2010 and Fisher \& Drory 2011 for statistics).

\vss\vss
\cl {9.6.~\it Differences between S$+$Im Galaxies and dSph Galaxies}
\cl {\it Are a Relatively Minor, Second-Order Effect}
\vss

      The differences between S$+$Im galaxies ({\it blue points in Figure 9\/}) and Sph galaxies 
({\it green points in Figure 9\/}) are small.  In particular, both galaxy kinds have similarly decreasing 
baryon-to-DM density and mass ratios at lower galaxy luminosities.  And among extreme dwarfs, some Sphs 
contain no gas and no young stars;  other Sphs contain no gas but formed stars relatively recently, and 
dIm galaxies such as M{\ts}81 dwarf A contain modest amounts of gas and still form stars gently.  But all 
of these galaxies have similar structure (Figure 5 and Appendix A) and structural parameter correlations 
(Kormendy 1985, 1987c; KFCB; Kormendy \& Bender 2012; Figure 9, here).

      This circumstance extends to higher galaxy masses, too: Kormendy \& Bender (2012)
show that Sph galaxies are essentially bulgeless S0 galaxies, that S0 disks extend the Sph correlations
to higher luminosities, and that S0 disks and spiral galaxy disks also have similar structural parameter 
correlations. Like Kormendy \& Bender (2012) and more explicitly Kormendy (2014), we conclude that the 
differences between $z \sim 0$ galaxies that still contain gas (and that still can form stars) and those
that do not (and that can not) is relatively minor.  

\textBlue 

      This is consistent with the general picture advocated in papers like Dekel \& Silk (1986) and 
Kormendy \& Bender (2012) that the most important physical processes (probably more than one) that 
control the formation of dwarf galaxies are the ones that determine the baryon depletion.  Whether 
all or just most of the cold gas gets ejected, gets used up, or never gets accreted is secondary. \textBlack

     It is interesting to note that almost all galaxies detected in the HIPASS 
blind H\ts{\sc I} survey for galaxies with velocities \ltapprox~12,000 km s$^{-1}$ were also
detected in starlight.  Galaxies with H\ts{\sc I} gas and DM but no stars appear to be very rare in
the nearby Universe (Doyle et al.~2005) \textBlue
      
\vss\vss
\cl {9.7.~\it There May Exist a Large Population of Dwarf Galaxies}
\cl {\it That Are Completely Dark}
\vss

      We suggested in our earlier papers (Freeman 1987; Kormendy 1988, 1990; Kormendy \& 
Freeman 2004) that there may exist a large population of galaxies that are completely dark.  This was based 
on the observation that, as luminosity decreases, dwarf galaxies become much more numerous and much more nearly 
dominated by dark matter.  Similar suggestions were made for different reasons by Tully et al. (2002).  

      This hypothesis is greatly strengthened by Figure~10. \textBlack Rotation curve decompositions reveal a 
linear correlation between the maximum rotation velocities $V_{\rm circ,disk}$ of galaxy disks and the outer, 
circular-orbit rotation velocities $V_{\rm circ}$ in their halos.~It extrapolates to $V_{\rm circ,disk}$\ts=\ts0 
(i.{\ts}e., baryons become unimportant) at $V_{\rm circ} \sim 42$~km~s$^{-1}$.
\textBlue In Section 7, we conclude that dSph galaxies also live in halos with $V_{\rm circ} \simeq 42$ km s$^{-1}$. 
This means that the range of baryon content of such DM halos is very large, extending from galaxies that allow 
rotation curve decomposition to extreme dwarfs that can only be found by detecting tiny enhancements in star counts.

\textBlue

      Our results suggest that empty halos are likely to be darker versions of Draco, UMi and the ultrafaint dSphs. 
DM halos that contain H{\ts}{\sc I} gas but no stars are rarely detected.  Tiny dSphs have only 
a frosting~of baryons because they lost or never captured~more.  But there is nothing magic about 
owning a few percent~of the normal baryon content.~Draco~and~UMi~do~not~know~or care that they contain just enough
baryons to be discovered by us 12 billion years after they formed.  If many DM halos captured or kept even fewer 
baryons, they could escape detection and form a population of dark dwarfs.

      Undiscovered dark dwarfs could solve the problem that the spectrum of initial density fluctuations predicted by 
CDM predicts too many dwarf satellites of giant galaxies 
(Moore et al.~1999;
Klypin et al.~1999;
Bullock et al.~2000;
Bovill \& Ricotti 2009, 2011a, b;
Simon \& Geha 2007; 
Diemand et al.\ts2008, 2011;
Tollerud et al.\ts2008;
Springel~et al.\ts2008;
Kirby et al.\ts2008; 
Walsh et al.\ts2009).  
The favored explanation for why these dwarfs are not seen is that they virialized early, 
before or during the reionization of the Universe, and so lost or never captured 
the canonical fraction of baryons because the baryons were too hot to be confined in 
shallow DM potential wells.

      Our suggestion is borne out by the discovery of ultrafaint dwarfs with $M_B$ as faint as $-1$ 
(see Simon~\&~Geha~2007; Tolstoy{\ts}et{\ts}al.\ts2009 for reviews and Table\ts2 for references~on galaxies 
used here).  Large surveys such as the Sloan Digital Sky Survey (York{\ts}et{\ts}al.\ts2000) have made it 
possible to detect ever-darker dwarfs via ever-smaller star-count excesses. The DM densities of these galaxies 
are in the range seen for brighter dSphs and for the smallest galaxies with rotation-curve decompositions.  
They are real galaxies in the sense of Section 9.2.  They are darker than Draco and UMi.  And they hint that still 
darker galaxies probably exist. \textBlack

\vss\vss
\cl {\textBlue 9.8.~\it Not Too Big To Fail}
\vss

      Sandy Faber (private communication) points out that our results suggest at least a 
partial solution to another problem with $\Lambda$CDM galaxy formation{\ts}--{\ts}``Too Big To Fail'' (TBTF).  
It is closely related to the problem that the spectrum of initial density fluctuations
predicts more halo substructure than we observe in dwarf galaxy companions of (e.{\ts}g.)~our Galaxy 
or M{\ts}31.  For the smallest and
most numerous such halos, the solution may be that they never formed visible galaxies
(Section 9.7).  But many authors (e.{\ts}g.,
Boylan-Kolchin, Bullock, \& Kaplinghat 2011, 2012;
Garrison-Kimmel \etal 2013, 2014)
have pointed out that intermdiate-mass halos with $V_{\rm circ} \sim 30$ to 70 km s$^{-1}$
(i.{\ts}e., galaxies that are a little less massive than the Magellanic Clouds) are
TBTF in that they are too massive to fail to accrete or to hold onto discoverable
masses of baryons.  Whether this problem is or is not new and separate from other dwarf
galaxy formation problems depends in part on whether or not these dwarfs have cored or
cuspy (i.{\ts}e. Navarro, Frenk, \& White 1996, 1997) density profiles~(see 
Garrison-Kimmel \etal 2013, 2014).

      We conclude (\S\ts9.4) that dSph and dIm galaxies whose stellar and gas velocity
dispersions are $\sigma \sim 10$ km s$^{-1}$ really live in bigger halos with
$\sigma \sim 10^{1.5}$ km s$^{-1}$ and $V_{\rm circ} \sim 42$ km s$^{-1}$.  This is
probably an underestimate.  Our $\Delta \log{\sigma}$ estimate was based on an
extrapolation of correlations that, at the low-mass end, include galaxies
that are dominated by DM and that have mass-to-light ratios of $\sim 10^1$.
These objects (e.{\ts}g., DDO 154: Carignan \& Freeman 1988) almost certainly also
lost baryons.  If we could correct for this loss, then these galaxies would be
plotted at brighter $M_B$ and the derived DM parameter correlations would get
slightly steeper.  We would derive a slightly larger value of $\Delta \log{\sigma}$.  
This correction cannot be made too large or the total masses of these
dwarfs become uncomfortable large.  But if $\Delta \log{\sigma} \sim 0.5$ to 0.6, then
the larger dSph and dIm galaxies in our sample are in the mass range of
the TBTF ``missing galaxies''.  Therefore:

      We suggest that the solution to the TBTF problem is at least in part that
intermediate-mass dwarfs are present in the Local Group but are masquerading
as smaller galaxies.  Brook \& Di Cintio (2014) reach a similar conclusion via a
different route.

\vss\vsss
\cl {9.9.~\it Dwarf Spheroidal Galaxies Formed At Least $\Delta\,z_{\rm coll}$ \gapprox 7}
\cl {\it Earlier Than Spiral and Irregular Galaxies}
\vss

      Virialized DM density depends on collapse redshift $z_{\rm coll}$,
$\rho_\circ \propto (1 + z_{\rm coll})^3$.\textBlack\ The DM densities for classical dSph galaxies 
like Draco and UMi are about 100 times higher that those for the brightest spirals.\textBlue\ The collapse 
redshifts of the halos of Draco and UMi and those of the brightest spirals are then 
related by $(1 + z_{\rm dwarf})/(1 + z_{\rm spiral}) \simeq 5$. For the faintest dSph galaxies, 
this ratio is about 8.  If there is a correction for baryonic DM compression in the giant galaxies, 
then it would make their DM $\rho_\circ$ smaller, and this would increase the above ratios still further. 

\vss\vsss
\cl {9.10.~\it The Power Spectrum of Initial Density Fluctuations}
\vss

      Djorgovski (1992) compared an earlier version of the DM parameter 
correlations to the scaling laws predicted by hierarchical clustering (Peebles 
1974; Gott \& Rees 1975).  For a power spectrum of initial density fluctuations 
that is a power law in wavenumber $k$, $|\delta_k|^2 \propto k^n$, the size $R$,
density $\rho$, and velocity dispersion $\sigma$ of a bound object are related
approximately by

$$\eqalignno{
\noalign{\vskip -25pt}
  \rho   &\propto R^{-3(3+n)/(5+n)}\,;            &(48)\cr
  \rho   &\propto \sigma^{-6(3+n)/(1-n)}\,;       &(49)\cr
  \sigma &\propto R^{(1-n)/(10+2n)}\,.            &(50)\cr
\noalign{\vskip -5pt}
}$$
Here we have used the relation $\rho \propto \sigma^2 R^{-2}$ for an isothermal
sphere.  Djorgovski pointed out that the DM parameter correlations in Kormendy
(1990) imply that $n \simeq -2.45$, close to the value $n \simeq -2$ expected
for giant galaxies in CDM theory.  The more accurate fits in Equations 
(31) -- (33), 
$$\eqalignno{
\noalign{\vskip -25pt}
  \rho_\circ &\propto r_c^{-1.109 \pm 0.066}\,;     &(51)\cr
  \rho_\circ &\propto \sigma^{-1.821 \pm 0.274}\,;  &(52)\cr
  \sigma &\propto r_c^{0.528 \pm 0.035}\,,          &(53)\cr
\noalign{\vskip -5pt}
}$$
give $n = -1.83 \pm 0.17$, $n = -2.07 \pm 0.07$, $n = -2.08 \pm 0.17$,
respectively.  (These values are not independent.)  Their average,
$n = -2.0 \pm 0.1$, is significantly improved over earlier determinations.
It is remarkably close to the value $n \simeq -2.1$ expected
in $\Lambda$CDM theory at a halo mass of $10^{12}\;M_\odot$ (Shapiro \& Iliev 
2002).  Note that the slopes have not been corrected for baryonic DM compression.
The above comparison provides a measure of the slope of the fluctuation power spectrum 
on mass scales that are smaller than those accessible to most other methods. 

      Shapiro \& Iliev (2002) have made a more detailed comparison of the DM
parameter correlations published by Kormendy \& Freeman (1996) with their 
predictions based on {\it COBE\/}-normalized CDM fluctuation spectra.  They found that 
the agreement between predictions and observations was best for $\Lambda$CDM.

      It is interesting to note a consequence of the theoretical prediction
that the slope $n$ gets steeper at smaller mass scales.  If $n \simeq -2.6$ 
for the smallest dwarfs (Shapiro \& Iliev 2002; Ricotti 2003), then the straight
lines in the left panels of Figure 6 should curve downward toward the visible
matter parameters of dSph galaxies. It will be important to look for curvature in 
the correlations 
as more data become available for dwarf galaxies.

\vss
\cl{9.11.~\it Could Galaxy Disks Be Very Submaximal?}
\vss

      Some authors (e.{\ts}g., van der Kruit 2010)                              )
have suggested that late-type galaxy disks are very submaximal, with
mass-to-light ratios $M/L$ of $\sim$\ts0.5\ts--\ts0.7 of maximum-disk values.
We addressed this issue in Section 2.2.  Here, having derived DM correlations
based on maximum-disk decompositions, we can use these results to help to
justify our assumptions ``after the fact'', as follows:

If the disks of giant spirals ($M_B < -19$) are typically 2/3 maximal,
then DM core radii must be decreased and DM central densities must be
increased to maintain good fits to the rotation curves.  The inevitable
consequence is that $\rho_\circ$ for the biggest galaxies becomes almost
as big as the values for dwarf galaxies that we illustrate in Figure 6.
That is, the $\rho_\circ$\ts--\ts$M_B$ correlation would get very shallow,
with only a small range in $\rho_\circ$.  This has three uncomfortable or
even untenable consequences:

(1) The derived differences in the formation redshifts of giant and dwarf
    galaxies would get much smaller than the current values in Section 9.9.
    This would be inconsistent with current thinking that the
    smallest dwarfs known formed (i.{\ts}e., their halos
    collapsed and virialized) at $z$\ts\gapprox\ts10.

(2) Our derived $\Delta M_B$ of baryon loss would get much smaller: $\Delta M_B
    \ll 3.5$ to 4.  The inferred baryon deficiency \textBlue would then be too small to
    be consistent with the large mass-to-light ratios $M/L \sim 10^2$ that are
    almost universally measured for these dwarfs.

(3) Shallower DM parameter correlations would force us to derive a slope of the
    power spectrum of initial density fluctuations (\S\ts9.10) that is very
    different from $-2$.  That derivation is an important consistency check that
    nothing has gone very wrong with our derivation of the parameter correlations.
    If DM halos are so cuspy that our analysis is invalid, or if disks are so
    submaximal that our derivations of the correlation slopes are very wrong,
    then the \S\ts9.10 consistency with the slope of $\Lambda$CDM fluctuation
    spectrum vanishes.  The success of the comparison in \S\ts9.10 is an argument
    in favor of our assumptions.  This includes our omission of any corrections
    to the correlation slopes that result from taking account of the compression
    of DM halos in giant galaxies that is caused by the gravity of their baryons.

\textBlack

\vss
\cl {9.12.~\it Caveats and Future Work}
\vss

      We emphasize that the scatter in Figures 2\ts--\ts4 and 6\ts--\ts8 has surely been 
increased by problems with the data. 
 
(1) Distance errors are smaller than in previous derivations of the correlations,
but they are not negligible. 
 
(2) The Table\ts1 parameter set is very heterogeneous.  Inconsistencies in 
the rotation curve decomposition procedures used by different authors inflate the scatter.
For example, the Broeils (1992) galaxies define the same correlations with fewer
points but smaller scatter than the sample as a whole.
 
(3) If some disks are submaximal, then this affects the scatter in the correlations.
If the degree to which they are submaximal depends on $M_B$ 
(Kranz, Slyz, \& Rix 2003;
Bosma 2004;
van der Kruit \& Freeman 2011), then this affects the correlation slopes, too.  

(4) The assumption that DM halos have isothermal cores is challenged by CDM theory, 
although it is suppported by many observations.  It will be important to see how 
the correlations are affected if NFW halos are used. 

(5) DM compression by the visible matter (Sellwood \& McGaugh 2005) may affect some
galaxies more than others.

(6) We have consistently used the absolute magnitude $M_B$ as a measure of baryonic
mass against which to scale the DM halo parameters. For some of the fainter dIm galaxies, 
the baryonic mass is dominated by the H{\ts}{\sc I} mass, and it would make sense to use a total
baryonic mass as is done in deriving the baryonic Tully-Fisher relation. This would reduce
the $\Delta M_B$ shift needed for some of the dIm galaxies in Figure 7. We are grateful 
to J.~S.~Gallagher for this suggestion.

(7) In selecting dIm galaxies for this sample, we were probably over-cautious in
choosing galaxies that have small solid-body rotation. The Gaussian surface density and
related apparatus for estimating the core density of the DM halo core are expected
to work for regions of these galaxies in which the rotation is solid-body.

       We also acknowledge that the shifts $\Delta M_B$, $\Delta \log{r_c}$, and
$\Delta \log{\sigma}$ determined in Section 7 and discussed further in Section 9.4
are probably underestimated.  The smallest dS$+$dIm galaxies that have 
rotation curve decompostion results in Table 1, in Figures 2 -- 4, and in the 
correlation derivations in Equations (21)\ts--\ts(33) already have mass-to-light ratios
of order $M/L_V \sim 10^1$.  Consistent with the rest of our discussion, they probably 
also have lost a substantial fraction of their baryons.  Certainly they are
faint enough to be on the baryon depletion sequence illustrated in Figure 9.  
If we could correct for this baryon loss, then the correlations that we derive
would get steeper.  Consequently, all parameter shifts for the extreme dSph and
dIm galaxies would get somewhat larger.  None of our conclusions would change
qualitatively, but correlations and shifts both would get tweaked.  We would infer
that the smallest detected DM halos have particle velocity dispersions that are slightly
bigger than $\sigma \sim 30$ km s$^{-1}$.

       We will address these issues in future papers.

\vss\vskip 2pt
\cl {\sc ACKNOWLEDGMENTS}
\vss

      This work has been in progress for many years; we owe our warm and sincere thanks 
to many people for their help and support.  JK is grateful to the Directors and staff 
of Mt.~Stromlo Observatory for their hospitality during four visits when parts of this
work were done.  JK also thanks the Alexander von Humboldt-Stiftung (Germany) for the 
Research Award that made possible his early visits to the Universit\"ats-Sternwarte, 
Ludwig-Maximilians-Universit\"at, Munich, where other parts of this work were done. 
More recently, JK has visited the Max-Planck-Institut f\"ur Extraterrestrische Physik (MPE)
as an External Member.  The hospitality of the Sternwarte the MPE and its Directors, 
R.~Bender, R.-P. Kudritzki, and R.~Genzel, is much appreciated.  We thank 
S.~Djorgovski, 
S.~M.~Faber, 
S.~M.~Fall, 
J.~F.~Gallagher, and 
P.~Shapiro 
for helpful discussions.  We also thank R.~Bender for making available the program 
(Kormendy \& Bender 2009) that was used to construct the symmetric least-squares fits.  
Finally, JK warmly acknowledges the productive collaboration with A.~Bosma on DDO 127 
whose results in Table~1 are quoted from Kormendy \& Freeman (2004).

      KCF is grateful to the Aspen Physics Center (NSF grant \#1066293) and the MPE and the
European Southern Observatory for hospitality on several occasions during which parts of 
this work were done. 

      This work would not have been practical without extensive use of the NASA/IPAC 
Extragalactic Database (NED), which is operated by the Jet Propulsion Laboratory and
the California Institute of Technology under contract with NASA.  We also used the 
HyperLeda database, {\tt http://leda.univ-lyon1.fr} (Paturel \etal 2003).
Finally, we made extensive use of NASA's Astrophysics Data System bibliographic services.

JK's work on this subject was supported by NSF grants AST--9219221 and AST-0607490, by 
the Alexander von Humboldt-Stiftung, and by Sonderforschungsbereich 375 of the German Science 
Foundation.  The visits to Mt.~Stromlo were made possible by the long-term support provided 
to JK by the Curtis T.~Vaughan, Jr.~Centennial Chair in Astronomy.  We are most sincerely 
grateful to Mr.~and Mrs.~Curtis T.~Vaughan, Jr.~for their support of Texas astronomy. 

\vfill\eject

\singlecolumn

\vskip -20pt

\cl {\sc APPENDIX A}
\vss

\cl{\sc BRIGHTNESS PROFILES OF DWARF Im AND Sph GALAXIES}
\vss

      Figures 11 and 12 show the surface brightness profiles of the dSph and dIm galaxies used in this paper. The Gaussian fits on which our DM $\rho_\circ$ determinations are based are also illustrated.

\vfill

\includegraphics{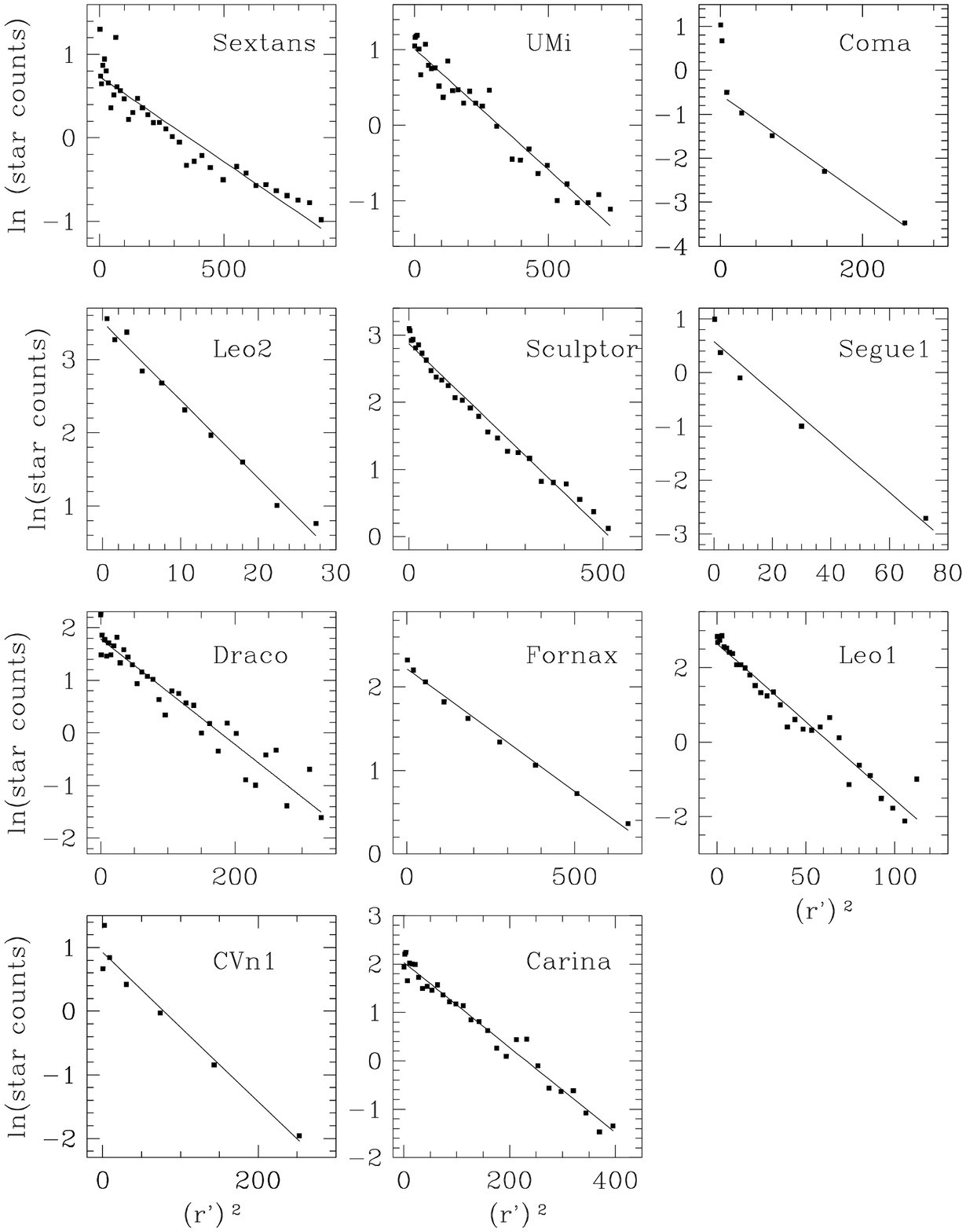}

      Fig.~11 -- Surface brightness profiles of the dSph galaxies used in this paper.  They are plotted against
radius$^2$ so that the Gaussian fits shown are straight lines. 

\eject

\cl{\null} \vskip -10pt

\vfill

\includegraphics{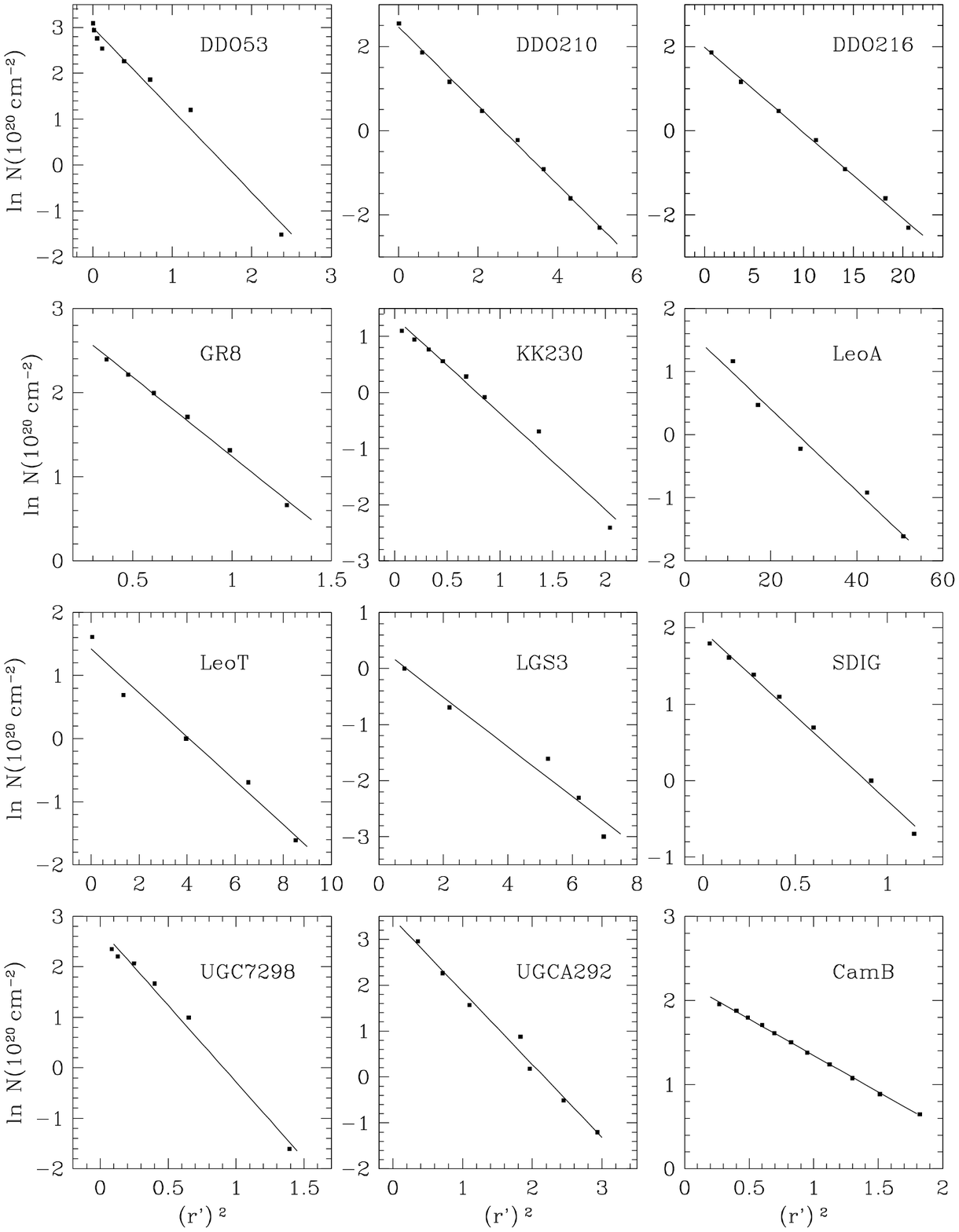}

      Fig.~12 -- H{\ts{\sc I} column density profiles of the dIm galaxies used in this paper.  They are plotted against
radius$^2$ so that the Gaussian fits shown are straight lines.  
\textBlack

\eject

\cl{\sc REFERENCES}

\doublecolumns

\frenchspacing

\def\etal{\vsc et al.\ }

{\vsc\smlbaselines

\nnhi Aaronson, M.~1983, ApJ, 266, L11

\nhi Aaronson, M., \& Mould, J.~1985, ApJ, 290, 191

\nnhi Aaronson, M., \& Olszewski, E.~1987, in IAU Symposium 117, Dark Matter in 
     the Universe, ed.~J.~Kormendy \& G.~R.~Knapp (Dordrecht: Reidel), 153
     
\nhi Amorisco, N. C., \& Evans, N. W.~2012, MNRAS, 739, L47     

\nhi Andredakis, Y. C., \& Sanders, R. H. 1994, MNRAS, 267, 283

\nnhi Athanassoula, E.~2004, in IAU Symposium 220, Dark Matter in Galaxies,
      ed.~S.~D.~Ryder, D.~J.~Pisano, M.~A.~Walker, \& K.~C.~Freeman (San Francisco: ASP),
      255

\nnhi Athanassoula,{\ts}E.,{\ts}Bosma,{\ts}A.,\ts\&{\ts}Papaioannou,{\ts}S.\ts1987,{\ts}A\&A,\ts179,\ts23\ts(ABP)

\nhi Bahcall, J.~N., \& Casertano, S.~1985, ApJ, 293, L7

\nnhi Barnes, J.~E., \& Hernquist, L.~1992, Nature, 360, 715

\nhi Beauchamp, D., Hardy, E., Suntzeff, N.~B., \& Zinn, R.~1995, AJ, 109, 1628

\nnhi Begeman, K. 1987, PhD Thesis, Rijksuniversiteit te Groningen

\nnhi Begeman, K.{\ts}G., Broeils, A.{\ts}H., \& Sanders, R.{\ts}H. 1991, MNRAS, 249, 523

\nnhi Begum, A., \& Chengalur, J. N. 2003, A\&A, 409, 879 (B03B)

\nnhi Begum, A., \& Chengalur, J.~N.~2004, in IAU Symposium 220, Dark Matter in Galaxies,
      ed.~S.~D.~Ryder, D.~J.~Pisano, M.~A.~Walker, \& K.~C.~Freeman (San Francisco: ASP), 347

\nnhi Begum, A., Chengalur, J. N., \& Hopp, U. 2003, New Astron., 8, 267 (B03A)

\nnhi Begum, A., Chengalur, J. N., Karachentsev, I. D., Kaisin, S. S., \& Sharina, M. E. 2006, MNRAS, 365, 1220 (B06)

\nnhi Bell, E.~F, \& de Jong, R.~S.~2001, ApJ, 550, 212

\nnhi Bender, R., Kormendy, J., Cornell, M.~E., \& Fisher, D.~B.~2014, ApJ, submitted

\nhi Bershady, M. A., Martinsson, T. P. K., Verheijen, M. A. W., \etal 2011, ApJ, 739, L47

\nnhi Binggeli, B., \& Cameron, L.~M.~1991, A\&A, 252, 27

\nnhi Blais-Ouellette, S., Carignan, C., Amram, P., \& C\^ot\'e, S.~1999, AJ, 
      118, 2123

\nhi Blumenthal, G.~R., Faber, S.~M., Flores, R., \& Primack, J.~R.~1986, 
       ApJ, 301, 27

\nnhi Borriello, A., \& Salucci, P.~2001, MNRAS, 323, 285

\nnhi Bosma, A.~1978, PhD Thesis, Rijksuniversiteit te Groningen

\nnhi Bosma, A.~1999, in Galaxy Dynamics: A Rutgers Symposium,
       ed.~D. Merritt, J.~A. Sellwood, \& M.~Valluri (San Francisco: 
       ASP), 339

\nnhi Bosma, A.~2004, in IAU Symposium 220, Dark Matter in Galaxies, ed.~S.~D.~Ryder, 
      D.~J.~Pisano, M.~A.~Walker, \& K.~C.~Freeman (San Francisco: ASP), 39

\nnhi Bottema, R.~1993, A\&A, 275, 16

\nnhi Bottema, R.~1997, A\&A, 328, 517

\nnhi Bovill, M. S., \& Ricotti, M. 2009, ApJ, 693, 1859

\nnhi Bovill, M. S., \& Ricotti, M. 2011a, ApJ, 741, 17

\nnhi Bovill, M. S., \& Ricotti, M. 2011b, ApJ, 741, 18

\nnhi Boylan-Kolchin, M., Bullock, J. S., \& Kaplinghat, M.~2011, MNRAS, 415, L40

\nnhi Boylan-Kolchin, M., Bullock, J. S., \& Kaplinghat, M.~2012, MNRAS, 422, 1203

\nnhi Broeils, A.~H.~1992, PhD Thesis, Rijksuniversiteit te Groningen

\nnhi Brodie, J., \& Romanowsky, A. 2014, in Lessons from the Local Group, A Conference 
in Honour of David Block and Bruce Elmegreen, ed. K. C. Freeman, B. G. Elmegreen, 
D. L. Block, \& M. Woolway (New York: Springer), 203

\nnhi Brook,{\ts}C.{\ts}B., \& Di{\ts}Cintio,{\ts}A. 2014, MNRAS, submitted (arXiv:1410.3825)

\nnhi Bullock, J. S., Kravtsov, A. V., \& Weinberg, D. H. 2000, ApJ, 539, 517

\nnhi Burkert, A.~1995, ApJ, 447, L25

\nnhi Carignan, C., Beaulieu, S., C\^ot\'e, S., Demers, S., \& Mateo, M.~1998, AJ, 116, 1690

\nhi Carignan, C., Beaulieu, S., \& Freeman, K.~C. 1990, AJ, 99,  178

\nhi Carignan, C., \& Freeman, K.~C.~1985, ApJ, 294, 494

\nhi Carignan, C., \& Freeman, K.~C.~1988, ApJ, 332, L33

\nnhi Carignan, C., \& Puche, D.~1990, AJ, 100, 394

\nnhi Carignan, C., \& Purton, C.~1998, ApJ, 506, 125

\nnhi Carignan, C., Sancisi, R., \& van Albada, T.~S.~1988, AJ, 95, 37 

\nhi Casertano, S., \& van Gorkom, J.~H.~1991, AJ, 101, 1231

\nhi Cattaneo,{\ts}A., Mamon,{\ts}G.{\ts}A., Warnick,{\ts}K., \& Knebe,{\ts}A.{\ts}2011,{\ts}A\&A,\ts553,\ts5

\nhi Chemin, L., Carignan, C., Drouin, N., \& Freeman, K.~C.~2006, AJ, 132, 2527 

\nnhi Chen, D.-M., \& McGaugh, S. 2010, Res. Astron. Astrophys., 10, 1215  

\nnhi Coleman, M. G., Da Costa, G. S., Bland-Hawthorn, J., \& Freeman, K. C. 2005, AJ, 129, 1443 (C05)

\nnhi Coleman, M. G., Jordi, K., Rix, H.-W., Grebel, E. K., \& Koch, A. 2007, AJ, 134, 1938 (C07)

\nnhi Col\'\i n, P., Klypin, A., Valenzuela, O., \& Gottl\"ober, S.~2004, ApJ, 612, 50  

\nnhi Corbelli, E.~2003, MNRAS, 342, 199

\nnhi C\^ot\'e, S., Carignan, C.,  \& Freeman, K.~C.~2000, AJ, 120, 3027 (C00)

\nnhi Courteau, S. 1996, ApJS, 103, 363

\nnhi Courteau, S., \& Rix, H.-W.~1999, ApJ, 513, 561

\nhi Cox, A. N., Ed. 2000, Allen's Astrophysical Quantities, Fourth Edition (New York: Springer)

\nnhi Da Costa, G.~S.~1994, in ESO/OHP Workshop on Dwarf Galaxies, 
       ed.~G. Meylan \& P.~Prugniel (Garching: ESO), 221

\nnhi Debattista, V.~P., \& Sellwood, J.~A.~1998, ApJ, 493, L5

\nnhi Debattista, V. P., \& Sellwood, J. A. 2000, ApJ, 543, 704

\nnhi de Blok, W. J. G. 2010, Adv. Astr., 2010, 789293  

\nnhi de Blok, W.~J.~G., \& Bosma, A. 2002, A\&A, 385, 816

\nnhi de Blok, W.~J.~G., \& McGaugh, S.~S.~1997, MNRAS, 290, 533

\nnhi de Blok, W.~J.~G., McGaugh, S.~S., \& Rubin, V.~C.~2001, ApJ, 122, 2396

\nnhi de Blok, W.~J.~G., Walter, F., Brinks, E., \etal 2008, AJ, 136, 2648  

\nnhi Dekel, A., \& Silk, J.~1986, ApJ, 303, 39

\nhi de Souza, R. S., Rodrigues, L. F. S., Ishida, E. E. O., \& Opher, R. 2011, MNRAS, 415, 2969

\nhi de Vaucouleurs, G., de Vaucouleurs, A., Corwin, H.~G., \etal 1991, Third Reference Catalogue of Bright 
     Galaxies (New York: Springer) (RC3)

\nnhi Dicaire, I., Carignan, C., Amram, P., \etal 2008, AJ, 135, 2038  

\nnhi Di Cintio, A., Brook, C. B., Macci\`o, A. V., \etal 2014, MNRAS, 437, 415

\nnhi Diemand, J., Kuhlen, M., Madau, P., \etal 2008, Nature, 454, 735  

\nnhi Diemand, J., Moore, B. 2011, Adv. Sci. Letters, 4, 297

\nnhi Diemand, J., Moore, B., \& Stadel, J. 2004, MNRAS, 353, 624

\nnhi Diemand, J., Moore, B., \& Stadel, J. 2004, MNRAS, 353, 624

\nnhi Diemand, J., Zemp, M., Moore, B., Stadel, J., \& Carollo, M. 2005, MNRAS, 364, 665  

\nnhi Djorgovski, S.~1992, in Cosmology and Large-Scale Structure in the 
           Universe, ed.~R.~R. de Carvalho (San Francisco: ASP), 19

\nnhi Djorgovski, S., \& Davis, M.~1986, in Galaxy Distances and 
           Deviations from Universal Expansion, ed. B.~F.~Madore \& R.~B.~Tully
           (Dordrecht: Reidel), 135
 
\nnhi Djorgovski, S., \& Davis, M.~1987, ApJ, 313, 59

\nnhi Djorgovski, S., de Carvalho, R., \& Han, M.-S.~1988, in The 
      Extragalactic Distance Scale, ed.~S.~van den Bergh \& C.~J.~Pritchet
      (San Francisco: ASP), 329

\nnhi Donato, F., Gentile, G., Salucci, P., \etal 2009, MNRAS, 397, 1169  

\nnhi Doyle, M.{\ts}T., Drinkwater, M.{\ts}J., Rohde, D.{\ts}J., \etal 2005,{\ts}MNRAS,\ts361,\ts34

\nhi Dressler, A., Lynden-Bell, D., Burstein, D., \etal 1987, ApJ, 313, 42

\nhi Drozdovsky, I.~O., \& Karachentsev, I.~D.~2000, A\&AS, 142, 425

\nnhi Elson, E. C., de Blok, W. J. G., \& Kraan-Korteweg, R. C. 2010, MNRAS, 404, 2061  

\nnhi Elson, E. C., de Blok, W. J. G., \& Kraan-Korteweg, R. C. 2013, MNRAS, 429, 2550  

\nhi Evans, N. W., An, J., \& Walker, M. G. 2009, MNRAS, 393, L50
 
\nhi Faber, S.~M.~1987, in IAU Symposium 117, Dark Matter in the Universe,
     ed.~J.~Kormendy \& G.~R.~Knapp (Dordrecht: Reidel), 1

\nhi Faber, S.~M., Dressler, A., Davies, R.~L., \etal 1987, in Nearly Normal
           Galaxies: From the Planck Time to the Present, ed.~S.~M.~Faber 
           (New York: Springer), 175

\nnhi Ferguson, H.~C., \& Binggeli, B.~1994, A\&AR, 6, 67

\nnhi Fisher, D. B., \& Drory, N. 2011, ApJ, 733, L47

\nhi Flores, R., Primack, J.~R., Blumenthal, G.~R., \& Faber, S.~M.~1993, ApJ,
     412, 443

\nhi Freedman, W.~L., Madore, B. F., Gibson, B. K., \etal 2001, ApJ, 553, 47

\nnhi Freeman, K.~C.~1970, ApJ, 160, 811

\nnhi Freeman, K.~C.~1987, in Nearly Normal Galaxies: From the Planck 
           Time to the Present, ed. S.~M.~Faber (New York: Springer), 317

\nnhi Garrison-Kimmel, S., Boylan-Kolchin, M., Bullock, J. S., \& Kirby, E. N.~2014, MNRAS (arXiv:1404.5313)

\nnhi Garrison-Kimmel, S., Rocha, M., Boylan-Kolchin, M., Bullock, J. S., \& Lally, J.~2013, MNRAS, 433, 3539

\nnhi Gebhardt, K., Richstone, D., Tremaine, S., \etal 2003, ApJ, 583, 92

\nnhi Gentile, G., Famaey, B., Zhao, H., \& Salucci, P. 2009, Nature, 461, 627  

\nhi Gentile, G., Salucci, P., Klein, U., \& Granato, G.~L.~2007, MNRAS, 375, 199 

\nhi Gentile, G., Salucci, P., Klein, U., Vergani, D., \& Kalberla, P.~2004, MNRAS, 351, 903

\nnhi Gilmore, G., Wilkinson, M. I., Wyse, R. F. G., \etal 2007, ApJ, 663, 948

\nnhi Gnedin, O. Y., \& Zhao, HS., 2002, MNRAS, 333, 299  

\nnhi Gott, J.~R., \& Rees, M.~J.~1975, A\&A, 45, 365

\nnhi Governato, F., Brook, C., Mayer, L., \etal 2010, Nature, 463, 203

\nnhi Graham, A. W., Merritt, D., Moore, B., Diemand, J., \& Terzi\'c, B., 2006, AJ, 132, 2711  

\nnhi Hayashi, E., Navarro, J. F., Power, C., \etal 2004, MNRAS, 355, 794

\nhi Herrmann, K., \& Ciardullo, R. 2009, ApJ, 705, 1686

\nnhi Irwin, M., \& Hatzidimitriou, D.~1995, MNRAS, 277, 1354 (IH95)

\nnhi Irwin, M.{\ts}J., Belokurov, V., Evans, N.{\ts}W., \etal 2007, ApJ,\ts656,{\ts}L13\ts(I07)

\nnhi Jardel, J. R. 2014, PhD Thesis, University of Texas at Austin

\nnhi Jardel, J. R., \& Gebhardt, K. 2012, ApJ, 746, 89                                                

\nnhi Jardel, J. R., \& Gebhardt, K. 2013, ApJ, 775, L30

\nnhi Jardel, J. R., Gebhardt, K., Fabricius, M. H., Drory, N., \& Williams, M. J. 2013, ApJ, 763, 91  

\nnhi Jardel, J.~R., Gebhardt, K., Shen, J., \etal 2011, ApJ, 739, 21

\nnhi Jobin, M., \& Carignan, C.~1990, AJ, 100, 648

\nhi Karachentsev, I.~D., \& Makarov, D.~A.~1996, AJ, 111, 794  

\nnhi Karachentsev, I. D., Sharina, M. E., Dolphin, A. E., \& Grebel, E. K. 2003, A\&A, 408, 111

\nhi Kennicutt, R.~C., Lee, J. C., Funes, J. G., Sakai, S., \& Akiyama, S. 2008, ApJS, 178, 247

\nhi Kent, S.~M.~1986, AJ, 91, 1301

\nhi Kent, S.~M.~1987, AJ, 93, 816

\nhi Kent, S.~M.~1989, AJ, 97, 1614  

\nnhi King, I.~R.~1966, AJ, 71, 64

\nnhi Kirby, E. N., Simon, J. D., Geha, M., Guhathakurta, P., \& Frebel, A. 2008, ApJ, 685, L43

\nnhi Klessen, R.~S., Grebel, E.~K., \& Harbeck, D.~2003, ApJ, 589, 798

\nnhi Klypin, A., Kravtsov, A. V., Bullock, J. S., \& Primack, J. R. 2001, ApJ, 554, 903

\nnhi Klypin,{\ts}A.,{\ts}Kravtsov,{\ts}A.{\ts}V.,{\ts}Valenzuela,{\ts}O.,\ts\&{\ts}Prada, {\ts}F.\ts1999,{\ts}ApJ,\ts522,\ts82

\nnhi Klypin, A. A., Trujillo-Gomez, S., \& Primack, J. 2011, ApJ, 740, 102

\nnhi Knapp, G.~R., Kerr, F.~J., \& Bowers, P.~F.~1978, AJ, 83, 360

\nnhi Koch, A., Wilkinson, M. I., Kleyna, J. T., \etal 2007, ApJ, 657, 241  

\nhi Komatsu, E., Dunkley, J., Nolta, M. R., \etal 2009, ApJS, 180, 330

\nhi Komatsu, E., Smith, K. M., Dunkley, J., \etal 2011, ApJS, 192, 18

\nnhi Kormendy, J.~1982, in Morphology and Dynamics of Galaxies, ed.~L. Martinet \& M.~Mayor (Sauverny: Geneva Observatory), 113

\nnhi Kormendy, J.~1984, ApJ, 287, 577

\nnhi Kormendy, J.~1985, ApJ, 295, 73

\nnhi Kormendy, J.~1987a, in IAU Symposium 117, Dark Matter in the
           Universe, ed. J.~Kormendy \& G.~R.~Knapp (Dordrecht: Reidel), 139

\nnhi Kormendy, J.~1987b, in IAU Symposium 127, Structure and Dynamics
           of Elliptical Galaxies, ed.~T.~de Zeeuw (Dordrecht: Reidel), 17

\nnhi Kormendy, J.~1987c, in Nearly Normal Galaxies: From the Planck Time 
           to the Present, ed. S.~M.~Faber (New York: Springer), 163

\nnhi Kormendy, J.~1988, in Guo Shoujing Summer School of Astrophysics;
      Origin, Structure and Evolution of Galaxies, ed.~Fang Li Zhi 
      (Singapore:~World Scientific), 252

\nnhi Kormendy, J. 1989, ApJ, 342, L63

\nnhi Kormendy, J.~1990, in The Edwin Hubble Centennial Symposium: The
      Evolution of the Universe of Galaxies, 
      ed.~R. G. Kron (San Francisco:~ASP), 33

\nnhi Kormendy, J. 2014, in Lessons from the Local Group, A Conference in Honour of 
      David Block and Bruce Elmegreen, ed. K. C. Freeman, B. G. Elmegreen, 
      D. L. Block, \& M. Woolway (New York: Springer), 319

\nnhi Kormendy, J., \& Bender, R.~1994, in ESO/OHP Workshop on
     Dwarf Galaxies, ed.~G.~Meylan \& P.~Prugniel (Garching: ESO), 161

\nnhi Kormendy, J., \& Bender, R.~2009, ApJ, 691, L142

\nnhi Kormendy, J., \& Bender, R.~2011, Nature, 469, 377

\nnhi Kormendy, J., \& Bender, R. 2012, ApJS, 198, 2

\nnhi Kormendy, J., \& Djorgovski, S.~1989, ARA\&A, 27, 235

\nhi Kormendy, J., Dressler, A., Byun, Y.-I., \etal 1994,
           in ESO/OHP Workshop on Dwarf Galaxies, ed.~G.~Meylan \& P.~Prugniel
           (Garching: ESO), 147

\nnhi Kormendy, J., Drory, N., Bender, R., \& Cornell, M.{\ts}E.~2010, ApJ, 723,~54

\nnhi Kormendy, J., Fisher, D. B., Cornell, M. E., \& Bender, R. 2009, ApJS, 182, 216 (KFCB)

\nnhi Kormendy, J., \& Freeman, K.~C.~1996, in Ringberg Proceedings 1996
           of Sonderforschungsbereich 375, Research in Particle Astrophysics,
           ed.~R.~Bender et~al. (Munich:~Technische Universit\"at M\"unchen,
           Ludwig-Maximilians-Universit\"at, Max-Planck-Institut f\"ur Physik,
           \& Max-Planck-Institut f\"ur Astrophysik), 13

\nnhi Kormendy, J., \& Freeman, K.~C.~2004, in IAU Symposium 220, Dark Matter
      in Galaxies, ed.~S.~D.~Ryder, D.~J.~Pisano, M.~A.~Walker, \& K.~C.~Freeman
      (San Francisco: ASP), 377

\nnhi Kormendy, J., \& Kennicutt, R. C. 2004, ARA\&A, 42, 603

\nnhi Kraan-Korteweg, R.~C.~1986, A\&AS, 66, 255

\nnhi Kranz, T., Slyz, A., \& Rix, H.-W.~2003, ApJ, 586, 143

\nnhi Kuhn, J.~R.~1993, ApJ, 409, L13

\nnhi Kuhn, J.~R., \& Miller, R.~H.~1989, ApJ, 341, L41

\nnhi Kuzio de Naray, R., McGaugh, S.~S., \& de Blok, W.~J.~G.~2008, ApJ, 676, 920   

\nnhi Lake, G., \& Feinswog, L.~1989, AJ, 98, 166

\nnhi Lauer, T.~R.~1985, ApJ, 292, 104
 
\nnhi Lauer, T.~R., Ajhar, E. A., Byun, Y.-I., et al.~1995, AJ, 110, 2622

\nhi Lee, M.~G., Freedman, W., Mateo, M., \etal 1993, AJ, 106, 1420 

\nnhi Madau, P., Shen, S., \& Governato, F. 2014, ApJ, 789, L17

\nnhi Marchesini, D., D'Onghia, E., Chincarini, G., \etal 2002, ApJ, 575, 801

\nnhi Martimbeau, N., Carignan, C., \& Roy, J.-R.~1994, AJ, 107, 543 

\nnhi Martin, N. F., de Jong, J. T. A., \& Rix, H.-W. 2008, ApJ, 684, 1075 (M08)

\nnhi Mateo, M.~1998, ARA\&A, 36, 435 (M98)

\nnhi Mateo, M., Olszewski, E.~W., Vogt, S.~S., \& Keane, M.~J.~1998, AJ, 116, 2315

\nnhi Meurer, G.~R., Carignan, C., Beaulieu, S.~F., \& Freeman, K.~C.~1996, AJ, 111, 1551

\nnhi Meurer, G.~R., Staveley-Smith, L., \& Killeen, N.~E.~B.~1998, MNRAS, 300, 705  

\nhi Mighell, K.~J.~1990, A\&AS, 82, 1

\nhi Mighell, K.~J., \& Butcher, H.~R.~1992, A\&A, 255, 26

\nnhi Milgrom, M.~1983a, ApJ, 270, 365

\nnhi Milgrom, M.~1983b, ApJ, 270, 371

\nnhi Milgrom, M.~1983c, ApJ, 270, 384

\nnhi Milgrom, M., \& Bekenstein, J.~1987, in IAU Symposium 117, Dark 
      Matter in the Universe, ed.~J.~Kormendy \& G.~R.~Knapp (Dordrecht: 
      Reidel), 319

\nnhi Miller, B.~W., \& Rubin, V.~C.~1995, AJ, 110, 2692  

\nnhi Moore, B.~1994, Nature, 370, 629

\nnhi Moore, B., Ghigna, S., Governato, F., et al.~1999, ApJ, 524, L19

\nnhi Moore, B., Calc\'aneo-Rold\'an, C., Stadel, J., \etal 2001, PhRevD, 64, 063508  

\nnhi Moore, B., Governato, F., Quinn, T. R., Stadel, J., \& Lake, G. 1998, ApJ, 499, L5

\nhi Mould, J.~R., Huchra, J. P., Freedman, W. L., \etal 2000, ApJ, 529, 786

\nnhi Napolitano, N. R., Romanowsky, A. J., Tortora, C. 2010, MNRAS, 405, 2351  

\nhi Navarro, J.~F., Eke, V.~R., \& Frenk, C.~S.~1996, MNRAS, 283, L72

\nnhi Navarro,{\ts}J.{\ts}F., Frenk,{\ts}C.{\ts}S., \& White,{\ts}S.{\ts}D.{\ts}M. 1996,{\ts}ApJ,\ts462,\ts563~(NFW)

\nnhi Navarro,{\ts}J.{\ts}F., Frenk,{\ts}C.{\ts}S., \& White,{\ts}S.{\ts}D.{\ts}M.~1997,{\ts}ApJ,\ts490,\ts493~(NFW)

\nnhi Navarro, J. F., Hayashi, E., Power, C., \etal 2004, MNRAS, 349, 1039  

\nnhi Navarro, J. F., Ludlow, A., Springel, V., \etal 2010, MNRAS, 402, 21    

\nnhi Noordermeer, E., van der Hulst, J. M., Sancisi, R., Swaters, R. S., \& van Albada, T. S. 2007,
             MNRAS, 376, 1513

\nnhi Noordermeer, E. 2008, MNRAS, 385, 1359

\nnhi O'Brien,{\ts}J.{\ts}C.,{\ts}Freeman,{\ts}K.{\ts}C.,\ts\&{\ts}van{\ts}der{\ts}Kruit,{\ts}P.{\ts}C.\ts2010,{\ts}A\&A,\ts515,{\ts}A63

\nnhi Oh, K.~S., Lin, D.~N.~C., \& Aarseth, S.~J.~1995, ApJ, 442, 142

\nnhi Oh, S.-H., Brook, C., Governato, F., \etal 2011a, AJ, 142, 24

\nnhi Oh, S.-H., de Blok, W.~J.~G., Brinks, E., Walter, F., \& Kennicutt, R.~C.~2011b, 
       AJ, 141, 193

\nnhi Oh, S.-H., de Blok, W.~J.~G., Walter, F., Brinks, E., \& Kennicutt, R.~C.~2008, 
      AJ, 136, 2761

\nnhi Palma, C., Majewski, S. R., Siegel, M. H., et al.~2003, AJ, 125, 1352

\nhi Paturel, G., Petit, C., Prugniel, P., \etal 2003, A\&A, 412, 45

\nnhi Peebles, P.~J.~E.~1974, ApJ, 189, L51

\nnhi Persic, M., Salucci, P., \& Stel, F.~1996, MNRAS, 281, 27   

\nnhi Persic, M., Salucci, P., Bertola, F., \& Pizzella, A.~1997, in
       The Second Stromlo Symposium:~The Nature of Elliptical~Galaxies,
       ed. M.~Arnaboldi, G.~S.~Da Costa, \& P.~Saha (San Francisco:~ASP), 151

\nnhi Piatek, S., \& Pryor, C.~1995, AJ, 109, 1071

\nnhi Piatek, S., Pryor, C, Armandroff, T.~E., \& Olszewski, E.~W.~2001,
      AJ, 121, 841

\nnhi Piatek, S., Pryor, C, Armandroff, T.~E., \& Olszewski, E.~W.~2002,
      AJ, 123, 2511

\nnhi Plana, H., Amram, P., Mendes de Oliveira, C., \& Balkowski, C. 2010,
      AJ, 139, 1  

\nnhi Puche, D., Carignan, C., \& Wainscoat, R.~J.~1991, AJ, 101, 447

\nnhi Puglielli, D., Widrow, L. M., \& Courteau, S. 2010, ApJ, 715, 1152  

\nnhi Regan, M. W., Thornley, M. D., Helfer, T. T., \etal 2001, ApJ, 561, 218

\nnhi Rhee, M.-H. 1996, PhD Thesis, Rijksuniversiteit te Groningen

\nnhi Richstone, D., Gebhardt, K., Aller, M., \etal 2004, ArXiv:astro-ph/ 0403257

\nnhi Ryan-Weber, E. V., Begum, A., Oosterloo, T., \etal 2008, MNRAS, 384, 535 (R08) 

\nhi Ryden, B.~S., \& Gunn, J.~E.~1987, ApJ, 318, 15  

\nnhi Ryder, S.~D., Pisano, D.~J., Walker, M.~A., \& Freeman, K.~C., eds.~2004, IAU
      Symposium 220, Dark Matter in Galaxies (San Francisco: ASP)

\nnhi Sackett, P.~D.~1997, ApJ, 483, 103

\nhi Salucci, P., \& Burkert, A.~2000, ApJ, 537, L9

\nhi Salucci, P., \& Persic, M.~1999, A\&A, 351, 442

\nnhi Salucci, P., Wilkinson, M. I., Walker, M. G., \etal 2012, MNRAS, 420, 2034

\nhi Sancisi, R., \& van Albada, T.~S.~1987, in IAU Symposium 117,
              Dark Matter in the Universe, ed.~J.~Kormendy \&
              G.~R.~Knapp (Dordrecht: Reidel), 67

\nhi Sanders, R. H., \& McGaugh, S. S. 2002, ARA\&A, 40, 263

\nnhi Schlegel, D. J., Finkbeiner, D. P., \& Davis, M. 1998, ApJ, 500, 525  

\nnhi Schwarzschild, M. 1979, ApJ, 232, 236

\nnhi Schwarzschild, M. 1982, ApJ, 263, 599

\nnhi Sellwood, J.~A.~2004, in IAU Symposium 220, Dark Matter in Galaxies,
      ed.~S.~D.~Ryder, D.~J.~Pisano, M.~A.~Walker, \& K.~C.~Freeman (San Francisco:
      ASP), 27

\nnhi Sellwood, J. A., 2009, in IAU Symposium 254, The Galaxy Disk in Cosmological Context,
      ed. J. Andersen, J. Bland-Hawthorn \& B. Nordstr\"om (Cambridge: Cambridge Univ. Press.), 73

\nnhi Sellwood, J. A., \& McGaugh, S. S. 2005, ApJ, 634, 70  

\nnhi Sellwood, J.~A., \& Moore, E.~M.~1999, ApJ, 510, 125

\nnhi Sellwood, J.~A., \& Pryor, C.~1998, Highlights Astron., 11, 638

\nnhi Shapiro, P.~R., \& Iliev, I.~T.~2002, ApJ, 565, L1

\nnhi Sicotte, V., \& Carignan, C.~1997, AJ, 113, 609  

\nhi Simon, J. D., \& Geha, M. 2007, ApJ, 670, 313

\nnhi Siopis, C., Gebhardt, K., Lauer, T. R., \etal 2009, ApJ, 693, 946

\nhi Sofue, Y.~1996, ApJ, 458, 120

\nnhi Sofue, Y., Honma, M., \& Omodaka, T. 2009, PASJ, 61, 227  

\nnhi Spano, M., Marcelin, M., Amram, P., \etal 2008, MNRAS, 383, 297  

\nnhi Springel, V., Wang, J., Vogelsberger, M., \etal 2008, MNRAS, 391, 1685  

\nnhi Swaters, R.~A., Madore, B.~F., \& Trewhella, M.~2000, ApJ, 531, L107

\nnhi Swaters, R.{\thinspace}A., Madore, B.{\thinspace}F., 
      van den Bosch, F.{\thinspace}C., \& Balcells, M.~2003, ApJ, 583, 732

\nnhi Taga, M., \& Iye, M.~1994, MNRAS, 271, 427

\nhi Thomas, J. 2010, Astron. Nachr., 88, 1 (arXiv:1007.3591) 

\nnhi Thomas, J., Saglia, R. P., Bender, R., \etal 2004, MNRAS, 353, 391

\nnhi Thomas, J., Saglia, R. P., Bender, R., \etal 2005, MNRAS, 360, 1355

\nnhi Thomas, J., Saglia, R. P., Bender, R., \etal 2009, ApJ, 691, 770  

\nnhi Tollerud, E. J., Bullock, J. S., Strigari, L. E., \& Willman, D. 2008, ApJ, 688, 277

\nhi Tolstoy, E., Hill, V., \& Tosi, M. 2009, ARA\&A, 47, 371

\nnhi Tonry, J.~L., Dressler, A., Blakeslee, J.~P., \etal 2001, ApJ, 546, 681

\nnhi Toomre, A.~1981, in The Structure and Evolution of Normal Galaxies,
     ed.~S.~M.~Fall \& D.~Lynden-Bell (Cambridge: Cambridge University Press), 111

\nnhi Tremaine, S., Gebhardt, K., Bender, R., \etal 2002, ApJ, 574, 740

\nhi Tully, R.~B.~1988, Nearby Galaxies Catalog (Cambridge: Cambridge University Press)

\nnhi Tully, R.~B., \& Fisher, J.~R.~1977, A\&A, 54, 661

\nnhi Tully, R.~B., \& Fouqu\'e, P.~1985, ApJS, 58, 67

\nhi  Tully, R.~B., \& Pierce, M.~J.~2000, ApJ, 533, 744

\nnhi Tully, R.~B., Somerville, R.~S., Trentham, N., \& Verheijen, M.~A.~W.~2002, 
      ApJ, 569, 573

\nnhi van Albada, T.~S., Bahcall, J.~N., Begeman, K., \& Sancisi, R.~1985, 
      ApJ, 295, 305

\nnhi van Albada, T.~S., \& Sancisi, R.~1986, Phil.~Trans.~R.~Soc.~London A, 320, 447

\nhi van der Kruit, P.~C. 2010, in AIP Conference Proceedings, Volume 1240,
     Hunting for the Dark: The Hidden Side of Galaxy Formation,
     ed. V. P. Debattista \& C. C. Popescu (New York: AIP), 387

\nhi van der Kruit, P.~C., \& Freeman, K.~C.~2011, ARA\&A, 49, 301

\nnhi van Zee, L. 2000, AJ, 119, 2757 (Z00)

\nnhi van Zee, L., Haynes, M.~P., Salzer, J.~J., \& Broeils, A.~H.~1997, AJ, 113, 1618 

\nnhi Verdes-Montenegro, L., Bosma, A., \& Athanassoula, E. 1997, A\&A, 321, 754  

\nnhi Verheijen, M.~A.~W.~1997, PhD Thesis, Rijksuniversiteit te Groningen

\nhi Visser, H.~C.~D.~1980, A\&A, 88, 159


\nnhi Walker, M. G., Mateo, M., Olszewski, E. W., \etal 2009, ApJ, 704, 1274; Erratum: 2010, ApJ, 710, 886 (W09)

\nnhi Walker, M. G., McGaugh, S. S., Mateo, M., Olszewski, E. W., \&  Kuzio de Naray, R. 2010, ApJ, 717, L87  

\nnhi Walsh, S. M., Willman, B., \& Jerjen, H. 2009, AJ, 137, 450

\nnhi Weiner, B.~J.,~2004, in IAU Symposium 220, Dark Matter in Galaxies,
      ed.~S.~D.~Ryder, D.~J.~Pisano, M.~A.~Walker, \& K.~C.~Freeman (San Francisco: ASP),
      265

\nnhi Weiner, B.~J., Sellwood, J.~A., \& Williams, T.~B.~2001, ApJ, 546, 931

\nnhi Weisz, D. R., Dalcanton, J. J., Williams, B. F., \etal 2011, ApJ, 739, 5

\nnhi Weldrake,{\ts}D.{\ts}T.{\ts}F., de{\ts}Blok,{\ts}W.{\ts}J.{\ts}G., \& Walter,{\ts}F. 2003, MNRAS, 340, 12

\nnhi Wilkinson, M.~I., Kleyna, J. T., Evans, N. W., \& Gilmore, G. F.~2004, in IAU Symposium 220, 
      Dark Matter in Galaxies,
      ed.~S.~D.~Ryder, D.{\ts}J.{\ts}Pisano, M.{\ts}A.{\ts}Walker, \& K.{\ts}C.{\ts}Freeman (San Francisco:~ASP),~359

\nnhi Yang, Y., Hammer, F., Fouquet, S., \etal 2014, MNRAS, 442, 2419

\nhi York, D. G., Adelman, J., Anderson, J. E., \etal 2000, AJ, 120, 1579

\nnhi Yoshino, A., \& Ichikawa, T. 2008, PASJ, 60, 493

\nnhi Young, L. M., \& Lo, K. Y. 1996, ApJ, 462, 203 (Y96)

\nnhi Young, L. M., \& Lo, K. Y. 1997, ApJ, 490, 710 (Y97)

\nnhi Young, L. M., van Zee, L., Lo, K. Y., Dohm-Palmer, R. C., \& Beierle, M. E. 2003, ApJ, 592, 111 (Y03)
 
\vfill\eject

\end